\documentclass[a4paper,11pt]{article}
\usepackage{jcappub}
\usepackage{amsmath}
\usepackage{amssymb}

\DeclareRobustCommand{\ion}[2]{%
\relax\ifmmode
\ifx\testbx\f@series
{\mathbf{#1\,\mathsc{#2}}}\else
{\mathrm{#1\,\mathsc{#2}}}\fi
\else\textup{#1\,{\mdseries\textsc{#2}}}%
\fi}

\usepackage{enumitem}
\setlist[enumerate]{itemsep=1mm}

\newcommand{\lya}{Lyman-$\alpha$~}
\newcommand{\Lya}{Lyman-$\alpha$~}
\title{\boldmath Cosmological Constraints from the eBOSS \lya Forest using the PRIYA Simulations}

\author{M.A. Fernandez,}\emailAdd{mfern027@ucr.edu}
\author{Simeon Bird}\emailAdd{sbird@ucr.edu}
\author{and Ming-Feng Ho,}\emailAdd{mho026@ucr.edu}
\affiliation{Department of Physics and Astronomy, University of California Riverside, 900 University Ave, Riverside, CA 92521}

\abstract{
We present new cosmological parameter constraints from the eBOSS \lya forest survey.
We use a new theoretical model and likelihood based on the PRIYA simulation suite.
PRIYA is the first suite to resolve the \lya forest in a ($120$~Mpc/h~)$^3$ volume, using a multi-fidelity emulation technique.
We use PRIYA to predict \lya forest observables with $\lesssim 1\%$ interpolation error over an $11$ dimensional ($9$ simulated, $2$ in post-processing) parameter space.
We identify an internal tension within the flux power spectrum data. 
Once the discrepant data is removed, we find the primeval scalar spectral index measured at a pivot scale of $k_0 = 0.78$ Mpc$^{-1}$ to be $n_P = 1.009^{+0.027}_{-0.018}$ at 68\% confidence.
This measurement from the \lya forest flux power spectrum alone is in reasonable agreement with Planck, and in tension with earlier eBOSS analyses.
The amplitude of matter fluctuations is $\sigma_8 = 0.733^{+0.026}_{-0.029}$ at 68\% confidence, in agreement with Dark Energy Survey weak lensing measurements and other small-scale structure probes and in tension with CMB measurements from Planck and ACT.
The effective optical depth to \lya photons from our pipeline is in good agreement with earlier high resolution measurements.
We find a linear power at $z=3$ and $k = 0.009$ s/km of $\Delta_L^2 = 0.302^{+0.024}_{-0.027}$ with a slope $n_\mathrm{eff} = -2.264^{+0.026}_{-0.018}$.
Our flux power spectrum only chains prefer a low level of heating during helium reionization.
When we add IGM temperature data we find $n_P = 0.983\pm 0.020$ and $\sigma_8 = 0.703^{+0.023}_{-0.027}$.
Our chains prefer an early and long helium reionization event, as suggested by measurements from the helium \lya forest.
In the near future we will use our pipeline to infer cosmological parameters from the DESI \lya data.
}

\begin{document}
\maketitle
\flushbottom
% - Different seed cosmic variance chain added to section 3.3
% - Plot like loo_vs_emu_error_wlegend.pdf but for z=2.2 - 2.6 and use it as main result, rather than the shrunk corner plots.
% - Drop LOO from the 'full posteriors' figure. Add 'with mean temperature'? Or just leave it alone?
%
% Main conclusion:
% There is a discrepancy between the z=2.2-2.4 and z >= 2.6 bins. This drives tensions across the whole parameter space. Clear from the best fit.

\section{Introduction}\label{sec:intro}

The \lya forest \citep{1965ApJ...142.1633G, 1998ApJ...495...44C, 1998MNRAS.301..478T, 2000ApJ...543....1M, 2001ApJ...552...15H, 2002MNRAS.329..848V, 2006AJ....132..117F, 2006MNRAS.365..231V, 2006ApJS..163...80M} measures the distribution of neutral gas at relatively low densities.
This gas traces the growth of cosmic structure, making the \lya~forest an exceptionally powerful cosmological probe, sensitive to the distribution of dark matter deep in the matter dominated era.
Correlating absorption from different quasar sightlines has allowed detection of the baryon acoustic oscillations and constraints on the expansion of the universe \cite{2011JCAP...09..001S, 2013JCAP...04..026S, 2020ApJ...901..153D, 2022arXiv220913942C}.
The densities probed by the \lya forest from redshift $z=2-5$ are $\sim 1-100 \ \times$ the cosmological mean density.
For these redshifts and densities stellar winds and star formation effects are negligible, though feedback from black holes can be important \citep{2013MNRAS.429.1734V, 2020MNRAS.495.1825C}.
Thus the \lya forest is able to measure the primordial fluctuations on some of the smallest scales available, $k \sim 1$ h/Mpc \citep{2004MNRAS.354..684V, 2005ApJ...635..761M, 2006MNRAS.370L..51V, 2005PhRvD..71j3515S, 2006JCAP...10..014S, 2017JCAP...06..047Y, 2020JCAP...04..038P, 2021JCAP...03..049G}.
In addition, the \lya~forest is sensitive to the thermal and ionization history of the intergalactic medium (IGM) \citep{2008MNRAS.386.1131B,2014MNRAS.438.2499B, 2016MNRAS.463.2335N,2019ApJ...872...13W, 2019ApJ...872..101B, 2019MNRAS.490.3177W,2021MNRAS.506.4389G, 2022ApJ...933...59V}, and by constraining the smallest structures, the mass scale of thermal relic dark matter \citep{2005PhRvD..71f3534V,  2013PhRvD..88d3502V, 2017PhRvD..96b3522I, 2020JCAP...04..038P, 2021MNRAS.502.2356G, 2021PhRvL.126g1302R, 2022arXiv220914220V}.

The extended Baryon Oscillation Sky Survey (eBOSS), part of the Sloan Digital Sky Survey (SDSS) \cite{2019JCAP...07..017C}, has computed the 1D flux power spectrum along quasar sight lines for over $43,000$ quasars, with a statistical error $\sim 1\%$ at some redshifts.
This exceptional statistical error means that the error budget is dominated by systematic uncertainty, especially uncertainty in the resolution of the spectrograph on small scales \cite{2019JCAP...07..017C}.
The Dark Energy Spectroscopic Instrument (DESI) has improved the spectrograph resolution by a factor of two \cite{2022AJ....164..207A}.
Thus, early data from DESI has measured the flux power spectrum at smaller scales ($k \gtrsim 0.035$ km$^{-1}$ s) than SDSS \cite{2023arXiv230606316G, 2023arXiv230606311R}.
Future releases will measure higher redshifts ($z>4.6$) and increase the number of \lya forest quasar spectra by a factor of four over SDSS \cite{2016arXiv161100036D}.

There are other high resolution, small sample datasets of quasar spectra, from which \lya forest flux power measurements have been made \cite{2017MNRAS.466.4332I, 2019MNRAS.489.2536D, 2022MNRAS.509.2842K, 2022MNRAS.515..857E}. Ref.~\cite{2022MNRAS.509.2842K} used spectra from multiple surveys (XQ-100, KODIAQ, and SQUAD) to measure the \lya forest flux power at redshifts $z=2-4.6$ and scales $k\approx0.005-0.1$ km$^{-1}$ s (albeit with larger uncertainty than eBOSS), and Ref.~\cite{2022MNRAS.515..857E} presents recent cosmological constraints from these datasets.
%Combining these higher resolution observations with the larger sample size observations will allow inference based on a broader range of scales than either data set allows separately.

Modeling the \lya forest requires numerical simulations that are able to follow the distribution of gas on small scales.
In this paper we present cosmological parameter inference using a new likelihood built on the PRIYA simulation suite \cite{2023simsuite}.
The PRIYA simulations are in $120$ Mpc/h boxes, and are comprised of \textbf{$60$} simulations with $2\times 1536^3$ particles (mean inter-particle spacing of $78$ kpc/h), as well as $3$ simulations with $2\times 3072^3$ particles (mean inter-particle spacing of $39$ kpc/h).
The higher of these two resolutions exceeds the resolution of state-of-the-art galaxy formation simulations such as Illustris-TNG \cite{2018MNRAS.475..676S}.
PRIYA is run with the same highly scalable MP-Gadget code as the ASTRID simulation \cite{2022MNRAS.512.3703B,2022MNRAS.513..670N}.
PRIYA contains full hydrodynamic simulations with models of galaxy formation and black hole feedback to $z=2.2$.
PRIYA is thus the first cosmological simulation suite which achieves, in a single box, the required box size of $120$ Mpc/h, capable of minimising sample variance in the \lya forest \cite{2014JCAP...07..005B}, and a resolution high enough that it includes the pressure smoothing scale.\footnote{Our model boosts gas temperature during reionization, increasing the smoothing scale \cite{2023simsuite}.}.
Importantly, this removes the need for the `splicing' correction used in earlier work to combine different boxsizes into a single whole \cite{2014JCAP...07..005B,2020JCAP...04..038P}.

Here, the PRIYA simulations are used to build multi-fidelity emulators \cite{2019JCAP...02..050B, 2022MNRAS.509.2551H, 2022MNRAS.517.3200F} for the flux power spectrum and the IGM temperature at mean density.
Each emulator is a surrogate model, able to reproduce the 1D flux power spectrum or mean IGM temperature for cosmological parameters (within the prior simulation volume) to $\sim 1 \%$ accuracy.
A multi-fidelity emulator combines two different resolution training samples.
Many low fidelity samples are used to explore parameter space, and their output is corrected with a few high fidelity samples.
A multi-fidelity emulator makes predictions for the highest resolution simulation at a fraction of the computational cost of a single fidelity emulator \cite{10.1093/biomet/87.1.1, 2022MNRAS.509.2551H}.
Emulators have been used to study various cosmological probes: the matter power spectrum \citep{Heitmann:2009, Heitmann:2014, Lawrence:2017, Giblin:2019, Euclid:2021, Arico:2021, Giri:2021}, weak lensing shear \citep{Harnois:2019, Davies:2021}, the halo mass function \citep{McClintock:2019, Nishimichi:2019, Bocquet:2022}, the 21-cm signal \citep{Kern:2017, Cohen:2020, Bevins:2021, Bye:2022} and the \lya forest \citep{2019JCAP...02..050B, Rogers:2019, 2021JCAP...05..033P, 2021JCAP...04..059W, Rogers:2021a,2021PhRvL.126g1302R, 2023MNRAS.tmp.2406C}.

Here, we present the first fully resolved multi-fidelity emulator based likelihood framework for the eBOSS \lya forest and the first cosmological constraints derived from it.
Our multi-fidelity emulator is similar to that described in Ref.~\cite{2022MNRAS.517.3200F}, but the simulation volume has been increased by a factor of $64$, and the spatial resolution has been improved by a factor of $1.5$.
We also use mean IGM temperature data \cite{2021MNRAS.506.4389G} to constrain the parameters of helium reionization, data which is ultimately derived from higher resolution quasar surveys \citep{2017MNRAS.466.4332I, 2022MNRAS.509.2842K, 2019MNRAS.489.2536D}.

In summary, our method is: (1) Construct an emulator for the 1D \lya flux power spectrum and mean IGM temperature using the PRIYA simulations \cite{2023simsuite}, Section~\ref{sec:emulator}.
(2) Augment observational errors with estimates of the residual theoretical uncertainty to build a covariance matrix, and correct the flux power spectra for metal contamination as described in Section~\ref{sec:inference}.
(3) Use this emulator and likelihood to constrain cosmological parameters using Markov Chain Monte Carlo (MCMC), with results described in Section~\ref{sec:results}.
We discuss some caveats and compare to earlier work in Section~\ref{sec:discussion} and our conclusions are presented in Section~\ref{sec:conclusions}.

MCMC chains for all the results presented in this work along with files containing the training outputs used to construct the emulators\footnote{\url{https://github.com/mafern/InferenceLyaData}}, as well as the code\footnote{\url{https://github.com/sbird/lya_emulator}}, which includes the emulator, likelihood, and integration with the Cobaya MCMC package, are available publicly.

\section{Simulation Suite and Emulator}
\label{sec:emulator}
\label{sec:simulations}

\begin{table}
	\centering
     \begin{tabular}{|c|c|c|c|c|}
		\hline
		Simulation & Box Volume & N$_{\text{gas}}$ & M$_{\text{gas}}$ (M$_{\odot}$ h$^{-1}$)\\
		\hline
		LF & $(120$ Mpc h$^{-1})^3$ & $1536^3$ & $[5.29, 6.98]\times10^6$\\
		HF & $(120$ Mpc h$^{-1})^3$ & $3072^3$ & $[6.73, 7.97]\times10^5$\\
		\hline
	\end{tabular}
    \caption{\label{table:simulations}
    Low-Fidelity (LF) and High-Fidelity (HF) simulation suite details.
    N$_{\text{gas}}$ is the number of gas particles simulated, M$_{\text{gas}}$ is the resulting mass resolution of those particles.}
\end{table}

In this Section, we briefly describe the properties of the simulations and emulator, and refer the reader to Ref.~\cite{2023simsuite} for the full details.
The emulator allows predictions for the output of a simulation at an arbitrary set of cosmological parameters within our prior volume with an average interpolation error of $0.2\%$ at low fidelity and $1\%$ at high fidelity.
Our multi-fidelity emulator combines simulations at different resolutions, following the scheme outlined in Ref.~\cite{2022MNRAS.517.3200F}.
The emulator combines low fidelity (LF) and high fidelity (HF) simulations.
Box volume, number of gas particles, and gas particle mass resolution are reported in Table~\ref{table:simulations}.
We ran a total of $60$ low fidelity (LF) and $3$ high fidelity (HF) simulations.
For this work we have added $12$ new LF simulations to those of Ref.~\cite{2023simsuite}, which extend the simulated parameter range to better cover the posterior range allowed by SDSS DR14.
Low fidelity simulations have $1536^3$ particles, while high fidelity simulations have $3072^3$ particles.
Sampled parameters are chosen to maximise spread in parameter space, as described in Ref.~\cite{2023simsuite}. 

Our simulations still have limited resolution and a finite box. Ref.~\cite{2023simsuite} showed that on these scales resolution is important at the $1\%$ level. Compared to the literature, our resolution convergence is slightly better, due to the temperature boost we impart during H~{\sc i} reionization \cite{2019ApJ...874..154D}. The finite box scatters the modes of the 1D flux power spectrum at the $2\%$ level on large scales, due mostly to the limited number of helium reionization bubbles ($30$ Mpc/h across) that can fit into our volume. Section~\ref{sec:theoryerror} describes how we attempt to marginalise out residual cosmic variance.

\begin{figure}
    \centering
    \includegraphics[width=\columnwidth]{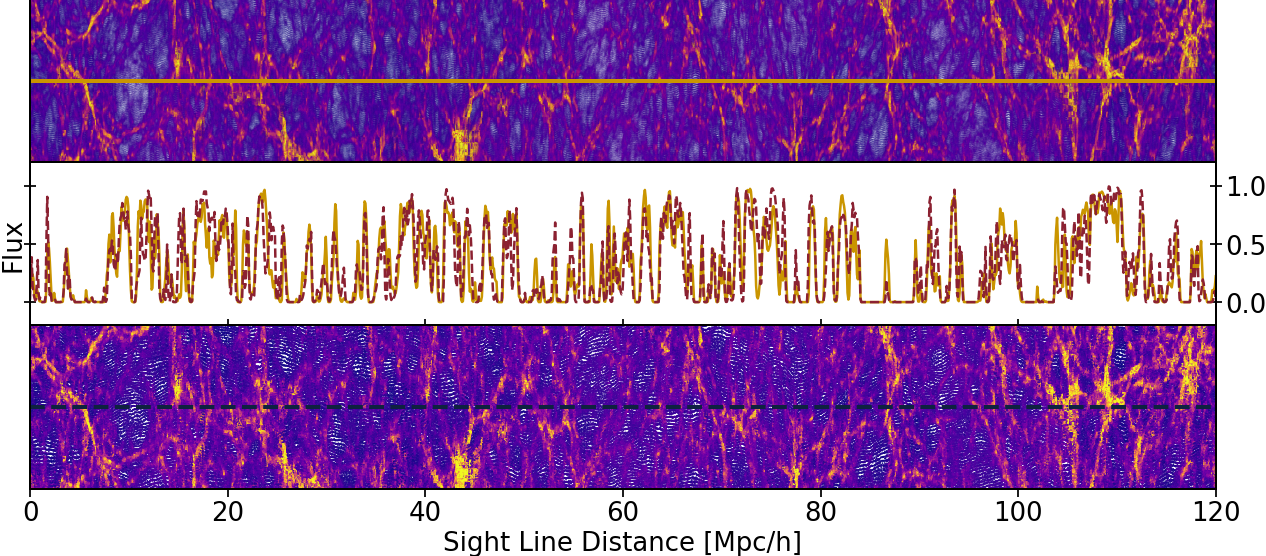}
    \caption{\label{fig:spec_sim}
    Example Lyman-$\alpha$ forest spectra and corresponding gas density and temperature (colors in top and bottom panels) from an LF and HF simulation at redshift $z=4$.
    The top panel shows high resolution and the bottom panel shows low resolution.
    The middle panel shows the \lya forest spectra for the skewers passing through the middle of the top panel (high resolution, yellow line) and bottom panel (low resolution, red dashed line).
    }
\end{figure}

The range given for the gas mass resolution is due to the varying value of $h$ in our simulation suite ($\Omega_b h^2$ is fixed at a value of $0.0224$).
We show in Ref.~\cite{2023simsuite} that this gas mass is sufficient for the scales and redshifts probed by the eBOSS flux power spectrum.
Our simulations include a full galaxy physics model with star formation, stellar and AGN feedback and inhomogeneous reionization models.
Simulations were performed using MP-Gadget\footnotemark, an N-body and smoothed particle hydrodynamics (SPH) code.\footnotetext{\url{https://github.com/MP-Gadget/MP-Gadget}}
MP-Gadget uses the gravitational timestepping algorithm from Gadget-4 \cite{Springel:2021}, and various other algorithmic improvements \cite{2020JCAP...06..002B}.
Simulations are initialised at $z=99$ and finish at $z=2.2$. The galaxy formation model is similar to the \texttt{ASTRID} simulation \cite{2022MNRAS.512.3703B, 2022MNRAS.513..670N} and is described fully in Ref.~\cite{2023simsuite}.

\subsection{Cosmological \& Astrophysical Parameters}\label{sec:parameters}

Table~\ref{tab:emulatorparams} summarises the parameters that are varied across our suite of simulations, as well as their limits.
We have expanded the limits from Ref.~\cite{2023simsuite} with $12$ additional LF simulations covering the parameter ranges $\alpha_q = 2.5 - 3.0$ and $n_P = 1.0 - 1.05$.
This was done so that the emulator covers the $2\sigma$ posterior range for $n_P$.
Simulated parameters were chosen using a Latin hypercube.
We model the primeval (that is, the pre-transfer function) power spectrum $P(k)$  using two parameters: a slope, $n_P$, and an amplitude, $A_P$:
\begin{equation}
    P(k) = A_P \left(\frac{k}{0.78\, \mathrm{Mpc}^{-1}}\right)^{n_P - 1}\,.
\end{equation}
The simulation initial conditions are then generated using a set of transfer functions from CLASS. This parameterization is the same as that used by Planck, but with a different pivot scale, $0.78\, \mathrm{Mpc}^{-1}$, rather than $0.05 \,\mathrm{Mpc}^{-1}$, reflecting the smaller scales probed by the forest.

We also vary the Hubble parameter $h$, and the total matter density through $\Omega_M h^2$, although we will see these are not strongly constrained by the \Lya~forest.
We add three parameters for the He~{\sc ii} reionization model \cite{2020MNRAS.496.4372U}: $z_{Hei}$ and $z_{Hef}$ are the redshifts for the start and end of He~{\sc ii} reionization, and $\alpha_q$ is the quasar spectral index (which scales the peak temperature during He~{\sc ii} reionization).
$z_{Hi}$ is the midpoint redshift of H~{\sc i} reionization.
Finally, $\epsilon_{AGN}$ is the black hole feedback factor, to which the \Lya~forest is insensitive.

\begin{table*}
\begin{centering}
  \begin{tabular}{llll}
  \hline
  Parameter & Minimum & Maximum & Description \\
    \hline
    $n_P$  &  $0.8$  & $1.05$ & Scalar spectral index \\
    $A_P$  &  $1.2 \times 10^{-9}$  & $2.6 \times 10^{-9}$ & Power amplitude at $k = 0.78 \,\mathrm{Mpc}^{-1}$ \\
    $h$    & $0.65$  & $0.75$ & Hubble parameter \\
    $\Omega_0 h^2$ & $0.14$ & $0.146$ & Total matter density \\
    $z_{Hei}$      & $3.5$  & $4.1$  & Start redshift of HeII reionization \\
    $z_{Hef}$      & $2.6$  & $3.2$  & End redshift of HeII reionization \\
    $\alpha_q$     & $1.3$  & $3.0$ & Quasar spectral index during HeII reionization  \\
    $z_{Hi}$        & $6.5$ & $8$   & Median redshift of HI reionization \\
    $\epsilon_{AGN}$ & $0.03$ & $0.07$ & Thermal efficiency of black hole feedback \\
    $\tau_0$ & $0.75$ & $1.25$ & Mean optical depth at $z=3$ in Eq.~\ref{eq:meanflux}.\\
    $d \tau_0$ & $-0.4$ & $0.25$ & Mean optical depth redshift evolution in Eq.~\ref{eq:meanflux}. \\
    \hline
  \end{tabular}
  \caption{Summary of likelihood function parameters, together with the ranges covered by the emulator. We vary a total of $11$ parameters: $4$ for cosmology, $3$ for helium reionization, $1$ for hydrogen reionization, $1$ for the strength of AGN feedback and $2$ for the mean optical depth.}
  \label{tab:emulatorparams}
  \end{centering}
\end{table*}

There are two further parameters for the \Lya~effective optical depth, varied by post-processing the artificial spectra.\footnote{We have freedom to vary the \Lya~mean flux as it is degenerate with the amplitude of the ultraviolet background (UVB).}
We parameterize the mean flux $\mathcal{F} = \exp(-\tau)$ by modifying the power law redshift evolution from Ref.~\cite{2007MNRAS.382.1657K}, as
\begin{align}
\tau^{\text{Kim}}_{\text{H~{\sc i}}}(z) &= 0.0023 \times (1+z)^{3.65}\,, \\
\tau^{\text{eff}}_{\text{H~{\sc i}}}(z) &= \tau_0 \left(\frac{\tau^{\text{Kim}}_{\text{H~{\sc i}}}(z)}{\tau^{\text{Kim}}_{\text{H~{\sc i}}}(3)}\right)^{d\tau_0} \tau^{\text{Kim}}_{\text{H~{\sc i}}}(z)\,.
 \label{eq:meanflux}
\end{align}
The parameters varied are $\tau_0$ and $d\tau_0$, with $(1, 0)$ corresponding to the redshift evolution of Ref.~\cite{2007MNRAS.382.1657K}.
$\tau_0$ is normalised at $z=3$ so that $d\tau_0 > 0$ corresponds to a higher optical depth at $z > 3$ and a lower optical depth at $z < 3$.
$\tau_0$ changes the normalisation of the optical depth at $z=3$.
We choose prior ranges for $\tau_0$ and $d\tau_0$ that comfortably cover the measurement error: $0.75 < \tau_0 < 1.25$ and $-0.4 < d\tau_0 < 0.25$.
As the mean flux is chosen in post-processing, we can dramatically over-sample these parameters.
We sample $10$ linearly spaced values of the mean flux.
Sampling values are independent of redshift and are chosen so that they include the value of $\tau_0(z)$ implied by our priors at the extreme ($z=2.2$ and $z=4.6$) redshifts included in eBOSS.
For concreteness, we generate flux power spectra scaling the mean optical depth in each redshift bin by a factor between $0.66$ and $1.3$.
This implies that $\tau^{\text{eff}}_{\text{H~{\sc i}}}(z=3)$ varies between $ 0.238$ and $0.47$ and $\tau^{\text{eff}}_{\text{H~{\sc i}}}(z=4)$ varies between $0.54$ and $1.06$.
We thus produce a total of $600$ LF and $30$ HF simulated flux power spectra in each redshift bin, ten times the number of simulations.

% --------------------------------------------------------------------------------------------------

\subsection{Summary Statistics: Flux Power and IGM Temperature}\label{sec:sim_fps}

Figure~\ref{fig:spec_sim} shows an example of the gas density and temperature (colors) at $z=4$ for both high and low resolution in our simulations, demonstrating how spectra connect to the matter density field.
We generate a total of $3\times 480^2 = 691,200$ spectra from each snapshot of each simulation, from $z=4.6$ to $z=2.2$ in increments of $\Delta z=0.2$, with a pixel resolution of $10$ km s$^{-1}$.
We generate \lya forest absorption spectra using Fake Spectra Flux Extractor \cite{2017ascl.soft10012B}\footnotemark, described in Ref.~\cite{2015MNRAS.447.1834B}.
\footnotetext{\url{https://github.com/sbird/fake_spectra}}
We compute the 1D flux power spectrum of the \Lya~forest flux, averaged over all sightlines.
The flux power is defined as 
\begin{equation}
 P_F(k) = |L^{-1}\tilde{\delta}^2_F(k)|\,.   
\end{equation}
$\tilde{\delta}^2_F(k)$ is the Fourier transform of the flux excess, $\delta_F(v) = F(v)/\langle F(v) \rangle - 1$, and $L$ is the length of the sightline. 

Our simulations contain a realistic population of DLAs, which we mask as in the observational pipeline.
We extract the IGM temperatures at mean density directly from the simulation snapshots.
First, the temperature and density for all the gas particles in the simulation are retrieved, then all particles that are within $5\%$ of the critical density are retained.
The median temperature of these retained particles is the IGM temperature at mean density.
All of the \lya forest flux power spectra, IGM temperatures, trained emulators, as well as select MCMC chains are available \footnote{\url{https://github.com/mafern/InferenceLyaData}}.

%--------------------------------------------------------------------------------------------------
% --------------------------------------------------------------------------------------------------
% --------------------------------------------------------------------------------------------------
% --------------------------------------------------------------------------------------------------
% --------------------------------------------------------------------------------------------------

\subsection{Gaussian Process Emulators}\label{sec:gps}

We use the Gaussian Process (GP) emulators described in Refs.~\cite{2022MNRAS.517.3200F, 2023simsuite} for the 1D flux power spectra and mean IGM temperature extracted from our simulations.
The emulators interpolate over simulation outputs and make predictions for arbitrary parameter sets within the parameter limits shown in Table~\ref{tab:emulatorparams}.
We use a multi-fidelity model, which allows simulations with different particle loads, and thus costs, to be combined together.
Specifically, we combine simulations run at two different resolutions, high fidelity (HF) and low fidelity (LF) specified in Table~\ref{table:simulations}. 
The multi-fidelity prediction for the HF outputs (shown here for the \lya forest flux power spectrum, but equally valid for the IGM temperature) is given by a linear multi-fidelity model, at each redshift:
\begin{equation}
    P_F^{^\mathrm{HF}}(k, \boldsymbol{\theta} | z) = \rho_z \cdot P_F^{^\mathrm{LF}}(k, \boldsymbol{\theta} | z) + \delta(k, \boldsymbol{\theta} | z),
    \label{eq:ko_model}
\end{equation}
where $\rho_z$ is a constant parameter, and $\delta(k, \boldsymbol{\theta} | z)$ is a GP.
We have simplified the model from Ref.~\cite{2022MNRAS.517.3200F} by dropping the $k$ dependence in $\rho_z$.
All cosmology and scale dependence is thus in the additive GP $\delta(k, \boldsymbol{\theta} | z)$.
We tested a GP emulator that included scale dependence in the $\rho$ term, but found that it was harder to train and did not significantly improve the accuracy of the predictions.
We train the GP emulators separately for each redshift. We implement our multi-fidelity models using Emukit \cite{2021arXiv211013293P}.

Ref.~\cite{2023simsuite} quantifies the accuracy of our emulator using a leave-one-out technique, in which one simulation is chosen as a `left out' sample.
A smaller emulator built excluding all samples from this simulation is used to predict the summary statistic for the left out sample.
After computing leave-one-out interpolation errors for all potential test samples, we found on average $0.2\%$ accuracy for the low-fidelity simulations and $1\%$ for the high fidelity simulations.
The last is likely a significant over-estimate of the actual error since the leave-one-out procedure in this case is missing $1/3$ of the total training data.
The interpolation errors from the mean IGM temperature emulator are significantly smaller than the $7\%$ average uncertainty in mean IGM temperature measurements.
Our likelihood function and emulator code is publicly available.\footnote{\url{https://github.com/sbird/lya_emulator}}

% --------------------------------------------------------------------------------------------------
\section{Inference Scheme and Likelihood Function}\label{sec:inference}

In this Section, we describe the inference scheme and likelihood function by which our cosmological parameter constraints are derived from our emulator, the eBOSS flux power spectrum \cite{2019JCAP...07..017C} and the mean IGM temperature.
The overall inference scheme is:
\begin{enumerate}
    \item Use the emulator to predict the flux power spectrum and IGM temperature at mean density for a set of input parameters (see Table~\ref{tab:emulatorparams}).
    \item Calculate a likelihood comparing these predictions to their observational counterparts from eBOSS \cite{2019JCAP...07..017C} and Ref.~\cite{2021MNRAS.506.4389G}.
    \item Use Cobaya \cite{2021JCAP...05..057T, 2019ascl.soft10019T} to run MCMC chains and compute posterior parameter constraints.
\end{enumerate}
Section~\ref{sec:fpsdata} discusses the flux power spectrum data, while Section~\ref{sec:t0data} discusses the IGM temperature data.
We derive our covariance matrix in Section~\ref{sec:theoryerror}.
Details of the likelihood calculation used in the MCMC sampling are given in Section~\ref{sec:likelihood}.
We validate our pipeline on simulated data in Section~\ref{sec:simdat}.

% ------------------------------------------------------------------------------------
% ------------------------------------------------------------------------------------

\subsection{Flux Power Spectrum Data}
\label{sec:fpsdata}
% percent error for four surveys (calculated from their available data):
%                z_min     z_max     k_min      k_max      overall
% chabanier:     6%(2.2) 18%(4.6)    6%(0.001)  14%(0.02)     7% (4-18% over z, 5-14% over k)
% day:          15%(2)   14%(4.2)   15%(0.003)   14%(0.1)     11%
% karacayli:     7%(2)   27%(4.6)   12%(0.005)   9%(0.1)      9% (5-27% over z, 7-13% over k)
% irsic:        12%(3)   15%(4.2)   10%(0.003)  16%(0.06)     10%

We use the observed \lya forest flux power spectrum from \cite{2019JCAP...07..017C}, which is based on the Baryon Oscillation Spectroscopic Survey (BOSS) and extended-BOSS (eBOSS) quasar samples \cite{2013AJ....145...10D, 2016AJ....151...44D}.
In \cite{2019JCAP...07..017C}, the BOSS/eBOSS quasar samples are refined to remove spectra that have not been visually inspected, and to remove spectra with broad absorption lines.
Sky lines and damped \lya absorbers (DLAs) are masked.
Our simulations include a realistic population of DLAs, which are masked in the same way.

The sample of \lya forests from the set of remaining quasar spectra is then further refined based on cuts to the spectral resolution, signal-to-noise ratio, number of masked pixels, and forest length, with a final sample of about $43,000$ spectra.
%Along with a windowing function to correct for the spectral resolution of the instrument, estimates for the noise and metal power (estimated using longer wavelength segments of the quasar spectra) are then subtracted to obtain the final \lya forest flux power spectrum.
The redshifts and scales covered by these observations set the redshift range and scales we use in our flux power spectrum emulator, namely $z=2.2-4.6$ (redshift bin size of $\Delta z = 0.2$), and $k\approx0.001-0.02$ s/km (over $35$ linearly spaced bins, $\Delta k = 5.42\times10^{-4}$ s/km).
Our emulator can easily be re-trained for the smaller scales probed by DESI.
The uncertainty in the eBOSS 1D flux power varies with $k$ and $z$, ranging from $> 10 \%$ at $z > 4$ to $\sim 2\%$ at $z \leq 3$, and is often dominated by systematic uncertainty \cite{2019JCAP...07..017C}.

We apply correction terms to the \lya forest flux power spectrum predicted by our emulator to model DLAs and metal contamination.
We correct for DLAs using the template from Ref.~\cite{2018MNRAS.474.3032R}. This allows us to account for differences in the DLA masking between our simulated pipeline and the observed pipeline.
An example would be DLAs, or Lyman limit systems (LLS), which are not detected in the observational pipeline due to low spectral signal-to-noise.
Note that our simulation includes a model that produces realistic populations of LLSs and DLAs, so the marginalised template allows for aspects in which the simulated model differs from the real Universe.
In \cite{2018MNRAS.474.3032R}, there are four parameters, with sub-DLAs separate from LLSs, and DLAs divided into two categories.
For each of the parameters, a redshift and scale dependent correction is applied, where a positive (negative) value for the parameter implies that our simulation has underestimated (overestimated) the number of absorbers in that category. 
We found that in practice our dataset was unable to measure separately all four of the column density bins.
We thus simplify our likelihood by using only two additional free parameters, one parameter covering sub-DLAs and LLS, $\alpha_{\mathrm{lls}}$, and one parameter covering DLAs, $\alpha_{\mathrm{dla}}$.
$\alpha_{\mathrm{lls}}$ covers column densities between $1.6\times10^{17} - 10^{20}$ cm$^{-2}$, and $\alpha_{\mathrm{dla}}$ covers $10^{21}-10^{22.5}$  cm$^{-2}$. 

We account for correlated Si~{\sc{iii}} absorption within the \lya~forest following Ref.~\cite{2006ApJS..163...80M}.
Our likelihood includes an additional nuisance parameter, $f_\mathrm{SiIII}$, which measures the amplitude of the metal contamination. An improved model could use the metal line distribution in a cosmological simulation.
% -------------------------------------------------------------------------------------------------

\subsection{IGM Temperature at Mean Density Data}\label{sec:t0data}

We use the IGM temperatures at mean density from Ref.~\cite{2021MNRAS.506.4389G}, derived from simulation modeling of high resolution quasar spectra from the KODIAQ survey \cite{2017AJ....154..114O}.
Ultimately the dataset is a relatively small, visually inspected set of high resolution quasar spectra.
Importantly, these spectra are independent of the eBOSS quasar sample, justifying our choice of separate likelihood functions.
We include IGM temperature data for $z=2.2-3.8$, for consistency with the available \lya forest flux power data.
The average uncertainty for this data set is $\approx10\%$, whereas our IGM temperature emulator has an average uncertainty of $\sim 1\%$.
Ref.~\cite{2021MNRAS.506.4389G} provides IGM temperatures derived from four different statistics: the \lya forest flux power spectrum, curvature, wavelet decomposition, and Doppler width distribution.
We use the \lya forest flux power spectrum derived temperatures in the main body of this work, but show results using these other data sets in Appendix~\ref{sec:t0-only}.

%In \cite{2021MNRAS.506.4389G}, their co-added quasar spectra are manually checked, and the sample is refined to remove spectra that do not contain \lya forest, do contain DLAs or sub-DLAs, have large gaps in the spectra, or do not meet a signal-to-noise ratio cut.
% A cut is also made to remove regions of the spectra that may lie near to a quasar proximity zone.
%Metal lines are removed by fitting Voigt profiles to the spectra and removing features with Doppler widths $b \leq 8$ km s$^{-1}$, which are assumed to be metal lines.
To derive temperatures from the observed quasar spectra, Ref.~\cite{2021MNRAS.506.4389G} calculated several summary statistics and compared them to those derived from spectra drawn from simulations.
The simulations they used were similar in resolution to our HF suite (gas mass resolution of $\sim10^5$ M$_{\odot}$), though much smaller in volume ($10$ Mpc/h box side length).
These observed IGM temperatures are themselves derived using a suite of simulations assuming fixed $\sigma_8$ and $n_P$ taken from a Planck prior \cite{2021MNRAS.506.4389G}. If the structure on \Lya~forest scales differs significantly from the predictions of the Planck model, the derived IGM temperatures may be biased or inaccurate. In this case, combining the IGM temperatures with the flux power spectrum data could lead to inconsistent constraints. An improved analysis would directly model the observed small-scale 1D flux power spectrum in combination with the flux power spectrum from eBOSS. We will attempt this in future work.

We do not include data on the IGM temperature-density relation, $\gamma$. The constraints from Ref.~\cite{2021MNRAS.506.4389G} are $\gamma(z=3) = 1.22 \pm 0.12$ at $68\%$ confidence. A reasonable theoretical prior would be $1.1 \lesssim \gamma \lesssim 1.6$, so this measurement does not represent a strong constraint on our model. In addition, $\gamma$ varies within our simulations, depending on the redshift at which a particular region of the box undergoes helium reionization \cite{2020MNRAS.496.4372U}.

% ---------------------------------------------------------------------------

\subsection{Covariance Matrix}
\label{sec:theoryerror}

\begin{figure}
    \centering
    \includegraphics[width=0.45\textwidth]{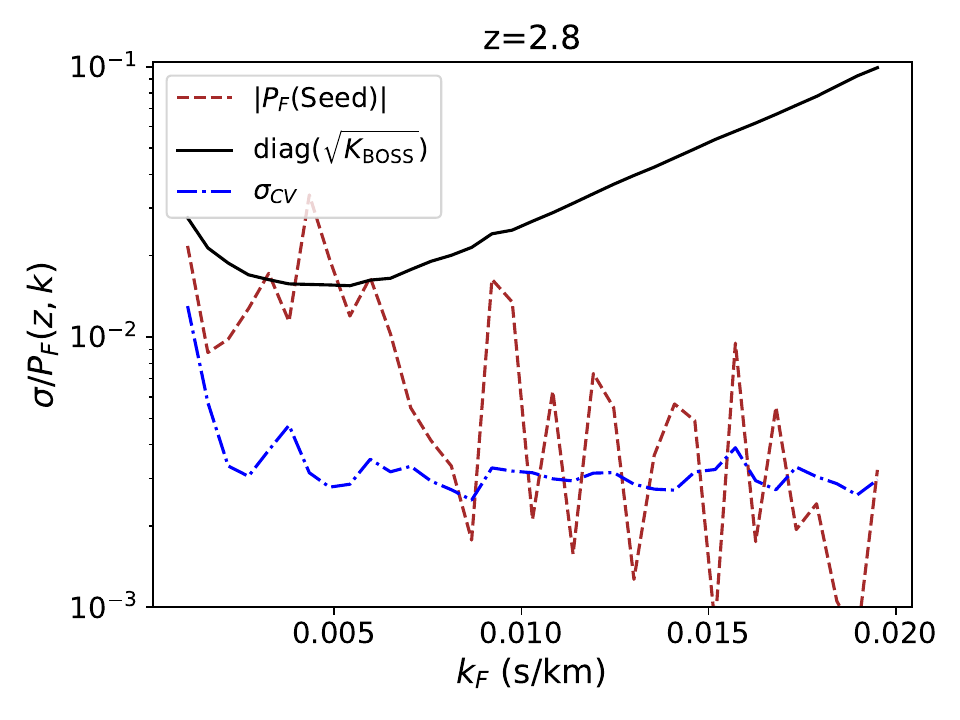}
    \includegraphics[width=0.45\textwidth]{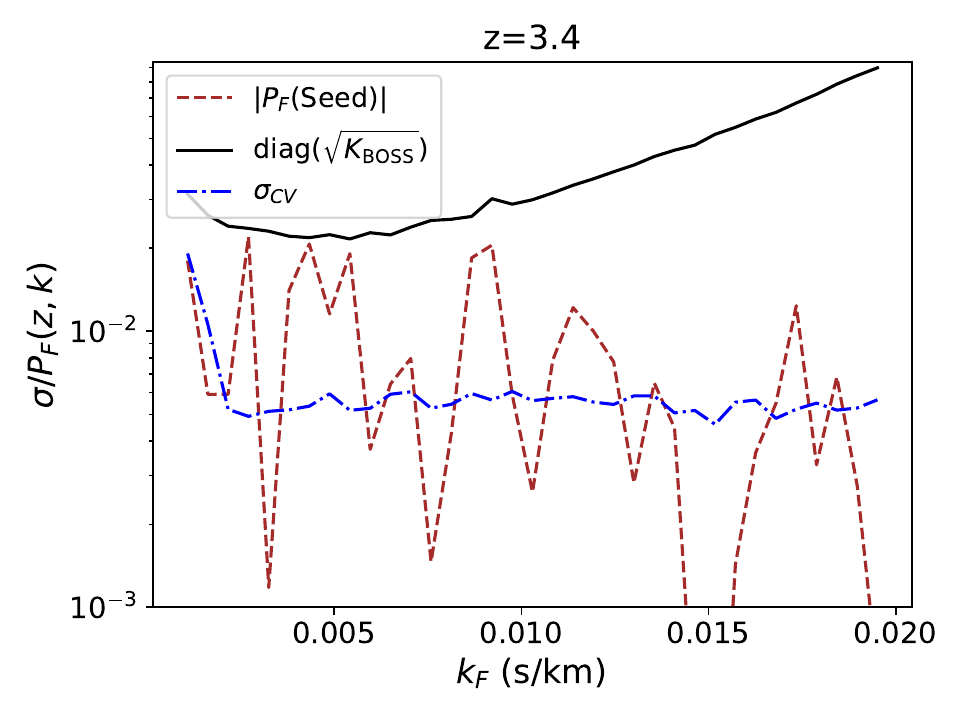}
    \caption{\label{fig:covariance_loo}
     Elements of the covariance matrix, as a function of scale and for selected redshift bins $z=2.8$ (left) and $z=3.4$ (right). We have selected these redshifts as they are those where $\sigma_{CV}$ is most important relative to $K_{BOSS}$. 
    Shown are:  1) The ratio of the 1D flux power spectra between the two LF simulations with different seeds discussed in Ref.~\protect\cite{2023simsuite} ($|P_F(\mathrm{Seed})|$, brown, dashed).
    2) the square root of the diagonal elements of the eBOSS covariance matrix $K_{BOSS}$ (black solid) 3) the diagonal cosmic variance error estimated from leave-one-out errors, $\boldsymbol{\sigma}_{CV}$ (blue, dot-dashed). The y-axis label $\sigma/P_F(z,k)$ signifies that all errors are normalised by the 1D flux power spectrum, $P_F(k, z)$. Hence $\sigma_{CV}$ is $\sigma_{CV}/ P_F(z,k)$ and $|P_F(\mathrm{Seed})|$ is $|P_F(\mathrm{Seed}, z,k)/P_F(\mathrm{Default Seed}, z, k)-1|$.}
\end{figure}

In this Section, we derive the covariance matrix, $\boldsymbol{K}$, that is used for our inference.
We decompose $\boldsymbol{K}$ as:
\begin{equation}
    \boldsymbol{K} = \boldsymbol{K}_\mathrm{BOSS} + \boldsymbol{\sigma}_{GP}(\boldsymbol{p}) \cdot \boldsymbol{\sigma}_{GP}^T (\boldsymbol{p}) + \boldsymbol{\sigma}_{CV} \cdot \boldsymbol{\sigma}_{CV}^T \,.
    \label{eq:covariance}
\end{equation}
Here, $\boldsymbol{K}_\mathrm{BOSS}$ is the covariance matrix from the eBOSS pipeline \cite{2019JCAP...07..017C}, and is the largest term in the covariance matrix on most scales.
We also add two extra terms which model theoretical error in our model.
$\boldsymbol{\sigma}_{GP}(\boldsymbol{p})$ is the parameter dependent estimate of the interpolation error from the Gaussian process.
Using mocks (see Figure~\ref{fig:simdat_posteriors}), we found that the GP error can sometimes unphysically drive the chain away from the edges of parameter space, where the expected interpolation error is large.
We thus choose to omit it from the overall covariance matrix.
Appendix~\ref{sec:loovsgperr} shows that its addition has a small effect on our final results.

The second theoretical error in our simulation suite (which dominates) is $\boldsymbol{\sigma}_{CV}$, which models residual sample variance from the finite box size, analogous to cosmic variance from the finite cosmological horizon\footnote{Ref.~\cite{2023ApJ...944..223P} reduced sample variance by interpolating the parameters of a higher order polynomial, rather than fitting the binned flux power spectrum directly.
Our emulator is much less affected by sample variance as our simulated volume is $8$ times larger.}.
We include an estimate of sample variance using the leave-one-out errors discussed in Ref.~\cite{2023simsuite}, a technique made possible by the inclusion of $h$ in our simulation suite.
The Hubble parameter does not directly affect the gravitational evolution in our simulations due to Gadget's use of Mpc/h units, and Ref.~\cite{2023simsuite} showed that the effect on the thermal history is small on the scales probed by eBOSS.
However, in our parameterization, $h$ also changes $\Omega_M$ \footnote{$\Omega_M h^2$ is a separate parameter and so kept fixed when varying $h$.} and so the conversion of wavenumbers from h/Mpc to s/km.
Individual Fourier modes thus move between bins depending on the value of $h$, mimicking the sample variance from different initial phases.
We thus approximate $\boldsymbol{\sigma^2}_{CV}$ with the averaged variance of the leave-one-out errors using the low fidelity simulations.
Leave-one-out errors are found by building a reduced emulator, which is trained on all but a single sample, then evaluating the prediction accuracy for that left-out sample using the reduced emulator.
This is then repeated, such that every sample takes a turn being left out (see Figure~2 of \cite{2023simsuite}):
\begin{equation}
    \boldsymbol{\sigma}^2_{CV}  = \frac{1}{N_{LF}}\Sigma_i \left(P_F^\mathrm{Predict}(k, z, p_i) - P_F^\mathrm{True}(k, z, p_i)\right)^2\,.
\end{equation}
Here the sum runs over all simulated low-fidelity parameter sets $p_i$ and $N_{LF}$ is the number of low-fidelity simulations.

Figure~\ref{fig:covariance_loo} shows the magnitude of the $\boldsymbol{\sigma}_{CV}$ term compared to the eBOSS errors.
$\boldsymbol{\sigma}_{CV}$ is significant only on the largest scales, $k < 2.5 \times 10^{-3}$ s/km, as expected from an effect due to finite box size.
In addition, there is a significant redshift dependence: $\boldsymbol{\sigma}_{CV}$ is important relative to the eBOSS errors only for $2.8 < z < 3.4$.
For clarity, Figure~\ref{fig:covariance_loo} shows only $z=2.8$ and $z=3.4$, which bracket the largest effect. For the redshift range $4.2 > z>3.4$, $\boldsymbol{\sigma}_{CV}$ remains approximately constant, but becomes less significant as the eBOSS statistical errors increase.
These details reveal the physical source of this large-scale variance.
The relevant scale is close to the $20$ Mpc/h size of the helium reionization bubbles, and the relevant redshift range is when our model performs helium reionization.
Helium reionization bubbles are placed randomly around rare large halos, which creates sample variance in a finite box. 

Figure~\ref{fig:covariance_loo} also shows the absolute value of the ratio of the flux power spectrum between two LF simulations with different initial structure seeds. If this ratio is consistently larger than $\boldsymbol{\sigma}_{CV}$, it would indicate an extra source of cosmic variance not included by varying $h$. The average flux power spectrum ratio is similar to $\boldsymbol{\sigma}_{CV}$ on the largest scales measured, where it is most important. We have deliberately chosen to show the redshift bins near the ends of helium reionization where the effects of cosmic variance are most important. To evaluate the effect of a potential under-estimation of cosmic variance from these k-bins, we performed two extra chains where the last term in Eq.~\ref{eq:covariance} was $2\times \boldsymbol{\sigma}_{CV}$, and where $\boldsymbol{\sigma}_{CV} = 0.02$.
Our posterior constraints were almost unchanged in both cases, likely because $K_{BOSS}$ is generally larger than $2\%$.

\subsection{Likelihood}\label{sec:likelihood}

We use a log normal likelihood summed over all redshifts and, for the flux power, all scale bins:
\begin{equation}
    \mathrm{log}\mathcal{L} = -\frac{1}{2} \sum_{z=2.2}^{z=4.6} \left(\left(\boldsymbol{P}_F^{\mathrm{diff}}\right)^\top \cdot \boldsymbol{K}^{-1} \cdot \boldsymbol{P}_F^{\mathrm{diff}} + \log\left( \mathrm{det}(\boldsymbol{K})\right)\right)
    \label{eq:likelihood}
\end{equation}
where $\boldsymbol{P}_F^{\mathrm{diff}} = \boldsymbol{P}_F^{\mathrm{sim}} - \boldsymbol{P}_F^{\mathrm{obs}}$ is the vector difference between the simulation prediction and the observation.
The covariance matrix, $\boldsymbol{K}$, is described in Equation~\ref{eq:covariance}. 

The likelihood for the IGM temperature is similar, but single valued per redshift.
We compute the \lya forest flux power and IGM temperature likelihoods separately and add the log likelihoods. We make use of the Cobaya package \cite{2021JCAP...05..057T, 2019ascl.soft10019T, 2013PhRvD..87j3529L, 2002PhRvD..66j3511L} to run MCMC chains using this likelihood.
The MCMC sampler uses the Metropolis method discussed in \cite{2013PhRvD..87j3529L}, and uses a Gaussian + exponential proposal distribution that dynamically learns the proposal covariance.
Convergence is determined using the Gelman-Rubin statistic, $R$, also detailed in \cite{2013PhRvD..87j3529L}.
The chains presented here were run until a convergence of $R-1 < 0.01$, with results plotted for those chains for samples at $R-1 < 1$.

\subsubsection{Priors}

We use the parameter limits shown in Table~\ref{tab:emulatorparams}.
As we showed in Ref.~\cite{2023simsuite}, the AGN feedback parameter $\epsilon_{AGN}$ has minimal effect on the \Lya forest 1D flux power spectrum.
Preliminary chains indicated that it is indeed poorly constrained by the data and has minimal correlations with other parameters.
We use a strong Gaussian prior with $\mu = 0.05$ and $\sigma = 0.005$, which dominates over data constraints, and will omit constraints on $\epsilon_{AGN}$ from our results.
We also place a weak Gaussian prior on the Hubble parameter, $h$, with $\mu = 0.70$ and $\sigma = 0.015$, as it is weakly constrained and this prior avoids the inference straying into areas near the edge of parameter volume where the emulation is less accurate.
For all other parameters we use uniform priors within the parameter limits.

\subsection{Inference Using Simulation Data}\label{sec:simdat}

In this section we test our inference framework with simulation outputs in place of the observational data, confirming that we recover the known input parameters.
We first used the flux power spectrum from one of the three high fidelity simulations, and confirmed that the maximum likelihood was indeed at the input parameter values for all parameters.
All input parameters were recovered to better than one sigma.

\begin{figure}
    \centering
    \includegraphics[width=\textwidth]{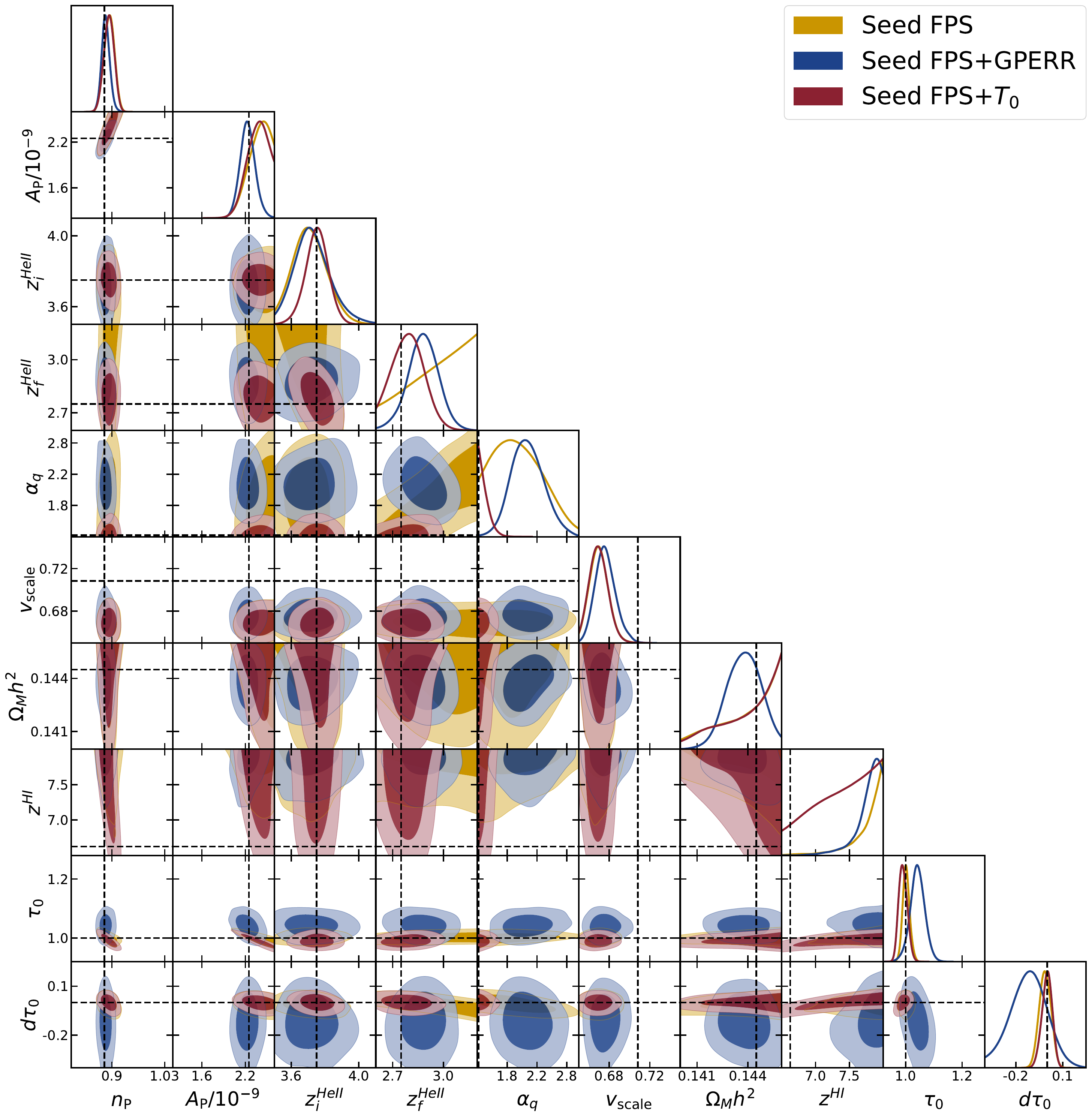}
    \caption{\label{fig:simdat_posteriors}
    Posteriors using mock data, a simulation output with a different initial seed to the main PRIYA suite.
    The true parameter values for the input are indicated by the black dashed lines.
    Three chains are used: `Seed FPS' uses the default error model, with the eBOSS covariance and leave-one-out errors.
    `Seed FPS + GPERR' adds the covariance from the Gaussian Process.
    `Seed FPS + $T_0$' uses the default error model but supplements the flux power spectrum data with information from the IGM temperature.
    $v_{scale}$ is the Hubble parameter $h$, re-labelled for reasons explained in the text.}
\end{figure}

We next ran chains using data from a low-fidelity simulation with a different random seed from the main PRIYA suite.
For these runs only, we used an emulator built using the low fidelity suite.
This test was designed to quantify whether the finite box size of our simulations can affect our parameter constraints.
Figure~\ref{fig:simdat_posteriors} shows the results, with dashed black lines indicating the correct parameters. Note that Figure~\ref{fig:simdat_posteriors} shows an estimate of the potential bias from cosmic variance on our results.

We have performed three runs.
The first (Seed FPS) is our preferred error model, using the eBOSS covariance and a leave-one-out error term.
The second adds an error term for the expected interpolation error from the Gaussian Process (Seed FPS + GPERR).
Both of these runs use only information from the eBOSS flux power spectrum and thus do not provide strong constraints on the parameters of helium reionization.
We therefore run another chain including constraints from the IGM temperature.

In all three runs, the optical depth parameters, $\tau_0$ and $d\tau_0$, are tightly constrained around the true value in our pipeline, despite the effect of a different structure seed. The GPERR chain increases the uncertainty, especially on $d\tau_0$, but does not bias the measurement.
The best estimate comes from the flux power spectrum data alone (Seed FPS).
We also consistently recover the true values of the cosmological parameters $n_P$ and $A_P$ to better than $1-\sigma$.
Note that we deliberately constructed our test data, with a different structure seed, to be different from the training data.
$\Omega_M h^2$ is poorly constrained in all chains, as expected given that our prior volume includes only a narrow range for $\Omega_M h^2$, motivated by Planck results.
All parts of the prior range are within the $1-\sigma$ posteriors.

The redshift of hydrogen reionization, $z_{HI}$, is estimated from the IGM temperature at $z > 3.6$ or from a large-scale increase in the flux power spectrum at $z > 4$ (see Ref.~\cite{2023simsuite}).
The second effect is due to a scale-dependent bias arising from placement of the reionization bubbles \cite{Montero:2019}.
Figure~\ref{fig:simdat_posteriors} indicates that this bias is sensitive to sample variance from the finite box, and so the hydrogen reionization redshift is not well measured by the flux power spectrum data alone.
The three parameters which govern helium reionization, $z_i^{HeII}$, $z_f^{HeII}$ and $\alpha_q$, are well constrained by the IGM temperature data.
The runs which do not include IGM temperature data have a slight preference for a larger $\alpha_q$ than the input value.
As discussed above, the main effect of a different structure seed is through the placement of helium reionization bubbles.
$\alpha_q$ is thus measured using a similar scale-dependent bias as $z_{HI}$, and so is slightly sensitive to the finite box size in the same way.
However, the IGM temperature is sensitive to $\alpha_q$ through the peak temperature during helium reionization, and thus the chains including it correctly infer $\alpha_q$.

The chain including Gaussian Process errors produces some incorrect parameter inferences, notably in $\alpha_q$.
This is because of the specific choice of simulated data, for which $\alpha_q$ is at the lower boundary of the emulator parameter range.
The GP expects the emulator error to be larger near the boundary of the space, which penalises the fit in this region when the constraints from the data are weak.
Notice that $\alpha_q$ is poorly measured by the flux power spectrum alone.
In fact, our leave-one-out error calculation reveals that the flux power spectrum is reasonably accurately modelled even near the emulator boundaries, but, with few simulations in this region, there is not enough information present for the GP to learn this and reduce the expected error.
We show in Appendix~\ref{sec:loovsgperr} that the posteriors from eBOSS data are not significantly changed by including GP error.
Nevertheless, to be conservative our main results are reported with chains run omitting GP errors from the likelihood.

As discussed above, the Hubble parameter, $h$, does not affect the evolution of our simulations except through its effect on $\Omega_M$ (at fixed $\Omega_M h^2$) and thus the scaling between velocity (km/s) and comoving (Mpc/h) units.
Constraints on $h$ ($v_\mathrm{scale}$) are incorrect in the chains shown in Figure~\ref{fig:simdat_posteriors}, driven by sample variance in the finite box.
We confirmed that gradients of the likelihood with respect to $h$ are largest for the largest scales, particularly the first $4$ $k$-bins.
We computed the $\chi^2$ per degree of freedom, which was $\sim 0.9$ for $h = 0.65$ and $\chi^2 \sim 1$ for the true input value, indicating over-fitting to the noise created by different structure seeds.
We confirmed that fixing $h$ to the known true value results in very small changes to the other parameters and their confidence intervals.
There is thus zero cosmological information in our $h$ constraints.
In order to avoid unwarranted conclusions, we will henceforth relabel $h$ as $v_\mathrm{scale}$, emphasising that it merely controls the mapping between the native Fourier-space binning of the simulations and the observed velocity space of the spectra, and its inference is dominated by modelling error from sample variance. 

The posterior $v_\mathrm{scale}$ from the simulation with a different seed does not match the input value. However, this is to be expected: $v_\mathrm{scale}$ is completely degenerate with the initial structure seed. The measured quantity is thus schematically a linear combination of $h$ and a `cosmic variance parameter' set by the initial distribution of structures. When the value of the cosmic variance parameter expected by the emulator differs from the true input, the emulator will measure a different posterior value for $h$, as shown in Figure~\ref{fig:simdat_posteriors}.

\section{Results}\label{sec:results}

In this Section, we report posterior constraints on the parameters listed in Table~\ref{tab:emulatorparams}.
Section~\ref{sec:cosmo} discusses the results for those parameters which are most strongly constrained by the flux power spectrum data.
These are: the optical depth parameters, $\tau_0$ and $d\tau_0$, the power spectrum parameters, $n_P$ and $A_P$, and the matter density $\Omega_M h^2$.
Section~\ref{sec:astro} then discusses the constraints on the other parameters, and shows the best fit to the IGM temperature at mean density data.
These are the three parameters defining the He~{\sc{ii}} reionization model, z$^{\text{He~{\sc ii}}}_i$, z$^{\text{He~{\sc ii}}}_f$, and $\alpha_q$; the parameter for the midpoint of H~{\sc{i}} reionization, z$^{\text{H~{\sc i}}}$, the strong absorber models, ($\alpha_{LLS}$ and $\alpha_{DLA}$), the Silicon III correction (fSiIII) and the velocity to distance scale parameter $v_{scale}$.
The same chains are used in all sections: we split parameters into two sections merely for readability.
We show the full corner plot, containing all constrained parameters, in Appendix~\ref{sec:full_posteriors}.
Table~\ref{table:parameters} shows posterior parameter constraints.
We also calculate the derived parameters $A_s$ and $\sigma_8$, $\Delta_L^2$ and $n_\mathrm{eff}$.

% --------------------------------------------------------------------------------------------------

\subsection{Cosmological Parameters}\label{sec:cosmo}

\begin{figure}
    \centering
    \includegraphics[width=0.75\textwidth]{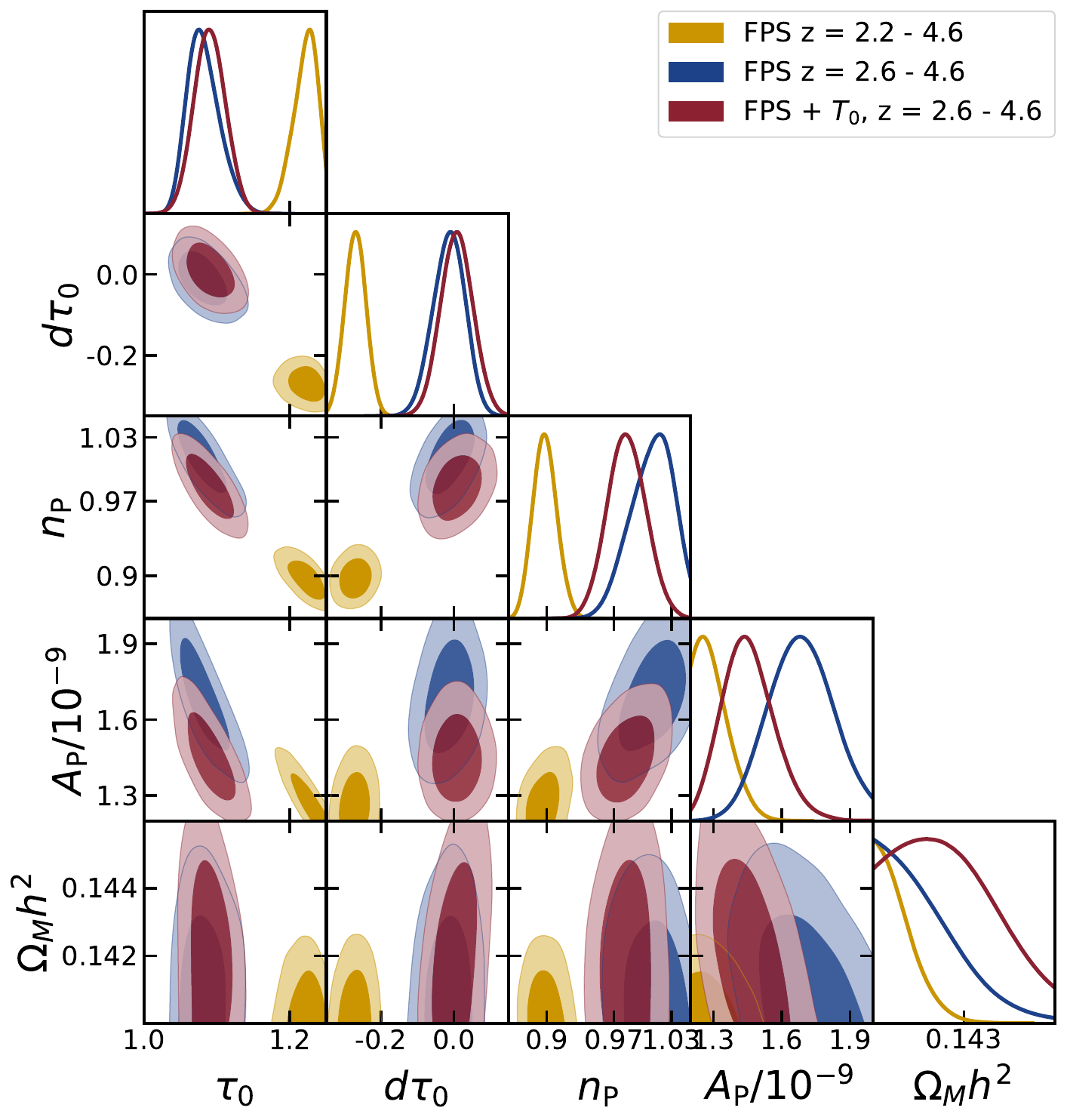}
    \caption{\label{fig:cosmo_corner}
    Posteriors for the optical depth and power spectrum parameters, $\tau_0$, $d\tau_0$, $n_P$, $A_P$, and $\Omega_M h^2$.
    Results are from three MCMC chains.
    `FPS $z=2.2-4.6$' (gold) uses the full redshift range eBOSS flux power spectrum dataset, `FPS $z=2.6-4.6$' (blue) uses a reduced redshift range eBOSS dataset flux power spectrum dataset, which removes the internal tension (Section~\protect\ref{sec:tension}).
    The third chain, `FPS $+ T_0, z=2.6-4.6$' (red), uses the limited range eBOSS dataset but adds the IGM temperature constraints.
    Our preferred cosmological constraints are from `FPS $z=2.6-4.6$'.
    }
\end{figure}

\begin{table}
	\centering
      \def\arraystretch{1.6}
\begin{tabular} {| l | c | c | c|}
\hline
 & FPS $z>2.6$ & FPS + $T_0$ & FPS $z>2.2$ \\
Parameter & 68\% (95\%) & 68\%, (95\%) & 68\%, (95\%)
\\
\hline
$d\tau_0        $ & $-0.013^{+0.047}_{-0.041}$  $\left(^{+0.083}_{-0.089}\right)$                   & $0.009\pm 0.045$            $\left(^{+0.089}_{-0.087}\right)$                   & $-0.270\pm 0.029$     $\left(^{+0.057}_{-0.057}  \right)$   \\
$\tau_0         $ & $1.082^{+0.018}_{-0.026}$  $\left(^{+0.046}_{-0.041}\right)$            & $1.090\pm 0.022$       $\left(^{+0.043}_{-0.042}\right)$                   & $1.221^{+0.021}_{-0.012}$   $(< 1.19)$   \\
$n_\mathrm{P}   $ & $1.009^{+0.027}_{-0.018}$  $\left(^{+0.040}_{-0.039}\right)$  & $0.983\pm 0.020$  $\left(^{+0.040}_{-0.039}\right)$    & $0.898^{+0.012}_{-0.013}$            $\left(^{+0.025}_{-0.023}\right)$   \\
$A_\mathrm{P}/10^{-9}$ & $1.69^{+0.14}_{-0.15}$        $\left(^{+0.30}_{-0.28}\right)$        & $1.46^{+0.099}_{-0.13}$   $\left(^{+0.22}_{-0.22}\right)$                   &           $< 1.33 $  ($< 1.44$)   \\
$\Omega_M h^2   $ & $< 0.142$ ($< 0.144$)                                              & $< 0.143$  ($--$)                                                                  & $< 0.141$ ($< 0.142$)   \\
$z^{HeII}_i     $ & $--                    $                                           & $> 4.00$   $(> 3.87 ) $                                                            & $> 4.07                    $  $(> 4.01)$   \\
$z^{HeII}_f     $ & $< 2.67                    $  $(< 2.80)$                   & $2.765^{+0.080}_{-0.093}$  $\left(^{+0.14}_{-0.16}\right)$           & $< 2.70                    $    $(< 2.83 )$   \\
$\alpha_{q}     $ & $> 2.86                    $  $(> 2.64)$                   & $1.74^{+0.18}_{-0.21}$  $\left(^{+0.37}_{-0.38}\right)$              & $> 2.85                    $    $(> 2.65 )                   $   \\
$z^{HI}         $ & $< 7.01                    $ ($< 7.57$)                    & $ 7.24\pm 0.38$ ($--$)                            & $7.28\pm 0.37$    $(--)$   \\
$v_\mathrm{scale}$ & $0.693\pm 0.0085$  $\left(^{+0.018}_{-0.017}\right)$    & $0.688^{+0.013}_{-0.0074}$  $\left(^{+0.018}_{-0.025}\right)$    & $0.695^{+0.0063}_{-0.0073}$  $\left(^{+0.015}_{-0.014} \right)$     \\
$\alpha_{lls}   $ & $0.193\pm 0.035$ $\left(^{+0.069}_{-0.069}\right)$          & $0.196\pm 0.033$  $\left(^{+0.063}_{-0.065}\right)   $     & $0.042\pm 0.018$  $\left(\pm 0.036\right)$   \\
$\alpha_{dla}/10^{-2}   $ & $-0.28\pm 0.64$ $\left(^{+1.2}_{-1.3}\right)$ & $-1.05\pm 0.58$ $\left(\pm 1.1\right)  $    & $-0.87\pm 0.34         $   $\left(^{+0.67}_{-0.66}\right)$  \\
$f_{SiIII}/10^{-3} $ & $9.63\pm 0.51$ $\left(^{+9.8}_{-1.0}\right)$              & $9.63\pm 0.51        $       $\left(^{+0.97}_{-1.0}\right) $          & $8.83\pm 0.39        $  $\left(\pm 0.76\right)$\\
\hline
$A_\mathrm{s}/10^{-9}      $ & $1.65^{+0.12}_{-0.14}$ $\left(^{+0.27}_{-0.25}\right)$ & $1.52^{+0.10}_{-0.13}$ $\left(^{+0.24}_{-0.22}\right)$  & $1.73^{+0.076}_{-0.11}$ $\left(^{+0.19}_{-0.17}\right)$   \\
$\sigma_8$ & $0.733^{+0.026}_{-0.029}$ $\left(^{+0.057}_{-0.053}\right)$ &  $0.703^{+0.023}_{-0.027}   $ $\left(^{+0.049}_{-0.047}\right)$ & $0.715^{+0.015}_{-0.022}$  $\left(^{+0.038}_{-0.034}\right)$  \\
$\Delta_L^2$     & $0.302^{+0.024}_{-0.027}$   $\left(^{+0.053}_{-0.048}\right)$ &  $0.267^{+0.018}_{-0.023}$    $\left(^{+0.042}_{-0.039}\right)$    & $0.2316^{+0.0082}_{-0.016}$ $\left(^{+0.027}_{-0.021}\right)$\\
$n_\mathrm{eff}$ & $-2.264^{+0.026}_{-0.018}$  $\left(^{+0.038}_{-0.042}\right)$ &  $-2.288\pm 0.020$     $\left(^{+0.040}_{-0.039}\right)$       & $-2.376^{+0.012}_{-0.013}$ $\left(^{+0.025}_{-0.024}\right) $ \\
\hline
\end{tabular}
\caption{\label{table:parameters}
Posterior parameter constraints, including the derived parameters $A_s$ and $\sigma_8$, as well as the predicted linear theory power, $\Delta_L^2$, and slope, $n_\mathrm{eff}$, evaluated at $k_P = 0.009$ s/km, $z_P=3$.
Maximum posterior values, and $68\%$ confidence limits are shown, with $95\%$ confidence intervals in brackets.
Each column shows a separate chain, from left to right: fits to the flux power spectrum alone from the reduced redshift range $z=2.6 - 4.6$, fits to the flux power spectrum from the reduced redshift range $z=2.6 - 4.6$ and the IGM temperature, and fits to the flux power spectrum alone from the full redshift range $z=2.2 - 4.6$.
Single sided limits are shown when one bound is larger than the prior volume of the emulator.
`$--$' denotes that both $68\%$ and $95\%$ constraints are wider than the prior volume of the emulator.
}
\end{table}

\begin{figure}
    \centering
    \includegraphics[width=\textwidth]{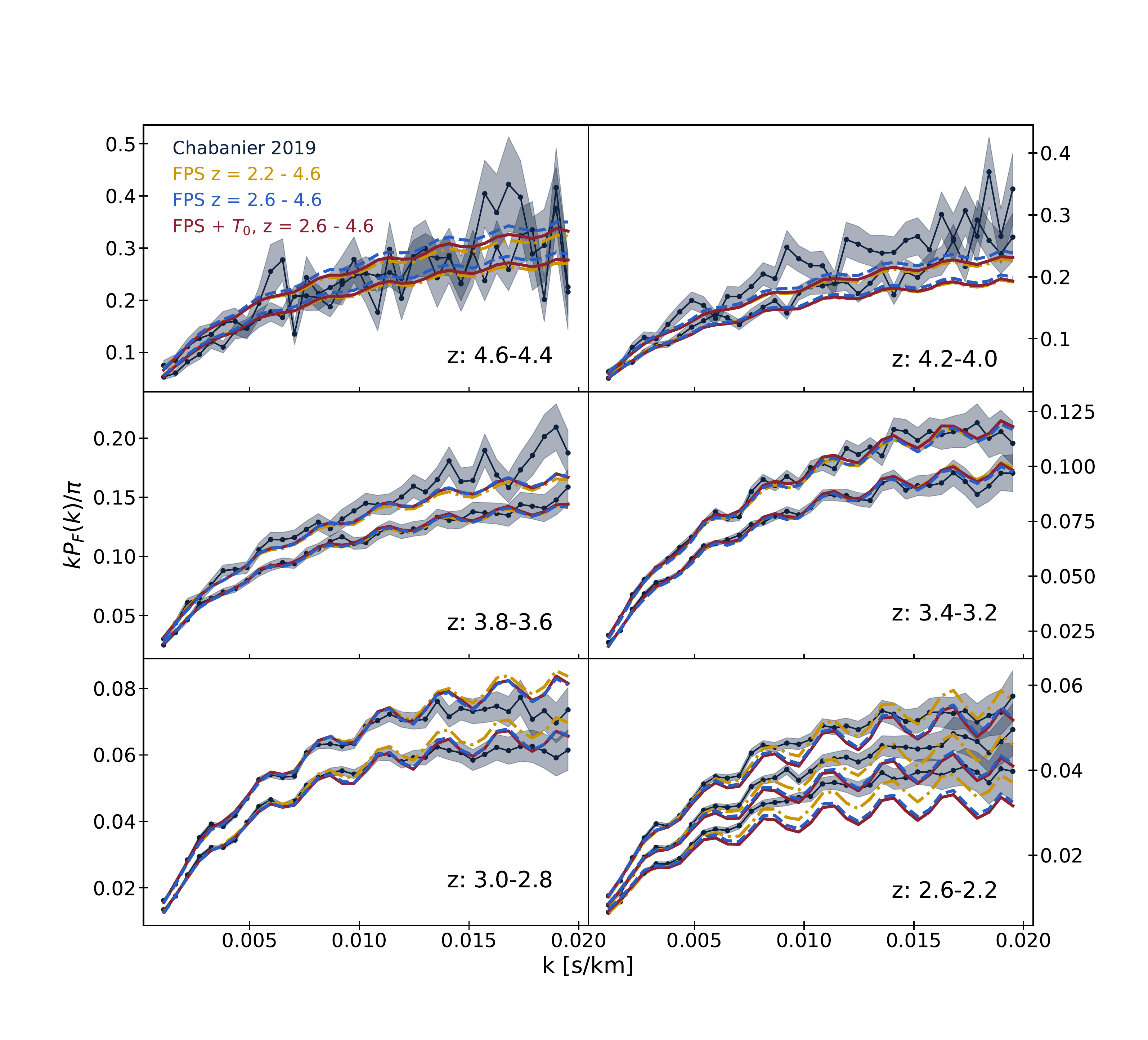}
    \caption{\label{fig:fps_data}
    Observed \lya forest flux power spectrum \cite{2019JCAP...07..017C}, from $z=4.6$ to $z=2.2$ (black lines and circles, with shading corresponding to one sigma uncertainty).
    Also shown are three predictions for the \lya forest flux power spectrum from our multi-fidelity emulator corresponding to the maximum posterior input parameters compiled in Table~\ref{table:parameters}.
    The negative of the log-likelihood for these fits is compiled in Table~\ref{table:chi2}.
    }
\end{figure}

Figure~\ref{fig:cosmo_corner} shows the results of our chains for the cosmological parameters.
We show three MCMC chains.
Two chains are fit to the eBOSS flux power spectrum data only.
The first fits to the full redshift range measured by eBOSS, $z=2.2 - 4.6$, while the second fits a limited redshift range $z=2.6 - 4.6$.
The third chain uses the limited redshift range eBOSS dataset but adds the IGM temperature likelihood.
The chain including the $z < 2.6$ data prefers lower $n_P$, lower $A_P$ and higher $\tau_0$ than the reduced redshift range.
Figure~\ref{fig:fps_data} shows that the shift in posterior parameters is driven by the fit.
The best-fit flux power spectrum to the data at $z \geq 2.6$ is a poor fit to the flux power spectra measured at $z=2.2$ and $z=2.4$.
Since the lowest redshift bins have the smallest statistical error, when they are included they drive the best-fit flux power spectrum to a region which is a poorer fit to the higher redshift data.
We confirmed that a chain which included the $z=2.4$ bin but not the $z=2.2$ bin produced posterior constraints mid-way between the chain including $z=2.2-4.6$ and the chain including $z=2.6-4.6$.
Table~\ref{table:chi2} shows this quantitatively.
The chains which fit to $z > 2.6$ are a poor fit to the lowest two redshift bins.
The chain fitting to $z=2.2$ is a better fit to these bins, at the cost of an overall worse fit to redshifts $z > 2.6$.
%The total $\chi^2$ per degree of freedom for the reduced redshift chains is close to $1.03$, indicating a good fit, while the full reduced range has a $\chi^2/$dof of $1.15$.
There is thus an internal tension in the eBOSS dataset, when compared to our model, driven by the lowest two redshift bins.
We discuss possible reasons for this tension in Section~\ref{sec:tension} and compare to the results of earlier analyses further in Section~\ref{sec:comparison}.
It is important not to over-interpret the posterior constraints from the $z=2.2-4.6$ chain.
When there is an internal tension in the data, the posteriors can be driven by noise in the dataset and may not be meaningful.

We also checked for other redshift bins where the fit is poor.
Visually, Figure~\ref{fig:fps_data} suggests that the fit is poor for $z=4.0$ and $z=4.2$.
The $\chi^2$ in Table~\ref{table:chi2} is moderately higher other bins, but not significantly so, as the statistical errors are also large.
It is possible that some element of the covariance matrix, theoretical or systematic, is moderately underestimated in these bins.
%We ran chains which fit to each redshift bin individually, finding that the lowest two redshift bins were still discrepant with the others.
Intriguingly, the fit to $z=2.6$ is better when the $z < 2.6$ bins are included, which may suggest that the cause of the low redshift tension still has a small effect at $z=2.6$.

\begin{table}
	\centering
     \def\arraystretch{1.2}
     \begin{tabular}{|c|c|c|c|c|c|c|c|}
		\hline
		Redshift & $4.6$ & $4.4$ & $4.2$ & $4.0$ & $3.8$ & $3.6$ & $3.4$\\
		\hline
        FPS $z= 2.2-4.6$ & $23.8$ & $22.1$ & $28.3$ & $33.9$ & $18.7$ & $8.7$ & $16.1$ \\
        FPS $z= 2.6-4.6$ & $27.8$ & $19.1$ & $23.0$ & $26.5$ & $15.6$ & $7.4$ & $16.5$\\
        FPS $+ T_0$ $z= 2.6-4.6$ & $24.9$ & $18.2$ & $26.0$ & $31.3$ & $16.7$ & $6.6$ & $14.4$ \\
        \hline
        Redshift & $3.2$ & $3.0$ & $2.8$ & $2.6$ & $2.4$ & $2.2$ & Total \\
		\hline
        FPS $z= 2.2-4.6$ & $19.6$ & $33.8$ & $33.7$ & $20.3$ & $19.9$ & $27.3$ & $306$ \\
        FPS $z= 2.6-4.6$ & $18.9$ & $25.6$ & $24.6$ & $29.1$ & $66.5$ & $97.6$ & $234$ \\
        FPS$ + T_0$ $z= 2.6-4.6$ & $18.2$ & $24.9$ & $25.2$ & $37.7$ & $76.0$ & $103$ & $244$\\
  \hline
	\end{tabular}
    \caption{\label{table:chi2}
    Negative log-likelihood ($\chi^2$) for the flux power spectrum for each redshift bin.
    %There are $35$ k-bins per redshift bin and $14$ parameters, and so a total of $441$ degrees of freedom, although neighbouring k-bins are highly correlated.
    Shown is the likelihood at the best fit parameters in each chain.
    We show chains fitting to the flux power spectrum only at $z=2.2-4.6$, fitting to the flux power spectrum only at $z=2.6-4.6$, and fitting to the flux power spectrum and IGM temperature at $z=2.6-4.6$.
    The column labelled `Total' excludes redshift bins not in the fit ($z=2.2$ and $z=2.4$ for the last two chains).
    % The column labelled `Total' is the total $\chi^2$ per degree of freedom for the redshift bins in the fit (i.e., excluding $z=2.2$ and $z=2.4$ for the last two chains).
    }
\end{table}

Posterior constraints from the reduced redshift range flux power spectrum data show a mean optical depth in good agreement with other measurements.
As a reminder, our $\tau_0$ and $d\tau_0$ parameters measure deviations from the mean flux relation of Ref.~\cite{2007MNRAS.382.1657K}, so a value of $\tau_0=1$ and $d\tau_0=0$ corresponds to agreement with that model.
The maximum posterior values are $\tau_0 = 1.1$, with a redshift variation $d\tau_0$ consistent with $0$.
This implies a mean optical depth at $z=3$ of $\tau^{\text{eff}}_{\text{H~{\sc i}}}(z=3) = 0.398$, which is extremely close to the best-fit value of $\tau^{\text{eff}}_{\text{H~{\sc i}}}(z=3) = 0.4$ from Ref.~\cite{2013MNRAS.430.2067B}, and consistent within the error bars with $\tau^{\text{eff}}_{\text{H~{\sc i}}}(z=3) = 0.36 \pm 0.1$ from Ref.~\cite{2007MNRAS.382.1657K}. 

The spectral index, $n_P$, is $n_P = 1.009^{+0.027}_{-0.018}$ (68\% confidence) using only the flux power spectrum. On inclusion of the IGM temperature data, we find slightly tighter constraints of $n_P = 0.983 \pm 0.02$ at 68\% confidence, lower by about $1-\sigma$.
Planck measures $n_s=0.965 \pm 0.004$, consistent with the IGM temperature chain at $1-\sigma$ and the flux power spectrum chain at $2.5-\sigma$ \cite{2020A&A...641A...6P}.
The matter density, $\Omega_M h^2$, is weakly constrained and not strongly affected by including IGM temperature data.
% The growth factor, $\Omega_M h^2$, is weakly constrained and not strongly affected by including IGM temperature data.
Planck found $\Omega_M h^2 = 0.1424\pm0.001$, which is close to the posterior of our chains.
The power spectrum amplitude is $A_P/10^{-9} = 1.69 \pm 0.14$ for the flux power spectrum reduced redshift result.
The inclusion of the IGM temperature shrinks the constraints moderately and shifts the posterior value down by about $1.5-\sigma$, driven by a correlation with $\alpha_q$.
Table~\ref{table:parameters} shows $A_s$, the power spectrum amplitude measured on large scales, which is related to $A_P$ via:
\begin{equation}
    A_s = \left(0.4/2\pi\right)^{n_P-1} A_P\,.
\end{equation}

We find $A_s = (1.65^{+0.12}_{-0.14}) \times10^{-9}$ for the flux power spectrum alone and $A_s = (1.53^{+0.10}_{-0.13})\times10^{-9}$ when including the IGM temperature.
Planck \cite{2020A&A...641A...6P} found a value of $A_s = \left(2.101^{+0.031}_{-0.034}\right)\times10^{-9}$. 
We also derived the value of $\sigma_8$ implied by our parameters by using CLASS in post-processing \cite{2011arXiv1104.2932L}. 
For the flux power spectrum alone, we find $\sigma_8 = 0.733^{+0.026}_{-0.029}$, and when the IGM temperature is included, $\sigma_8 = 0.703^{+0.023}_{-0.027}$ (see Table~\ref{table:parameters}).
The Planck result is $\sigma_8 = 0.811 \pm 0.006$ \cite{2020A&A...641A...6P}.
We thus measure a power spectrum amplitude around $3-\sigma$ lower than Planck or ACT CMB lensing \cite{2023arXiv230405202Q}.
Interestingly, the dark energy survey year 3 results measure $\sigma_8 = 0.733^{+0.039}_{-0.049}$ \cite{2022PhRvD.105b3520A}, in good agreement with our results.
Other small-scale structure probes vary \cite[e.g.~][]{2020JCAP...05..042I, 2022JHEAp..34...49A, 2023JCAP...04..057Y}. 

Figure~\ref{fig:fps_data} shows the \lya forest flux power spectrum from \cite{2019JCAP...07..017C}, along with their estimated one sigma uncertainty (black).
Also shown are predictions from our multi-fidelity emulator based on the maximum posterior input parameters from MCMC analysis with only the \lya forest flux power emulator in the full and reduced redshift ranges, and MCMC analysis using both the IGM temperature and flux power emulators.
The correlation between \lya and Si~{\sc iii} absorption is visible in the form of regular oscillations in the power spectrum (in Section~\ref{sec:likelihood} we describe the correction we make for Si~{\sc iii}).
The best-fit flux power spectrum is not significantly affected by the inclusion of the $T_0$ data in the likelihood, although the fit is slightly worse at $z=2.6$. %, as is expected if the $T_0$ data mainly breaks degeneracies.

Table~\ref{table:parameters} shows the constraints our chains imply for the linear theory power at $z=3$, $\Delta_L^2(k,z)$ and its slope, $n_\mathrm{eff}$ \cite{2005ApJ...635..761M, 2019JCAP...07..017C, 2023ApJ...944..223P}.
These are defined by:
\begin{align}
 \Delta_L^2(k_*, z_*) = k_*^3 P_L(k_*,z_*) / (2\pi^2) \\
 n_\mathrm{eff}(k_*, z_*) = \frac{d \ln P_L}{d \ln k} (k_*, z_*)\,.
\end{align}
Here $P_L(k,z)$ is the linear theory power spectrum, evaluated at $z_* = 3$ and $k_* = 0.0009\;\mathrm{s/km}$. We find $\Delta_L^2 = 0.302^{+0.024}_{-0.027}$ and $n_\mathrm{eff} = -2.264^{+0.026}_{-0.018}$ from our reduced redshift chains.
The correlation coefficient between $\Delta_L$ and $n_\mathrm{eff}$ is $0.42$ for the flux power spectrum only (FPS $z>2.6$) chain and $0.40$ when the mean IGM temperature data is included (FPS + $T_0$).

% --------------------------------------------------------------------------------------------------
% --------------------------------------------------------------------------------------------------

\subsection{Reionization and Other Parameters}\label{sec:astro}

\begin{figure}
    \centering
    \includegraphics[width=0.85\textwidth]{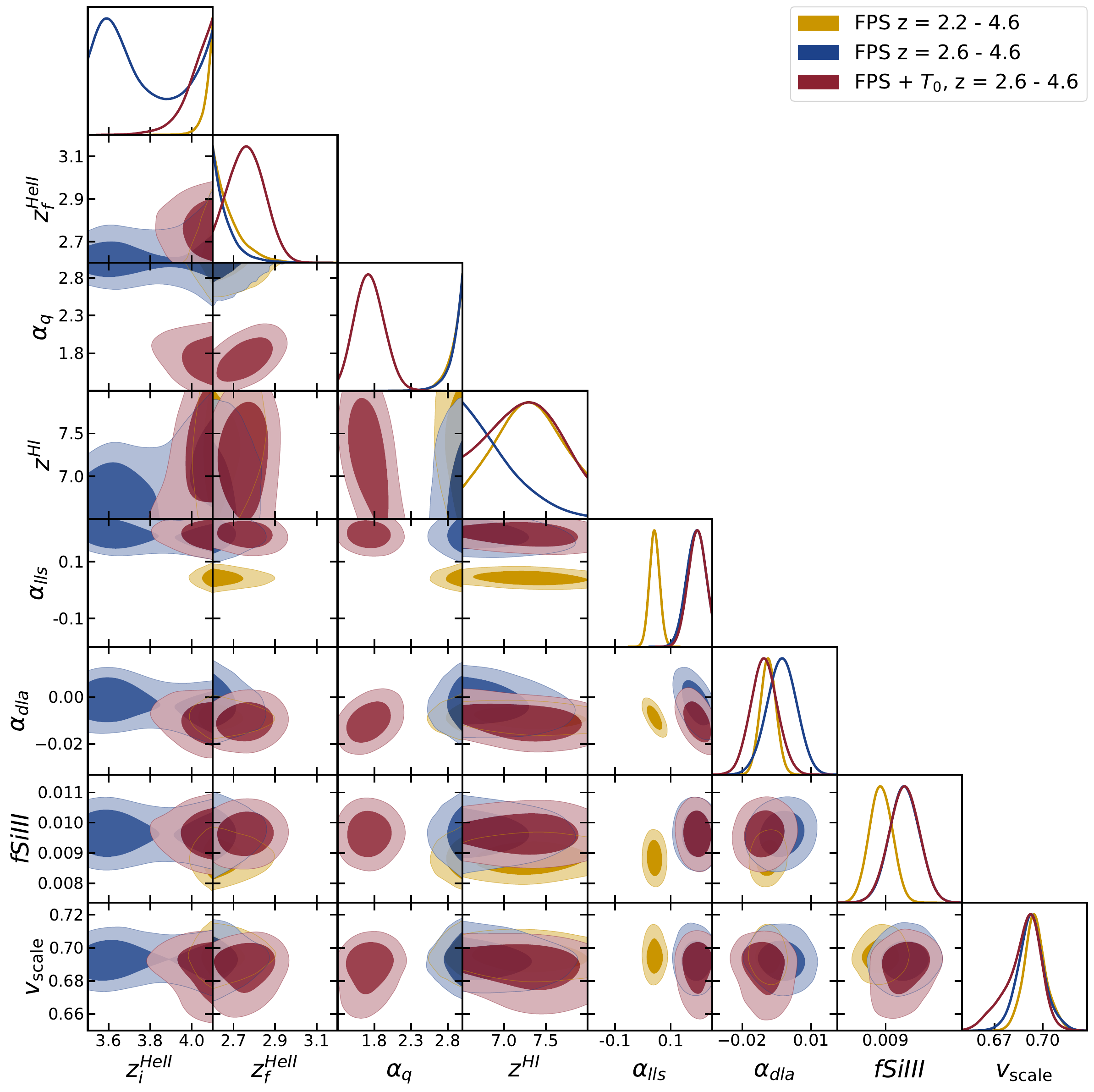}
    \caption{\label{fig:astro_corner}
    Posterior constraints for the parameters of the helium reionization model ($z_i^{HeII}$, $z_f^{HeII}$, $\alpha_q$), the hydrogen reionization model ($z^{HI}$), the strong absorber models ($\alpha_{LLS}$, $\alpha_{DLA}$), the Silicon III correction ($f_\mathrm{SiIII}$), and the velocity to distance scale parameter $v_{scale}$.
    Results are from the same three MCMC chains as Figure~\ref{fig:cosmo_corner}. `FPS $z=2.2-4.6$' (gold) uses the full redshift range eBOSS flux power spectrum dataset, `FPS $z=2.6-4.6$' (blue) uses a reduced redshift range eBOSS dataset flux power spectrum dataset, which removes the internal tension (Section~\ref{sec:tension}). The third chain, `FPS $+ T_0, z=2.6-4.6$' (red), uses the limited range eBOSS dataset but adds the IGM temperature constraints.
    }
\end{figure}

\begin{figure}
    \centering
    \includegraphics[width=0.75\textwidth]{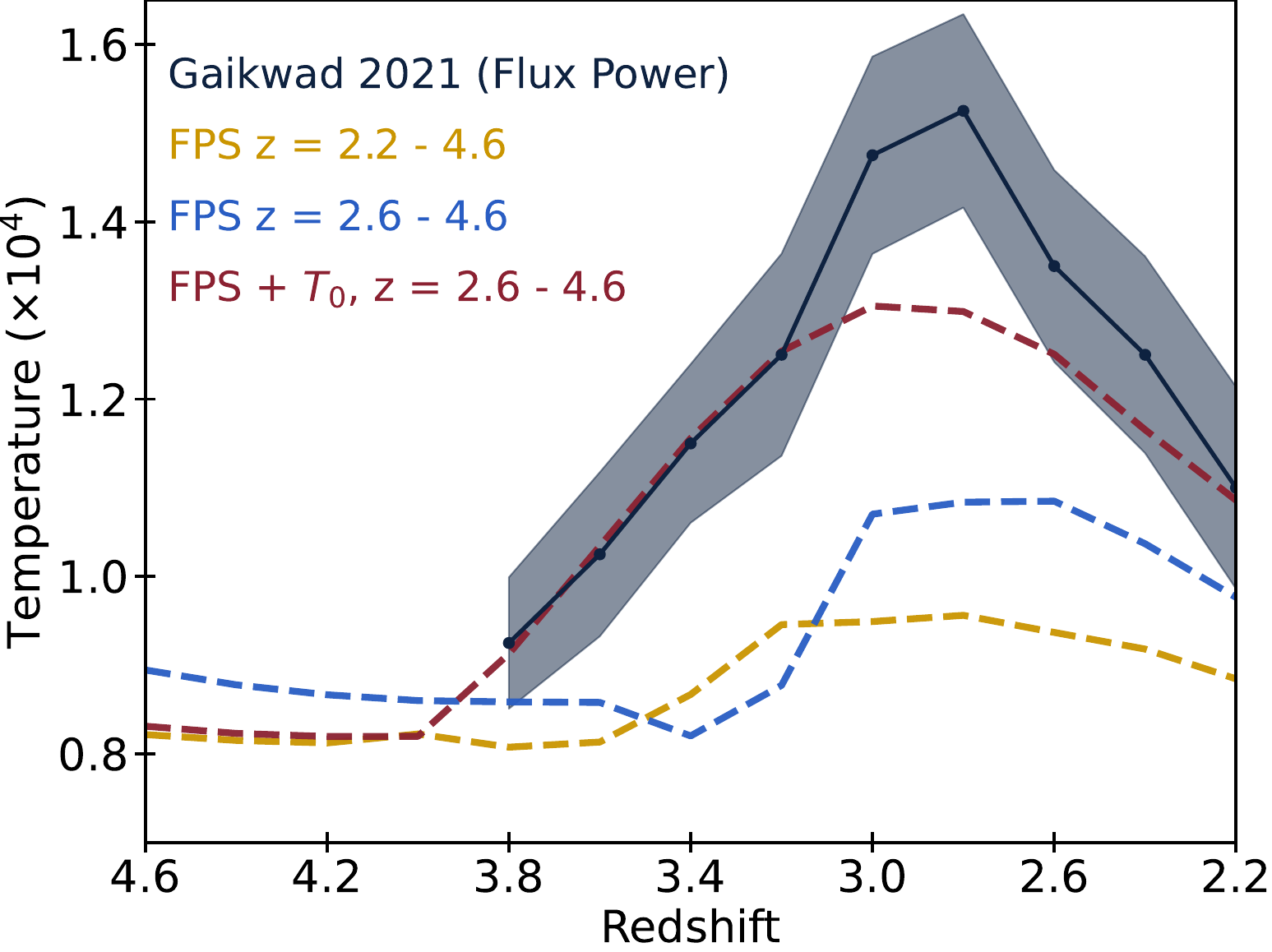}
    \caption{\label{fig:temp_data}
    IGM temperatures at mean density from \protect\cite{2021MNRAS.506.4389G} (black lines and circles, with shading corresponding to one sigma uncertainty).
    Specifically their temperatures derived from the flux power spectrum calculated using high resolution \lya forest spectra.
    Also shown are predictions for the IGM temperature from our multi-fidelity emulator corresponding to the same three maximum posterior input parameters used in Figure~\protect\ref{fig:fps_data}.
    }
\end{figure}

Figure~\ref{fig:astro_corner} shows the other parameters of our model from the same chains as Figure~\ref{fig:cosmo_corner}.
These are: three parameters of the helium reionization model (z$^{\text{He~{\sc ii}}}_i$, z$^{\text{He~{\sc ii}}}_f$, and $\alpha_q$), the midpoint of H~{\sc{i}} reionization z$^{\text{H~{\sc i}}}$, the parameters of the strong absorber model ($\alpha_{LLS}$, $\alpha_{DLA}$), the strength of the metal contamination $f_{SiIII}$ and the box velocity scale $v_{scale}$.

The combined data prefer an early start to helium reionization, $z_i^{HeII} > 3.87$ at $95\%$ confidence \footnote{The flux power only chain admits a solution where both helium and hydrogen reionization are late, but this is ruled out by the inclusion of IGM temperature data.}.
Interestingly, this is in agreement with constraints from the helium \lya forest, where regions of high transmission suggest that HeII reionization has already started at $z = 3.5$ \cite{2016ApJ...825..144W, 2021ApJ...912...38M}.
The end of helium reionization, $z_f^{HeII}$, is constrained by the flux power spectrum data alone to be $z_f^{HeII} < 2.8$.
Adding the IGM temperature data reinforces this and constrains it to finish by $z=2.6$, so that the IGM temperature can drop at lower redshifts.
This is consistent with the He~{\sc{ii}} \lya forest, which suggests an end at $z \leq 2.7$ \cite{2009ApJ...704L..89M, 2011ApJ...733L..24W, 2019ApJ...875..111W}.
Our constraints on the timing of helium reionization hit our prior volume.
Note this prior volume is set more by the redshift limits of our datasets than our simulation choices.
The flux power spectrum at $z\geq  2.6$ alone, for example, cannot determine whether the IGM is cooling by $z=2.2$.
In addition, the upper prior limit on $z^{HeII}_i$ is $4.1$, which is larger than the highest redshift IGM temperature data.

The most significant effect of the IGM temperature data is on the spectral index during helium reionization, $\alpha_q$.
Smaller values of $\alpha_q$ correspond to a larger heating rate.
The flux power spectrum data prefers a high value of $\alpha_q > 2.64$ at 95\% confidence, which corresponds to minimal heating during helium reionization.
However, as shown in Figure~\ref{fig:temp_data}, this high value of $\alpha_q$ produces an IGM temperature which is low and in disagreement with the data from Ref.~\cite{2021MNRAS.506.4389G}. The posterior constraint on $\alpha_q$ when the IGM temperature data is included is lower than that with the flux power spectrum data alone by about $4-\sigma$.
Figure~\ref{fig:fps_data} shows that the flux power spectrum is not significantly different at the maximum posterior parameters of either chain.
Appendix~\ref{sec:t0-only} shows the results of chains which include only the IGM temperature data.
They are consistent with the combined chains, and the helium reionization parameters are constrained at similar values, although the maximum posterior $\alpha_q$ is slightly lower than in the joint chains.

There are two possible explanations for this discrepancy in $\alpha_q$. The first is that the discrepancy is caused because the IGM temperature constraints assumed the Planck value of $\sigma_8$, which is inconsistent with our results. Assuming a larger value for $\sigma_8$ will increase the flux power spectrum and require a greater degree of thermal heating to smooth the gas. The IGM temperature constraints from Ref.~\cite{2021MNRAS.506.4389G} may thus be biased high. A consistent analysis of the flux power spectra from both eBOSS and XQ-100 varying $A_P$ and $\alpha_q$ simultaneously would resolve this question, and we will perform it in future work.
The second possibility is suggested by Table~\ref{table:chi2}, which reveals that the high $\alpha_q$ is driven by the fit to the $z=2.6$ flux power spectrum bin.
It is thus possible that the same effect which is responsible for the tension between the full redshift and reduced redshift chains is also present at $z=2.6$ in the flux power spectrum, exploiting partial parameter degeneracies to move the posterior confidence intervals on $\alpha_q$.
We will in future work examine this possibility with the higher resolution DESI data.

The midpoint of hydrogen reionization is poorly constrained in all models.
The redshift range explored here ($z=2.2-4.6$) is well after the completion of hydrogen reionization, even in models where it ends late. All our chains are consistent with a midpoint of z$^{\text{H~{\sc i}}} \sim 7$.
This is within the range allowed by other experiments: Planck suggests z$^{\text{H~{\sc i}}} = 7.7 \pm 0.7$, while an analysis of the \Lya emitter luminosity function finds z$^{\text{H~{\sc i}}} \sim 7.25$ \cite{2021ApJ...919..120M}.

We show results for the nuisance parameters associated with the strong absorbers and the \Lya-SiIII cross-correlation.
As discussed in Ref.~\cite{2023simsuite}, our simulation suite includes strong absorbers self-consistently using a galaxy formation model.
The strong absorber parameters measure the difference between the strong absorber model in the simulation and the model in the observed spectra, so that $\alpha_{LLS} = 0$ means that our galaxy formation model is a good match to the circumgalactic gas in the observed Universe.
DLAs are subtracted from both the observed and simulated spectra, and so $\alpha_{DLA}$ measures primarily the efficiency of the observational DLA finder. 
All our chains produce a posterior $\alpha_{DLA}$ tightly peaked and close to $0$.
The chains using the flux power alone are centered on $\alpha_{DLA} = 0$, while the chain including the IGM temperature prefers a slightly negative value, perhaps indicating that the eBOSS DLA finder includes some false positives\footnote{As DLAs finders are sensitive to reduced flux transmission, regions with more \Lya forest absorption are more likely than average to be flagged as a DLA. The efficiency of the DLA finder may thus affect the flux power spectrum.}.
Note that DESI includes improved DLA finding algorithms based on machine learning \cite{Ho:2020,Parks:2018, 2022arXiv220100827W}.

For the restricted redshift chains, we measure $\alpha_{LLS} \sim 0.19$, while for the full redshift range $\alpha_{LLS} \sim 0.04$.
The preference for a non-zero $\alpha_{LLS}$ in the reduced redshift chains suggest that our simulations have fewer LLS than the real Universe.
LLS are on the boundary of being optically thick and thus radiative transfer effects within the gas are important, making them the most difficult absorbers to model accurately.
$\alpha_{LLS}$ can affect the flux power spectrum normalisation, and so it is reasonable to interpret the preference of the full redshift chains for a low $\alpha_{LLS}$ as an artifact of the fit.
However, it is also possible that the low $\alpha_{LLS}$ points to the origin of the internal tension, a possibility we discuss further in Section~\ref{sec:tension}. 

The SiIII cross-correlation is $f_{SiIII} = 0.0096 \pm 0.001$ (95\% confidence) from our $z=2.6 - 4.6$ chains.
The full redshift range prefers a slightly lower value of $0.0085 \pm 0.0008$, which is in good agreement with the measurement of $0.008 \pm 0.001$ from DR9 by Ref.~\cite{2013A&A...559A..85P}.
The effect of $f_{SiIII}$ can be seen in the oscillations of the flux power spectrum in Figure~\ref{fig:fps_data}.
The results for $v_\mathrm{scale}$ are dominated by the prior, as expected \cite{2015JCAP...11..011P, 2020JCAP...04..038P}.
Constraints are weaker than for the simulated data, likely because the simulated data did not include noise and so was over-fitting.

Figure~\ref{fig:temp_data} shows the IGM temperature from the flux power spectrum on small scales from Ref.~\cite{2021MNRAS.506.4389G}.
Also shown in Figure~\ref{fig:temp_data} are predictions from our multi-fidelity emulator based on the maximum posterior input parameters from the same chains used in Figure~\ref{fig:fps_data}.
Once the IGM temperature data is included in the fit, the chains are in good agreement.
However, when it is not included the chains prefer a lower IGM temperature, as discussed above.
The thermal history preferred by the full redshift range of the flux power spectrum is similar to that preferred by the restricted range flux power spectrum.

\subsection{Parameter Correlations}
\label{sec:correlations}

\begin{figure}
    \centering
    \includegraphics[width=0.98\textwidth]{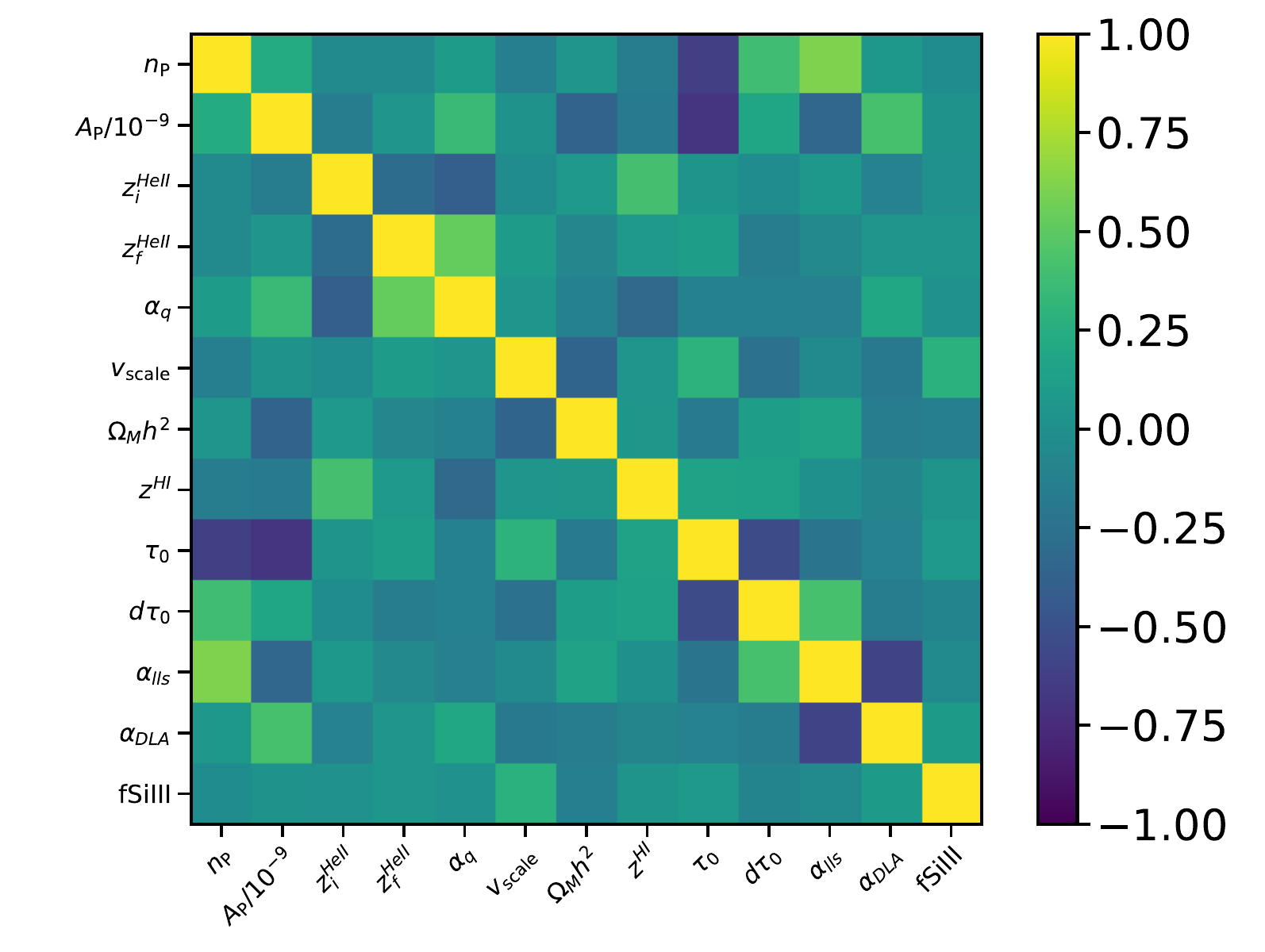}
    \caption{\label{fig:correlations}
    Correlation matrix between parameters for the chain using the flux power spectrum and IGM temperature for $z: 2.6-4.6$.
    }
\end{figure}

Figure~\ref{fig:correlations} shows the correlations between our parameters, for the chain using the flux power spectrum from $z=2.6 - 4.6$, as well as the IGM temperature.
Most correlations are weak.
We have deliberately chosen our pivot scale of $0.78$ Mpc$^{-1}$ to minimise the correlation between $A_P$ and $n_P$, and the correlation matrix confirms it is weak, with a correlation coefficient $r=0.2$. 
%Several of the stronger correlations arise from intrinsic degeneracies in the definition of the parameters and could in principle be reduced by modest redefinitions of pivot scales or pivot redshifts. 
There is a correlation between $\tau_0$ and $d\tau_0$ ($r=-0.54$), as the redshift bin which provides the strongest constraints on the optical depth is not exactly $z=3$.
The optical depth $\tau_0$ is anti-correlated with both $A_P$ ($r=-0.7$) and $n_P$ ($r=-0.62$) as its main effect is to change the amplitude of the flux power spectrum. 

There is a three-dimensional degeneracy between $\alpha_q$, $z_i^{HeII}$ and $z_i^{HeII}$ (see Figure~\ref{fig:correlations}), which allows a wide range of $\alpha_q$ to fit the flux power spectrum data, and is only broken by information from the thermal history.
Lower $\alpha_q$ corresponds to more heating from quasars during He~{\sc{ii}} reionization.
If He~{\sc{ii}} reionization starts earlier or ends later, the IGM requires more heating from quasars to match the observations, while the opposite is true for late starting, or early ending He~{\sc{ii}} reionization.
Appendix~\ref{sec:t0-only} shows the results of chains which include only the IGM temperature data, which clearly shows this three-dimensional degeneracy: a slightly later start to helium reionization would require less total heating and thus a higher value of $\alpha_q$. 
Several of these correlations could be broken by the inclusion of higher redshift thermal history data, or lower redshift flux power spectrum data. 

Finally, the abundance of Lyman Limit Systems, $\alpha_{LLS}$, exhibits several interesting correlations.
$\alpha_{LLS}$ is anti-correlated with $\alpha_{DLA}$ ($r= - 0.59$), as the flux power spectrum templates for strong absorbers have similar shapes in neighbouring column density bins.
$\alpha_{LLS}$ is also correlated with $n_P$ ($r=0.61$) and $d\tau_0$ ($r=0.42$), due to similarities in the shapes of their flux power spectrum templates.
The combination of the three-way correlation between $n_P$, $\alpha_{LLS}$ and $\tau_0$ is exploited by the chains to explain the inconsistent $z=2.2, 2.4$ flux power spectrum bins and drives the discrepant constraints these chains show.
This correlation may be reduced by the inclusion of extra small-scale data available in the DESI early data release.

%--------------------------------------------------------------------------------------------------
\section{Discussion}
\label{sec:discussion}

In this Section we discuss the implications of the results in Section~\ref{sec:results}. 
Section~\ref{sec:tension} discusses possible explanations for the internal tension in the data between $z = 2.2 - 2.4$ and $z \geq 2.6$.
Section~\ref{sec:comparison} compares our results to other datasets and earlier \Lya~analyses.
Section~\ref{sec:altlikelihood} discusses how our results are affected by modifications to our likelihood.

\subsection{The Tension in the Lowest Redshift Bins}
\label{sec:tension}

\begin{figure}
    \centering
    \includegraphics[width=0.45\textwidth]{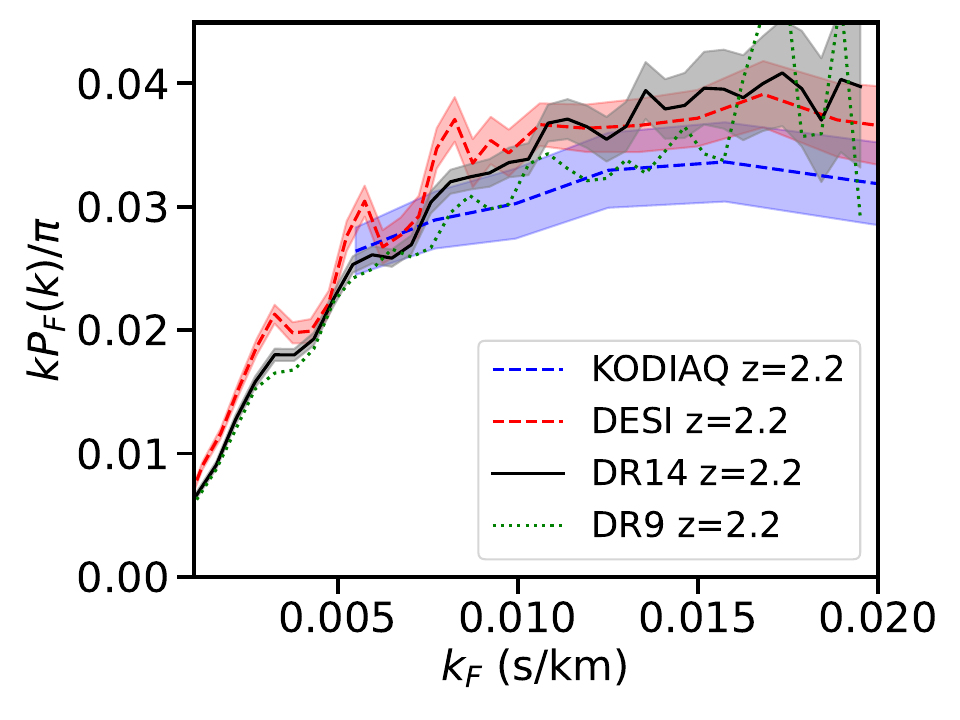}
    \includegraphics[width=0.45\textwidth]{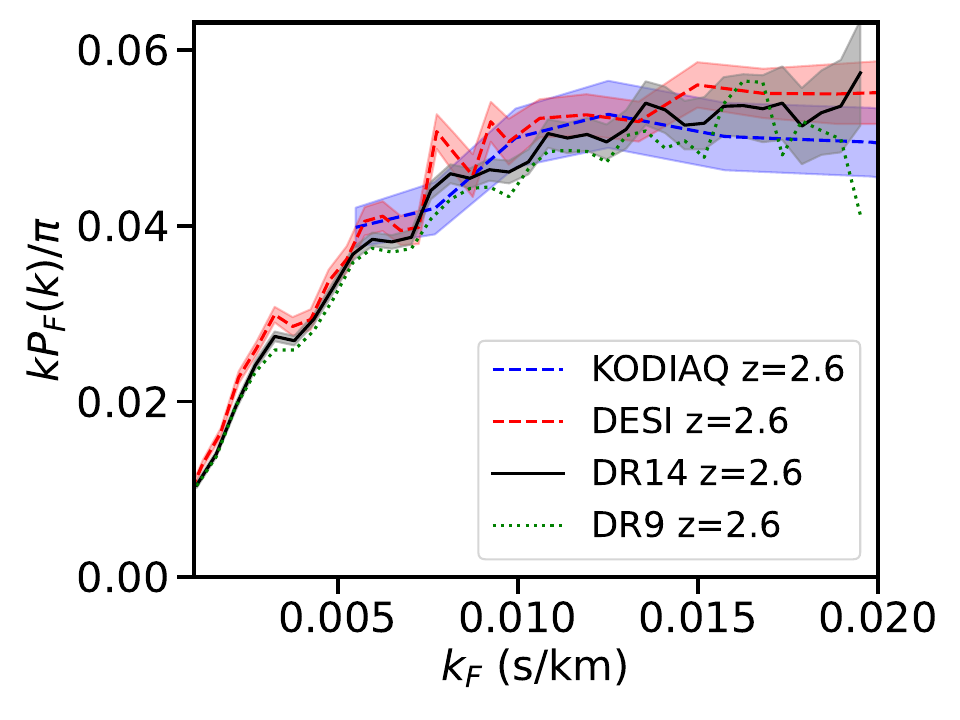}
    \caption{\label{fig:p1d_data}
    Observational 1D power spectrum data from SDSS DR14 (solid black) \protect\cite{2019JCAP...07..017C}, SDSS DR9 (green dotted) \protect\cite{2013A&A...559A..85P} and KODIAQ/SQUAD \protect\cite{2022MNRAS.509.2842K}.
    Filled bands show the range covered by diagonal elements of the covariance matrix.
    (Left) At $z=2.2$.
    (Right) At $z=2.6$.}
\end{figure}

In this Section, we discuss possible explanations for the internal tension between the flux power spectrum data at $z=2.2 - 2.4$ and $z \geq 2.6$.
There are two generic possibilities: either an important physical effect is missing from our simulation model, or there is a systematic in the dataset not captured by the systematic error budget. 

To evaluate the possibility of systematic error, we can look at independent measurements of the flux power spectrum on similar scales and at similar redshifts.
Figure~\ref{fig:p1d_data} shows different measurements of the 1D flux power, $P_F(k)$, at $z=2.2$ and $z=2.6$.
We show the results from SDSS DR14 \cite{2019JCAP...07..017C}, SDSS DR9 \cite{2013A&A...559A..85P}, a recent analysis using high resolution spectra from KODIAQ/SQUAD \cite{2022MNRAS.509.2842K}, and from DESI Early Data Release data (DESI EDR) \cite{2023arXiv230606316G}.
At $z=2.6$ (and higher redshift bins) all analyses are in reasonably good agreement, given their respective statistical errors.
However, this is not the case at $z=2.2$, where there is a discrepancy between SDSS (DR14 and DR9), DESI and KODIAQ.
SDSS DR14 and DR9 are in reasonable agreement.
Appendix \ref{sec:dr9_results} shows the posterior parameter constraints for DR9.

The KODIAQ flux power spectrum is lower by around $1-\sigma$ on the smallest scales measured by eBOSS, which could be due to the effect of continuum modelling in the KODIAQ data \cite{2022MNRAS.509.2842K}.
The DESI EDR data agrees well with eBOSS for $k > 0.01$ s/km, but is discrepant by $> 2 \sigma$ for $k < 0.01$ s/km.
This discrepancy is also present at $z=2.4$, and is discussed in the DESI EDR papers, see Ref.~\cite{2023arXiv230606311R}, Appendix D.
They ascribed $30\%$ of the difference between eBOSS and DESI to continuum fitting, but the origin of the rest is currently unclear.
Given the fairly large disagreements between different measurements at $z=2.2$, systematic error is a highly plausible explanation for the internal tension we find.
In future work we will combine our likelihood function with the DESI flux power spectra and investigate their cosmological implications.

We should also consider possible theoretical explanations.
Explanations rooted in alternative early Universe models seem a priori unlikely as they would have to cause an effect only for $z < 2.6$, when the Universe is known to be matter dominated. 
However, there are a few possible astrophysical explanations.
Feedback effects from AGN become increasingly important at low redshift.
It is possible that a stronger AGN feedback prescription than we use, or than is implemented in current cosmological simulation suites, could efficiently disrupt gas at $z < 2.6$ on small scales and explain these results.
Interestingly, such an AGN feedback model has recently been proposed as an explanation for the low value of $S_8 = \sigma_8 (\Omega_M/0.3)^{0.5}$ preferred by some weak lensing surveys \cite{2022MNRAS.516.5355A}, which matches our results.
Ref.~\cite{2020MNRAS.495.1825C} found that AGN feedback affects the forest at the $8\%$ level at $z=2$, but their AGN model has a different mean flux to their non-AGN simulation, and once this is corrected for the effect is much smaller. Ref.~\cite{2013MNRAS.429.1734V} finds a $10\%$ effect of AGN feedback, but their AGN model results in a star formation rate which is a factor of $10$ too low at $z=2$ \cite{2010MNRAS.402.1536S}. Existing models of AGN feedback which also match the galaxy stellar mass function thus do not seem able to affect the \Lya forest at $z \sim 2$ enough to explain this tension. We note, however, that more aggressive models are possible.
%Ref.~\cite{2023AJ....166..228T} examined a wide range of AGN feedback models, including some much more aggressive than those in PRIYA (or ASTRID), finding strong effects at $z < 1$, but the $z=2$ analysis is ongoing.

DLAs are important at low redshift, and do affect the slope of the flux power spectrum.
Our simulations include a population of DLAs in good agreement with observations \cite{2023simsuite}, which are masked using the same procedure as the observational pipeline.
We include a free parameter to model the residual power from any DLAs not detected by eBOSS, and the posterior value for this free parameter is consistent with zero.
We considered separate constraints on the DLA efficiency at low and high redshift: $\alpha_{DLA}(z < 2.6)$ and $\alpha_{DLA}(z \geq 2.6)$. The high redshift DLA efficiency $\alpha_{DLA}(z \geq 2.6)$ remained consistent with zero. However, the low redshift parameter was negative; $\alpha_{DLA}(z < 2.6) = -0.0134 \pm 0.0074$, indicating an excess of large-scale power in eBOSS at $z < 2.6$. While most cosmological parameters were unchanged, $n_P$ increased (by about $2-\sigma$) to $n_P = 0.916 \pm 0.014$, reducing the internal tension by $1/3$. Since this is only a partial resolution, the DLA parameter may be sensitive to a continuum fitting problem in the (relatively short) low redshift spectra.

At low redshift, the \Lya~forest is increasingly contaminated by metal lines.
We include a simple prescription for \ion{Si}{III} and the inclusion of low redshift data does not drive the best-fit parameter for this model, preferring slightly less metal contamination.
However, it is possible that a more sophisticated model could help reduce the tension.

One interesting but entirely speculative possibility is suggested by the LLS abundance, $\alpha_{LLS}$, which is lower in the full redshift chains.
Ref.~\cite{Prochaska:2009a} identified a systematic in the SDSS colour selection which causes quasar sightlines containing Lyman Limit Systems (LLS) to be preferentially selected for spectroscopic followup.
Refs.~\cite{Worseck:2011, Fumagalli:2013} showed that, due to the width of the $u$-band filter in SDSS, LLS are over-sampled for $z=2.5-3.6$ for all quasars in the redshift range $z=3.0-3.6$.
It is thus possible that $\alpha_{LLS}$ depends on the quasar (not absorber) redshift.
Note that the flux power spectrum we measure depends on the absorber redshift and so a simple redshift split would not detect this effect\footnote{We ran a chain with two $\alpha_{LLS}$ parameters for $z < 2.6$ and $z \geq 2.6$. The maximum likelihood for $\alpha_{LLS}$ at $z > 2.6$ was $\sim 2-\sigma$ larger than in the full redshift chain, and $n_P$ increased by $\sim 0.5\sigma$, but all other parameters were unchanged.}.
A check for colour selection systematics would involve the flux power spectrum being computed from two different quasar redshift bins.
%The \Lya forest probes marginally higher over-densities at low redshifts, so it is possible that a novel self-interacting dark matter model could be constructed which explains these observations, but it would have trouble matching cluster data (commented because I think the model would be a bit contrived).

\subsection[Comparison of the Posterior Constraints to Other Lyman-alpha Analyses]{Comparison of the Posterior Constraints to Other \Lya Analyses}
\label{sec:comparison}

%Shift between DR9 and DR14 found in PD2020 but we don't (perhaps because the DLA model absorbs it).

%Other results: can compare only for the full redshift range.
%PD2020: ns low, sigma8 high, tau0 normal. We find tau0 high, sigma8 low, ns low. Quantitatively our results are more discrepant with Planck than theirs. Perhaps this is because their Taylor expansion tends to flatten gradients far from the central value, perhaps because we are measuring nP on small scales, not ns on large scales.
%PD2020 explicitly tested excluding the two lowest redshift bins and found no change in cosmological parameters. Not clear why.

It is interesting to compare the results of our chains to those of Ref.~\cite{2020JCAP...04..038P} (for DR14) and Ref.~\cite{2015JCAP...11..011P} (for DR9).
The most notable difference is that Ref.~\cite{2020JCAP...04..038P} tested excluding the lowest two redshift bins and found minimal change in the posteriors of their cosmological parameter.
This disagrees with our results.
We believe this discrepancy can be ascribed to our different treatment of nuisance parameters. Ref.~\cite{2020JCAP...04..038P} employed correction functions for supernova feedback from the OWLs simulation suite \cite{2013MNRAS.429.1734V} and for AGN feedback from the Horizon-AGN suite \cite{2020MNRAS.495.1825C}.
Each correction function is most significant at low redshift, and is included with a free amplitude parameter which is marginalised over.
In addition, earlier models were forced by computational limits to use the `splicing' technique of Ref.~\cite{2014JCAP...07..005B}.
In this model multiple simulation boxes with over-lapping scale ranges are combined to model the scales probed by the \Lya~forest.
A single larger simulation with $2048^3$ particles was used to generate a scale and redshift dependent correction function, and the amplitude of this correction was marginalised over with a Gaussian prior.

Our larger simulations can instead model all relevant scales in a single simulation and so do not need a free parameter for splicing.
In addition, we self-consistently incorporate models for stellar and AGN feedback and star formation into our simulations.
Thus the analysis of Ref.~\cite{2020JCAP...04..038P} differs from ours in that it has three nuisance parameters, each of which affects the lowest redshift bins most strongly and each of which is marginalised over in the chains.
Ref.~\cite{2015JCAP...02..045P} mentions that removing splicing reduces $n_s$ significantly (although the posterior value of the splicing correction is not reported), which is what we would expect if the splicing correction were absorbing an internal tension. 
Thus we believe that the $z=2.2$ and $z=2.4$ redshift bins contribute only marginally to the cosmological constraints in Ref.~\cite{2020JCAP...04..038P} and instead constrain splicing and AGN feedback. 
We should therefore compare the quantitative results of Ref.~\cite{2020JCAP...04..038P} to our reduced redshift chains. 

Ref.~\cite{2020JCAP...04..038P} found $n_s = 0.954 \pm 0.006$ for DR14 and $n_s = 0.938 \pm 0.010$ for DR9.
Meanwhile the power spectrum amplitude as measured by $\sigma_8$ is $0.826 \pm 0.02$ in DR14 and similar in DR9.
As shown in Appendix~\ref{sec:dr9_results}, we do not reproduce this shift in $n_s$, although we do observe a smaller shift in $d\tau_0$.
Ref.~\cite{2020JCAP...04..038P} attribute this shift to the different catalogues for masking DLAs and BAL, and we marginalise out the DLA masking.

Ref.~\cite{2015JCAP...11..011P} find a mean optical depth of $\tau_{eff} (z=3) = 0.0025 \pm 0.0001$ and $d\tau = 3.734 \pm 0.015$ in DR9.
We define the mean optical depth relative to the power law $\tau_{eff} = 0.0023 (1+z)^{3.65}$, so that in our parameterization these constraints correspond to $\tau_0 = 1.09 \pm 0.05$ and $d\tau_0 = 0.084\pm 0.015$.
Their optical depth measurements from $z=2.2 - 4.4$ in DR9 are thus in good agreement with our measurements for the reduced redshift range of $z=2.6 - 4.6$ from DR14.
Appendix~\ref{sec:dr9_results} shows that DR14 generally prefers a lower $d\tau_0$ than DR9.
Ref.~\cite{2020JCAP...04..038P} do not report a value for $\tau_{eff}(z=3)$ from DR14. A higher $d\tau_0$ implies a lower $n_P$. It is possible that the Ref.~\cite{2020JCAP...04..038P} DR14 value of $d\tau_0$ is large and discrepant with other optical depth measurements, driving the change in $n_P$.

We find $n_P \sim 1.0$ and $\sigma_8 \sim 0.73$ from the reduced redshift flux power alone, a $3.5-\sigma$ tension in $\sigma_8$. Some of the differences between our constraints on $A_P$ and $\sigma_8$ are due to a lever arm effect: for $n_s < 1$ the power spectrum amplitude will be increased on larger scales and measuring $n_s$ will induce a correlation between $n_s$ and $\sigma_8$.
A direct comparison can be made by comparing our constraints on $\Delta_L^2$ and $n_\mathrm{eff}$ to those of Ref.~\cite{2019JCAP...07..017C}, who found $\Delta_L^2 = 0.31 \pm 0.02$ and $n_\mathrm{eff} = -2.339 \pm 0.006$.
Their measurement of $n_\mathrm{eff}$ is thus $ \sim 4 \sigma$ lower than the maximum posterior from our reduced redshift flux power spectrum chains, while their measurement of $\Delta_L^2$ is in agreement.
The discrepancy between our results thus lies in the measurement of the spectral slope. The ultimate source of this tension is not clear, but our simulation suite is substantially larger and more robust than that used by Ref.~\cite{2020JCAP...04..038P}. A possibility is that the splicing correction they use does not fully account for the correlation of small and large scales (for example, the accuracy of the splice may be cosmology dependent). Another possibility is that the emulator used is a polynomial expansion around a `best-fit' simulation, with $n_s = 0.9624$. This is quite far from the posterior constraints, and so the accuracy of the polynomial emulator may be reduced. A third possibility is the lack of an explicit model for LLS. There is a correlation between $n_P$ and $\alpha_{LLS}$, as shown in Appendix~\ref{sec:full_posteriors}, so that fixing $\alpha_{LLS} = 0$ could produce a low $n_P$.

%Our DR9 and DR14 constraints on \textbf{$A_P$} are in good agreement. Our simulations include self-shielding of the gas following Ref.~\cite{Rahmati:2013} and a realistic DLA model, masked from the flux power spectrum in the same way as the observational data.
%That our posterior values for $\sigma_8$ are not affected by the change in DLA catalogue is reassuring, as it suggests that our analysis is indeed correctly marginalising over the uncertainty from DLAs.

%Notice that the average bias of forest gas without self-shielding is not the same as the average bias of forest gas with self-shielding and DLA masking \cite{2023MNRAS.524.1464P}. Furthermore, galaxy formation model changes which only affect dense gas cannot affect the \Lya flux power spectrum in our modelling and thus we are insensitive to the details of supernova feedback.

Rather than use effective broken power laws for the IGM thermal history as a function of redshift, we have explicit physical models for hydrogen and helium reionization, which include the scale-dependent effects of patchy reionization.
As shown in Figure~\ref{fig:temp_data}, the preferred IGM thermal history shows a temperature peak at $z=2.8$ and thus $T_0(z)$ cannot be described by a power law broken at $z=3$ as assumed in Ref.~\cite{2020JCAP...04..038P} and many earlier works.
Figure~\ref{fig:fps_data} shows that this does not affect the flux power spectrum on the scales measured by eBOSS.
However, DESI data probes smaller scales, so it is not clear that this will be the case in future.

\subsection{Likelihood Modifications}
\label{sec:altlikelihood}

We also considered modifications to the likelihood not shown here.
First, we increased the observational uncertainty from eBOSS by a uniform factor of two.
This increased the posterior uncertainties, but did not significantly resolve the internal tension at low redshift (which is many $\sigma$).
We also considered removing the high redshift data, with $z > 3.8$, as is done in some earlier analyses \cite{2011MNRAS.413.1717B}.
We found that this made little difference as the statistical errors in the high redshift data are large and so they provide little information. 
%We did confirm that the high redshift bins are not in tension with the low redshift data. 
We considered removing the largest and smallest scales with cuts in $k$.
Several of the smallest scale bins are highly correlated, and so removing them either led to very poor constraints or had small effects, depending on the scale cut.
Removing the largest bins on the largest scales increased the posterior uncertainty, but did not noticeably shift the posteriors.
Thus none of these checks show any evidence for an internal tension between scales.

We tested whether a second mean flux rescaling slope would improve the fit to the observed \lya forest flux power (Figure~\ref{fig:fps_data}), especially at lower redshifts and smaller scales.
To do this, we added a second mean flux slope to the MCMC sampled parameters, and assigned each to a specific redshift range (we tested this using a redshift pivot of $z=3$ and $z=3.6$).
Posterior constraints on the other parameters from a chain run using the second mean flux slope were unaffected and the fit was not improved.

\section{Conclusions}\label{sec:conclusions}

We have developed a new likelihood and pipeline for the analysis of \Lya forest data and run MCMC chains using the Cobaya package \cite{2021JCAP...05..057T, 2019ascl.soft10019T}.
Our likelihood is built on a percent-level accurate emulator using the PRIYA simulations \cite{2023simsuite}.
We use the multi-fidelity emulation technique \cite{2022MNRAS.517.3200F} and a set of high resolution simulations to avoid the need to `splice' together multiple simulations resolving different scales.

We model the \Lya forest 1D flux power spectrum from eBOSS \cite{2019JCAP...07..017C}, and include several simulated and post-processed parameters.
Our main cosmological constraints are on the slope and amplitude of the primeval power spectrum on the scales probed by the \Lya forest ($n_P$ and $A_P$). 
We augment our \Lya flux power spectrum likelihood with information about the IGM thermal history \cite{2021MNRAS.506.4389G}.
With this information, we constrain the start, end, and heating rate for a patchy model of helium reionization (z$^{\text{He~{\sc ii}}}_i$, z$^{\text{He~{\sc ii}}}_f$, $\alpha_q$), as well as the mean optical depth and its evolution with redshift ($\tau_0$ and $d\tau_0$).
Our likelihood includes corrections to the \lya forest flux power spectrum from correlated Si~{\sc iii} absorption and from the presence of Damped \lya systems.

We found that the lowest redshift bins in the eBOSS flux power spectrum, at $z=2.2$ and $z=2.4$, produced results which were discrepant with those from higher redshifts.
The flux power spectrum from the DESI early data release is also discrepant with eBOSS at these redshifts.
It thus seems likely that this discrepancy is due to an as-yet unidentified systematic in the eBOSS pipeline at $z < 2.6$. We found that the discrepancy was reduced if the DLA finder efficiency was allowed to be redshift dependent. However, an unmodeled astrophysical effect, perhaps connected with AGN feedback, is still possible.  

We added data on the IGM temperature at mean density to our chains, improving constraints on the thermal history. Our best-fit parameters from the flux power spectrum alone prefer a peak IGM temperature data much lower than the constraints of Ref.~\cite{2021MNRAS.506.4389G}, although the eBOSS flux power spectrum only weakly constrains helium reionization. It is possible that these constraints are affected by assuming the Planck value of $\sigma_8$, which is inconsistent with our results. The low temperature is driven by improving the fit at $z=2.6$, so it is also possible some residual effect connected with the tension between the low redshift bins is biasing the flux power only chains. Our least constraining and so most conservative constraints are those from the eBOSS flux power spectrum alone.

When removing the lowest redshift bins from the analysis, we find, for a primeval power spectrum with a pivot scale of $0.78$ 1/Mpc:
\begin{itemize}
    \item A power spectrum slope of $n_P = 1.009^{+0.027}_{-0.018}$ from the eBOSS flux power alone and $n_P = 0.983\pm 0.020$ when adding the IGM temperature data, both in reasonable ($2-\sigma$) agreement with Planck.
    \item A power spectrum amplitude $A_P = 1.69^{+0.14}_{-0.15}$ from the eBOSS flux power spectrum, which translates to $A_s=\left(1.65^{+0.12}_{-0.14}\times10^{-9}\right)$ or $\sigma_8 = 0.733^{+0.026}_{-0.029}$, approximately $2-3\sigma$ lower than measurements from Planck, but in agreement with some other measurements from weak lensing or galaxy surveys.
    \item An early start and late finish to helium reionization, beginning at $z > 4.00$ and ending at $z = 2.765^{+0.080}_{-0.093}$ once IGM temperature data is included. Our data is consistent at 95\% confidence with helium reionization still being underway at the lowest redshift eBOSS data we use, $z=2.6$. When IGM temperature data is not included the chain prefers models where the IGM temperature does not increase substantially and thus cannot place strong constraints on the start and end of helium reionization.
    \item Weak constraints on hydrogen reionization and the matter density $\Omega_M h^2$.
\end{itemize}
In terms of the reduced likelihood of \cite{2005ApJ...635..761M, 2019JCAP...07..017C, 2023ApJ...944..223P}, we find a linear power of $\Delta_L^2 = 0.302^{+0.024}_{-0.027}$ and $n_\mathrm{eff} = -2.264^{+0.026}_{-0.018}$ from our $z> 2.6$ flux power chains. The $\Delta_L^2$ constraint is in agreement with earlier analyses, while $n_\mathrm{eff}$ is $4-\sigma$ larger. When the IGM temperature data is included, it reduces both the best-fit $\Delta_L^2$ and $n_\mathrm{eff}$ to $0.267^{+0.018}_{-0.023}$ and $-2.288\pm 0.020$, respectively. Ref.~\cite{2023arXiv231116377R} noted that earlier analyses were in tension with the values implied by Planck of $\Delta^2_L \sim 0.35$ and $n_\mathrm{eff} \sim -2.305$. In our analysis the tension in $n_\mathrm{eff}$ is largely eliminated, while the $\Delta^2_L$ tension remains.

%Future work
In future work, we will combine our \Lya likelihood with other cosmological information, in particular the Planck CMB and Baryon Acoustic Oscillation measurements, with which we can constrain several extensions to the $\Lambda$CDM model.
While we defer quantitative constraints to later papers, we are able to qualitatively discuss the constraints we expect.
We will be able to constrain the running of the spectral index, $\alpha_s = \frac{d n_s}{d \ln k}$.
Constraints on $\alpha_s$ come from the difference between the spectral index measured by the CMB on large scales and the spectral index measured by the \Lya forest on small scales.
% As our $n_P$ constraints are completely consistent with Planck, the combined constraints are likely to be consistent with zero running.
% It is also likely that the strong constraints on early dark energy found by Ref.~\cite{2023arXiv230300746G} from the low $n_s$ in earlier analyses will be weakened.
The sum of neutrino masses can be constrained via a comparison between the power spectrum amplitude on CMB and \Lya scales \cite{2020JCAP...04..025P}. Our constraints are in strong tension with those of Ref.~\cite{2020JCAP...04..038P}, which will certainly affect neutrino mass constraints from the \Lya forest.
Since our preferred power spectrum amplitude is lower than that of Planck, we will likely have a preference for a non-zero neutrino mass, although the strength of the preference and the value preferred is yet to be determined. 

We will also incorporate new data sets into our likelihood.
We will examine the posterior parameter constraints from the DESI EDR flux power spectrum.
The statistical power of DESI EDR is currently weaker than that of eBOSS, but it is able to measure smaller scales ($k_F < 0.05$ s/km rather than $k_F < 0.02$ s/km for eBOSS).
The higher resolution data may also improve the internal consistency of the dataset at $z < 2.6$.
Finally, we can perform a joint analysis of eBOSS and the high resolution \Lya~forest flux power spectra from Ref.~\cite{2022MNRAS.509.2842K}.
These smaller scales would directly measure the parameters of helium reionization, without the intermediate step of the IGM temperature, allowing an end-to-end validation of the consistency of our modelling of large and small scales.

\acknowledgments
MAF is supported by a National Science Foundation Graduate Research Fellowship under grant No. DGE-1326120.
SB was supported by NSF grant AST-1817256 and by NASA-80NSSC21K1840. 
MQ was supported by NSF grant AST-2107821. 
MFH is supported by a NASA FINESST grant No. ASTRO20-0022.
Computing resources were provided by Frontera LRAC AST21005.
The authors acknowledge the Frontera computing project at the Texas Advanced Computing Center (TACC) for providing HPC and storage resources that have contributed to the research results reported within this paper.
Frontera is made possible by National Science Foundation award OAC-1818253.
URL: \url{http://www.tacc.utexas.edu}. Analysis computations were performed using the resources of the UCR HPCC, which were funded by grants from NSF (MRI-2215705, MRI-1429826) and NIH (1S10OD016290-01A1).

\appendix

\section{Leave-one-out versus Emulator Error}
\label{sec:loovsgperr}
\begin{figure}
    \centering
    \includegraphics[width=\textwidth]{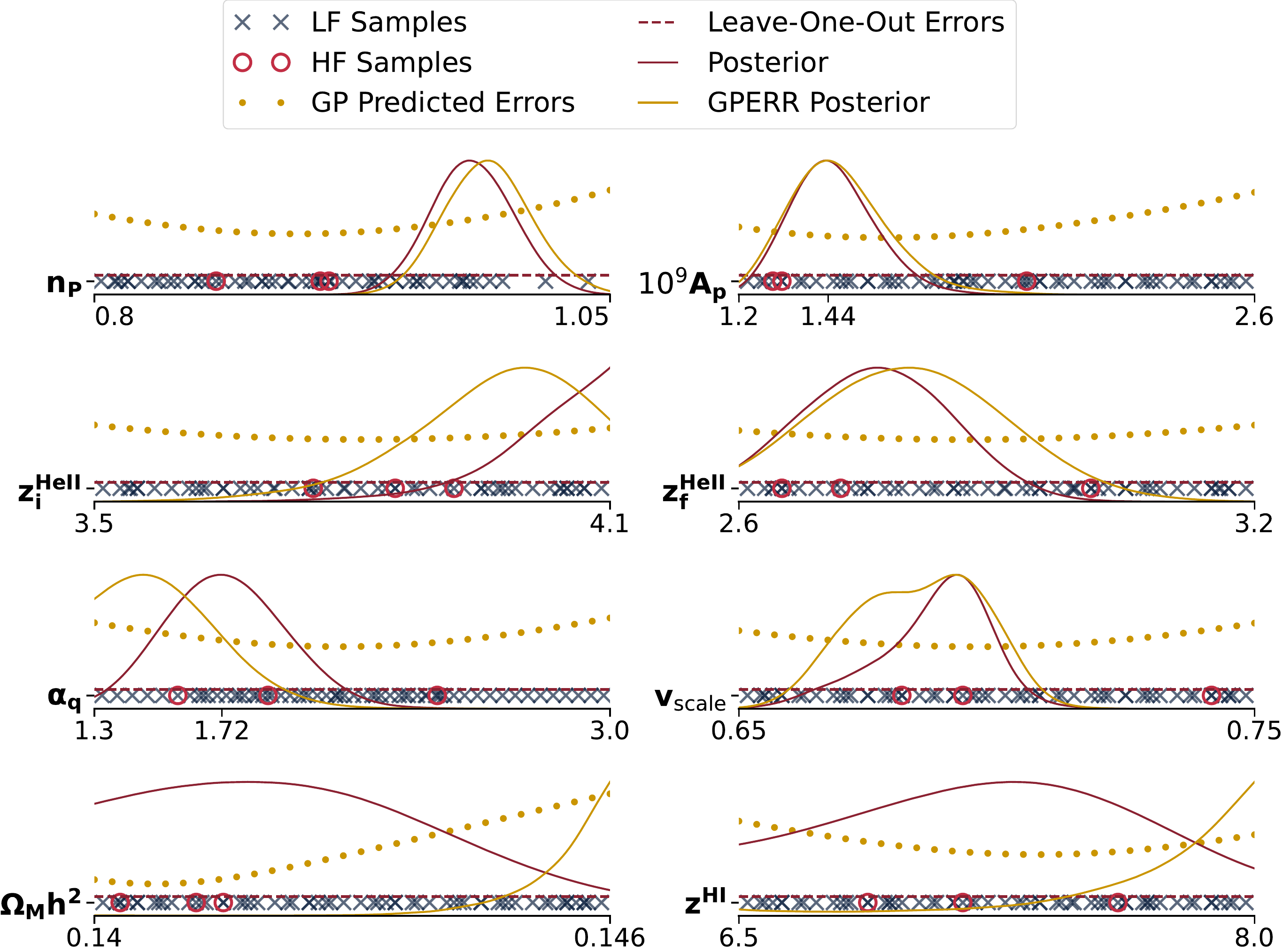}
    \caption{\label{fig:loo_v_emu}
    Emulator error and leave-one-out errors across parameter space.
    For eight of the input parameters, the training samples (grey crosses for LF, red circles for HF), GP emulator errors (yellow dots), and scale-averaged leave-one-out errors (red dashed) are shown. All chains contain the IGM temperature data, and the eBOSS flux power spectrum from $z=2.6 - 4.6$.
    Shown are 1D marginalised posteriors for the chains with the default likelihood (red) compared to chains run adding the GP emulator error to the likelihood (yellow).
    }
\end{figure}

In this Appendix we evaluate the impact of including the Gaussian Process interpolation error, $\boldsymbol{\sigma}_{GP}$, on the posterior parameters, as discussed in Equation~\ref{eq:covariance}.
Figure~\ref{fig:loo_v_emu} compares the effect of including emulator errors, showing the training samples, GP emulator errors and scale-averaged leave-one-out errors.
The leave-one-out error is independent of position in parameter space, whereas the GP error is larger towards the edge of parameter space.
The largest effect is on the matter density $\Omega_M h^2$ and the hydrogen reionization midpoint, which are very weakly measured by the data and so end up dominated by the GP error prior. There are smaller shifts in helium reionization, preferring a slightly later start and extra heating, albeit with shifts less than $1.5-\sigma$.
The cosmological parameters are unchanged.
\section{BOSS DR9 Data}\label{sec:dr9_results}
\begin{figure}
    \centering
    \includegraphics[width=\textwidth]{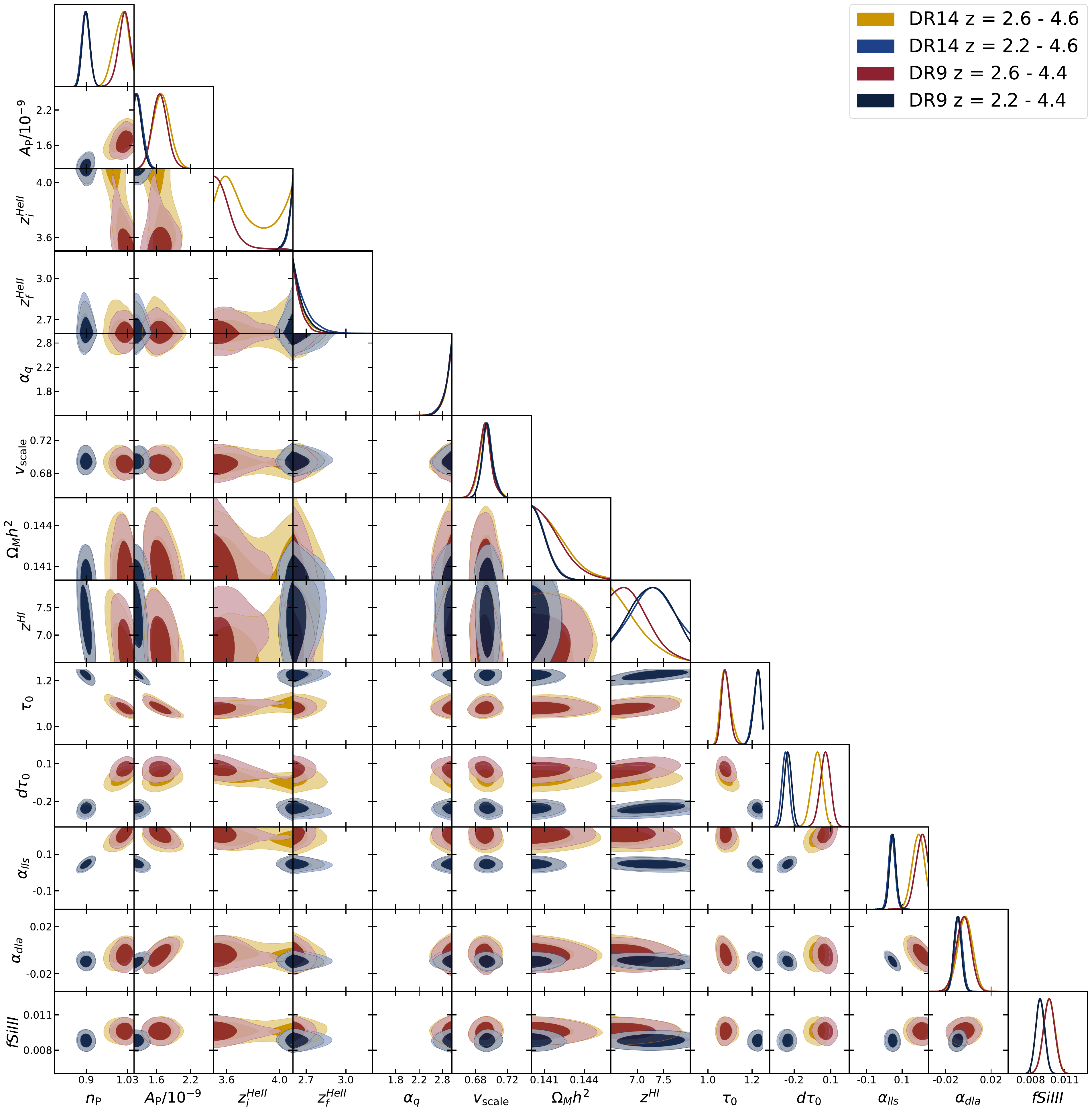}
    \caption{\label{fig:dr9_corner}
    Posteriors for chains run using observations from the earlier SDSS data release, DR9, for the reduced redshift range (gold) and full redshift range (blue), compared to our main chains using DR14 with the reduced redshift range (black) and full redshift range (red).
    }
\end{figure}

In this section we compare posteriors obtained using the flux power spectrum from Ref.~\cite{2013A&A...559A..85P}, which is based on BOSS DR9 quasar spectra.
Figure~\ref{fig:dr9_corner} shows the posteriors for chains run with DR14, with both reduced and full redshift ranges, and chains with DR9, again with the reduced and full redshift ranges. We show chains using only the flux power spectrum likelihood, to emphasise any differences between the datasets.

The tension between $z <2.6$ and $z \geq 2.6$ remains in DR9, and the parameter shifts by dropping the lowest redshift bins are similar. DR9 measures a similar $A_P$ to DR14, and thus also implies a low $\sigma_8$.
The $\tau_0$ posteriors are also similar. The best-fit helium reionization model is similar in both datasets. Although DR14 includes many more quasars, the covariance matrix is often dominated by systematic error. At $z=2.6$ DR14 and DR9 have very similar errors, while DR14 has smaller measurement uncertainty at $z \geq 3$. The posterior uncertainties for DR9 are for most parameters similar to those for DR14.
The largest parameter changes are in the posterior values of $d\tau_0$ and $z_i^{HeII}$, although both are less than $1-\sigma$. Particularly, DR9 prefers a low value of $z_i^{HeII}$, plausibly because of the inclusion of high redshift data in DR14. The good consistency between our posterior constraints in DR9 and DR14 is a validation of the robustness of our analysis.

\section{Posterior Constraints from the IGM Temperature Alone}\label{sec:t0-only}

\begin{figure}
    \centering
    \includegraphics[width=\textwidth]{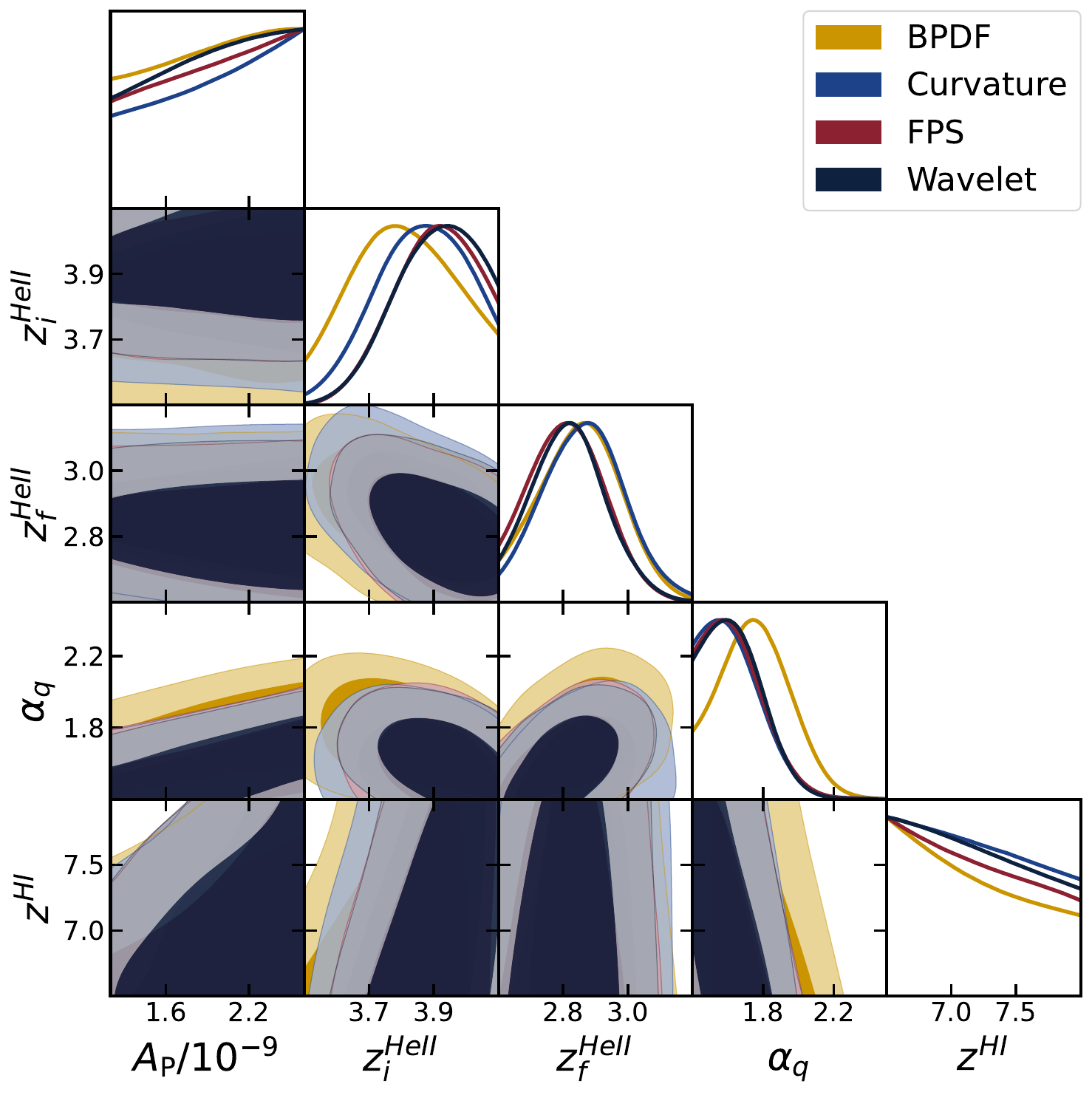}
    \caption{\label{fig:t0_datasets}
    Posteriors for chains run with only the IGM temperature likelihood.
    Shown are chains using each of the four observational measurements of the IGM temperature: using the flux power spectrum (blue), using the Doppler width distribution (BPDF, yellow), using the curvature statistic (red), and using a wavelet decomposition (black).
    The main results of this work use the flux power derived IGM temperatures.
    }
\end{figure}

In this section, we present results from chains run using only the IGM temperature likelihood.
Shown in Figure~\ref{fig:t0_datasets} are four chains, each using one of the observational measurements of the IGM temperature at mean density, derived from different \lya forest summary statistics: the flux power spectrum, the Doppler width distribution, the curvature statistic, and a wavelet decomposition.
Most of the cosmology parameters are entirely unconstrained by the IGM temperature and are omitted from the Figure.
We show $A_P$ for reference.
The three He~{\sc ii} reionization parameters are well constrained by the IGM temperature history.
There is very little difference between the different IGM temperature observations, with the BPDF derived temperature differing the most, specifically preferring a later start to He~{\sc ii} reionization, and less heating, corresponding to a lower IGM temperature.
However, differences are well within $1\sigma$.

The midpoint of H~{\sc i} reionization has a marginal preference for a late midpoint, but is very weakly constrained.
The IGM temperature provides information on $z_{HI}$ as the IGM cools from the completion of H~{\sc i} reionization, setting the temperature before the onset of helium reionization.
Incorporating measurements of the IGM temperature at $z > 3.8$  \cite[e.g.~][]{2023arXiv230402038G} could substantially improve these constraints and we may do so in future work.

\section{Full Posteriors}
\label{sec:full_posteriors}
\begin{figure}
    \centering
    \includegraphics[width=\textwidth]{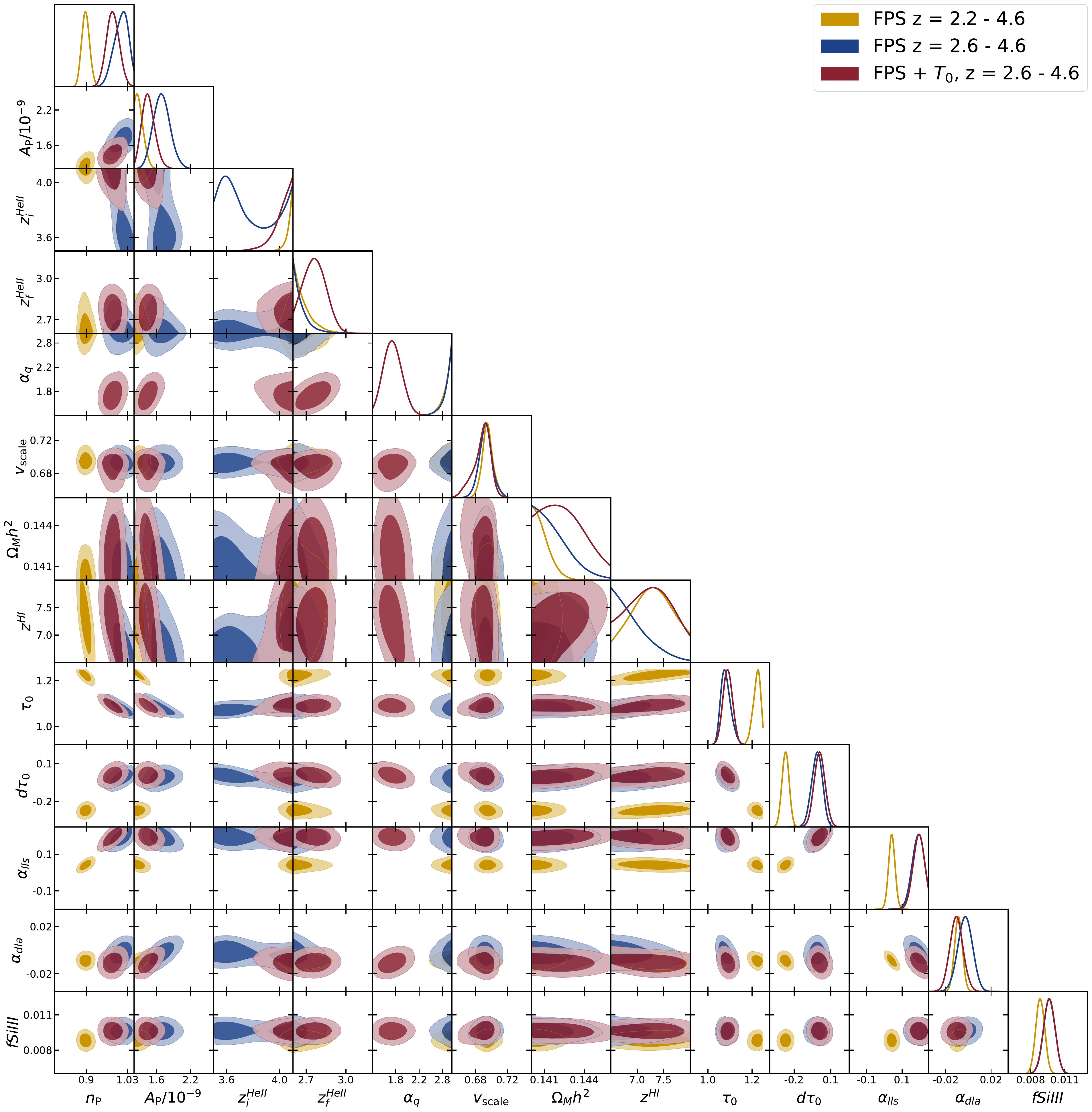}
    \caption{\label{fig:full_posterior}
    Posteriors for the full set of simulation parameters, both cosmological and astrophysical.
    Shown are the same chains from the figures in Section~\ref{sec:results}, thus the correlations between the parameters in Figure~\ref{fig:cosmo_corner} and those in Figure~\ref{fig:astro_corner} are the only new information here. `FPS $z=2.2-4.6$' (gold) uses the full redshift range eBOSS flux power spectrum dataset, `FPS $z=2.6-4.6$' (blue) uses a reduced redshift range eBOSS dataset flux power spectrum dataset, which removes the internal tension (Section~\ref{sec:tension}).
    The third chain, `FPS $+T_0, z=2.6-4.6$' (red), uses the limited range eBOSS dataset but adds the IGM temperature constraints.
    }
\end{figure}

Figure~\ref{fig:full_posterior} presents the full posteriors, including the correlations between the cosmology and astrophysics parameter sets, using the same chains discussed extensively in Section~\ref{sec:results}.
Correlations are discussed in Section~\ref{sec:correlations}.

\bibliographystyle{JHEP.bst}
\bibliography{refs}

\providecommand{\href}[2]{#2}\begingroup\raggedright\begin{thebibliography}{100}

\bibitem{1965ApJ...142.1633G}
J.E.~{Gunn} and B.A.~{Peterson}, \emph{{On the Density of Neutral Hydrogen in
  Intergalactic Space.}}, \href{https://doi.org/10.1086/148444}{\emph{ApJ}
  {\bfseries 142} (1965) 1633}.

\bibitem{1998ApJ...495...44C}
R.A.C.~{Croft}, D.H.~{Weinberg}, N.~{Katz} and L.~{Hernquist}, \emph{{Recovery
  of the Power Spectrum of Mass Fluctuations from Observations of the
  Ly{\ensuremath{\alpha}} Forest}},
  \href{https://doi.org/10.1086/305289}{\emph{ApJ} {\bfseries 495} (1998) 44}
  [\href{https://arxiv.org/abs/astro-ph/9708018}{{\ttfamily
  astro-ph/9708018}}].

\bibitem{1998MNRAS.301..478T}
T.~{Theuns}, A.~{Leonard}, G.~{Efstathiou}, F.R.~{Pearce} and P.A.~{Thomas},
  \emph{{P\^3M-SPH simulations of the Ly{\ensuremath{\alpha}} forest}},
  \href{https://doi.org/10.1046/j.1365-8711.1998.02040.x}{\emph{MNRAS}
  {\bfseries 301} (1998) 478}
  [\href{https://arxiv.org/abs/astro-ph/9805119}{{\ttfamily
  astro-ph/9805119}}].

\bibitem{2000ApJ...543....1M}
P.~{McDonald}, J.~{Miralda-Escud{\'e}}, M.~{Rauch}, W.L.W.~{Sargent},
  T.A.~{Barlow}, R.~{Cen} et~al., \emph{{The Observed Probability Distribution
  Function, Power Spectrum, and Correlation Function of the Transmitted Flux in
  the Ly{\ensuremath{\alpha}} Forest}},
  \href{https://doi.org/10.1086/317079}{\emph{ApJ} {\bfseries 543} (2000) 1}
  [\href{https://arxiv.org/abs/astro-ph/9911196}{{\ttfamily
  astro-ph/9911196}}].

\bibitem{2001ApJ...552...15H}
L.~{Hui}, S.~{Burles}, U.~{Seljak}, R.E.~{Rutledge}, E.~{Magnier} and
  D.~{Tytler}, \emph{{On Estimating the QSO Transmission Power Spectrum}},
  \href{https://doi.org/10.1086/320436}{\emph{ApJ} {\bfseries 552} (2001) 15}
  [\href{https://arxiv.org/abs/astro-ph/0005049}{{\ttfamily
  astro-ph/0005049}}].

\bibitem{2002MNRAS.329..848V}
M.~{Viel}, S.~{Matarrese}, H.J.~{Mo}, M.G.~{Haehnelt} and T.~{Theuns},
  \emph{{Probing the intergalactic medium with the Ly{\ensuremath{\alpha}}
  forest along multiple lines of sight to distant QSOs}},
  \href{https://doi.org/10.1046/j.1365-8711.2002.05060.x}{\emph{MNRAS}
  {\bfseries 329} (2002) 848}
  [\href{https://arxiv.org/abs/astro-ph/0105233}{{\ttfamily
  astro-ph/0105233}}].

\bibitem{2006AJ....132..117F}
X.~{Fan}, M.A.~{Strauss}, R.H.~{Becker}, R.L.~{White}, J.E.~{Gunn},
  G.R.~{Knapp} et~al., \emph{{Constraining the Evolution of the Ionizing
  Background and the Epoch of Reionization with
  z\raisebox{-0.5ex}\textasciitilde6 Quasars. II. A Sample of 19 Quasars}},
  \href{https://doi.org/10.1086/504836}{\emph{AJ} {\bfseries 132} (2006) 117}
  [\href{https://arxiv.org/abs/astro-ph/0512082}{{\ttfamily
  astro-ph/0512082}}].

\bibitem{2006MNRAS.365..231V}
M.~{Viel} and M.G.~{Haehnelt}, \emph{{Cosmological and astrophysical parameters
  from the Sloan Digital Sky Survey flux power spectrum and hydrodynamical
  simulations of the Lyman {\ensuremath{\alpha}} forest}},
  \href{https://doi.org/10.1111/j.1365-2966.2005.09703.x}{\emph{MNRAS}
  {\bfseries 365} (2006) 231}
  [\href{https://arxiv.org/abs/astro-ph/0508177}{{\ttfamily
  astro-ph/0508177}}].

\bibitem{2006ApJS..163...80M}
P.~{McDonald}, U.~{Seljak}, S.~{Burles}, D.J.~{Schlegel}, D.H.~{Weinberg},
  R.~{Cen} et~al., \emph{{The Ly{\ensuremath{\alpha}} Forest Power Spectrum
  from the Sloan Digital Sky Survey}},
  \href{https://doi.org/10.1086/444361}{\emph{ApJS} {\bfseries 163} (2006) 80}
  [\href{https://arxiv.org/abs/astro-ph/0405013}{{\ttfamily
  astro-ph/0405013}}].

\bibitem{2011JCAP...09..001S}
A.~{Slosar}, A.~{Font-Ribera}, M.M.~{Pieri}, J.~{Rich}, J.-M.~{Le Goff},
  {\'E}.~{Aubourg} et~al., \emph{{The Lyman-{\ensuremath{\alpha}} forest in
  three dimensions: measurements of large scale flux correlations from BOSS
  1st-year data}},
  \href{https://doi.org/10.1088/1475-7516/2011/09/001}{\emph{JCAP} {\bfseries
  2011} (2011) 001} [\href{https://arxiv.org/abs/1104.5244}{{\ttfamily
  1104.5244}}].

\bibitem{2013JCAP...04..026S}
A.~{Slosar}, V.~{Ir{\v{s}}i{\v{c}}}, D.~{Kirkby}, S.~{Bailey}, N.G.~{Busca},
  T.~{Delubac} et~al., \emph{{Measurement of baryon acoustic oscillations in
  the Lyman-{\ensuremath{\alpha}} forest fluctuations in BOSS data release 9}},
  \href{https://doi.org/10.1088/1475-7516/2013/04/026}{\emph{JCAP} {\bfseries
  2013} (2013) 026} [\href{https://arxiv.org/abs/1301.3459}{{\ttfamily
  1301.3459}}].

\bibitem{2020ApJ...901..153D}
H.~{du Mas des Bourboux}, J.~{Rich}, A.~{Font-Ribera}, V.~{de Sainte Agathe},
  J.~{Farr}, T.~{Etourneau} et~al., \emph{{The Completed SDSS-IV Extended
  Baryon Oscillation Spectroscopic Survey: Baryon Acoustic Oscillations with
  Ly{\ensuremath{\alpha}} Forests}},
  \href{https://doi.org/10.3847/1538-4357/abb085}{\emph{ApJ} {\bfseries 901}
  (2020) 153} [\href{https://arxiv.org/abs/2007.08995}{{\ttfamily
  2007.08995}}].

\bibitem{2022arXiv220913942C}
A.~{Cuceu}, A.~{Font-Ribera}, S.~{Nadathur}, B.~{Joachimi} and P.~{Martini},
  \emph{{Constraints on the Cosmic Expansion Rate at Redshift 2.3 from the
  Lyman-{\ensuremath{\alpha}} Forest}},
  \href{https://doi.org/10.1103/PhysRevLett.130.191003}{\emph{\prl} {\bfseries
  130} (2023) 191003} [\href{https://arxiv.org/abs/2209.13942}{{\ttfamily
  2209.13942}}].

\bibitem{2013MNRAS.429.1734V}
M.~{Viel}, J.~{Schaye} and C.M.~{Booth}, \emph{{The impact of feedback from
  galaxy formation on the Lyman {\ensuremath{\alpha}} transmitted flux}},
  \href{https://doi.org/10.1093/mnras/sts465}{\emph{MNRAS} {\bfseries 429}
  (2013) 1734} [\href{https://arxiv.org/abs/1207.6567}{{\ttfamily 1207.6567}}].

\bibitem{2020MNRAS.495.1825C}
S.~{Chabanier}, F.~{Bournaud}, Y.~{Dubois}, N.~{Palanque-Delabrouille},
  C.~{Y{\`e}che}, E.~{Armengaud} et~al., \emph{{The impact of AGN feedback on
  the 1D power spectra from the Ly {\ensuremath{\alpha}} forest using the
  Horizon-AGN suite of simulations}},
  \href{https://doi.org/10.1093/mnras/staa1242}{\emph{MNRAS} {\bfseries 495}
  (2020) 1825} [\href{https://arxiv.org/abs/2002.02822}{{\ttfamily
  2002.02822}}].

\bibitem{2004MNRAS.354..684V}
M.~{Viel}, M.G.~{Haehnelt} and V.~{Springel}, \emph{{Inferring the dark matter
  power spectrum from the Lyman {\ensuremath{\alpha}} forest in high-resolution
  QSO absorption spectra}},
  \href{https://doi.org/10.1111/j.1365-2966.2004.08224.x}{\emph{MNRAS}
  {\bfseries 354} (2004) 684}
  [\href{https://arxiv.org/abs/astro-ph/0404600}{{\ttfamily
  astro-ph/0404600}}].

\bibitem{2005ApJ...635..761M}
P.~{McDonald}, U.~{Seljak}, R.~{Cen}, D.~{Shih}, D.H.~{Weinberg}, S.~{Burles}
  et~al., \emph{{The Linear Theory Power Spectrum from the
  Ly{\ensuremath{\alpha}} Forest in the Sloan Digital Sky Survey}},
  \href{https://doi.org/10.1086/497563}{\emph{ApJ} {\bfseries 635} (2005) 761}
  [\href{https://arxiv.org/abs/astro-ph/0407377}{{\ttfamily
  astro-ph/0407377}}].

\bibitem{2006MNRAS.370L..51V}
M.~{Viel}, M.G.~{Haehnelt} and A.~{Lewis}, \emph{{The Lyman
  {\ensuremath{\alpha}} forest and WMAP year three}},
  \href{https://doi.org/10.1111/j.1745-3933.2006.00187.x}{\emph{MNRAS}
  {\bfseries 370} (2006) L51}
  [\href{https://arxiv.org/abs/astro-ph/0604310}{{\ttfamily
  astro-ph/0604310}}].

\bibitem{2005PhRvD..71j3515S}
U.~{Seljak}, A.~{Makarov}, P.~{McDonald}, S.F.~{Anderson}, N.A.~{Bahcall},
  J.~{Brinkmann} et~al., \emph{{Cosmological parameter analysis including SDSS
  Ly{\ensuremath{\alpha}} forest and galaxy bias: Constraints on the primordial
  spectrum of fluctuations, neutrino mass, and dark energy}},
  \href{https://doi.org/10.1103/PhysRevD.71.103515}{\emph{Phys. Rev. D}
  {\bfseries 71} (2005) 103515}
  [\href{https://arxiv.org/abs/astro-ph/0407372}{{\ttfamily
  astro-ph/0407372}}].

\bibitem{2006JCAP...10..014S}
U.~{Seljak}, A.~{Slosar} and P.~{McDonald}, \emph{{Cosmological parameters from
  combining the Lyman-{\ensuremath{\alpha}} forest with CMB, galaxy clustering
  and SN constraints}},
  \href{https://doi.org/10.1088/1475-7516/2006/10/014}{\emph{JCAP} {\bfseries
  2006} (2006) 014} [\href{https://arxiv.org/abs/astro-ph/0604335}{{\ttfamily
  astro-ph/0604335}}].

\bibitem{2017JCAP...06..047Y}
C.~{Y{\`e}che}, N.~{Palanque-Delabrouille}, J.~{Baur} and H.~{du Mas des
  Bourboux}, \emph{{Constraints on neutrino masses from Lyman-alpha forest
  power spectrum with BOSS and XQ-100}},
  \href{https://doi.org/10.1088/1475-7516/2017/06/047}{\emph{JCAP} {\bfseries
  2017} (2017) 047} [\href{https://arxiv.org/abs/1702.03314}{{\ttfamily
  1702.03314}}].

\bibitem{2020JCAP...04..038P}
N.~{Palanque-Delabrouille}, C.~{Y{\`e}che}, N.~{Sch{\"o}neberg},
  J.~{Lesgourgues}, M.~{Walther}, S.~{Chabanier} et~al., \emph{{Hints, neutrino
  bounds, and WDM constraints from SDSS DR14 Lyman-{\ensuremath{\alpha}} and
  Planck full-survey data}},
  \href{https://doi.org/10.1088/1475-7516/2020/04/038}{\emph{JCAP} {\bfseries
  2020} (2020) 038} [\href{https://arxiv.org/abs/1911.09073}{{\ttfamily
  1911.09073}}].

\bibitem{2021JCAP...03..049G}
M.~{Garny}, T.~{Konstandin}, L.~{Sagunski} and M.~{Viel}, \emph{{Neutrino mass
  bounds from confronting an effective model with BOSS
  Lyman-{\ensuremath{\alpha}} data}},
  \href{https://doi.org/10.1088/1475-7516/2021/03/049}{\emph{JCAP} {\bfseries
  2021} (2021) 049} [\href{https://arxiv.org/abs/2011.03050}{{\ttfamily
  2011.03050}}].

\bibitem{2008MNRAS.386.1131B}
J.S.~{Bolton}, M.~{Viel}, T.S.~{Kim}, M.G.~{Haehnelt} and R.F.~{Carswell},
  \emph{{Possible evidence for an inverted temperature-density relation in the
  intergalactic medium from the flux distribution of the
  Ly{\ensuremath{\alpha}} forest}},
  \href{https://doi.org/10.1111/j.1365-2966.2008.13114.x}{\emph{MNRAS}
  {\bfseries 386} (2008) 1131}
  [\href{https://arxiv.org/abs/0711.2064}{{\ttfamily 0711.2064}}].

\bibitem{2014MNRAS.438.2499B}
J.S.~{Bolton}, G.D.~{Becker}, M.G.~{Haehnelt} and M.~{Viel}, \emph{{A
  consistent determination of the temperature of the intergalactic medium at
  redshift z = 2.4}}, \href{https://doi.org/10.1093/mnras/stt2374}{\emph{MNRAS}
  {\bfseries 438} (2014) 2499}
  [\href{https://arxiv.org/abs/1308.4411}{{\ttfamily 1308.4411}}].

\bibitem{2016MNRAS.463.2335N}
F.~{Nasir}, J.S.~{Bolton} and G.D.~{Becker}, \emph{{Inferring the IGM thermal
  history during reionization with the Lyman {\ensuremath{\alpha}} forest power
  spectrum at redshift $z \simeq 5$}},
  \href{https://doi.org/10.1093/mnras/stw2147}{\emph{MNRAS} {\bfseries 463}
  (2016) 2335} [\href{https://arxiv.org/abs/1605.04155}{{\ttfamily
  1605.04155}}].

\bibitem{2019ApJ...872...13W}
M.~{Walther}, J.~{O{\~n}orbe}, J.F.~{Hennawi} and Z.~{Luki{\'c}}, \emph{{New
  Constraints on IGM Thermal Evolution from the Ly{\ensuremath{\alpha}} Forest
  Power Spectrum}}, \href{https://doi.org/10.3847/1538-4357/aafad1}{\emph{ApJ}
  {\bfseries 872} (2019) 13}
  [\href{https://arxiv.org/abs/1808.04367}{{\ttfamily 1808.04367}}].

\bibitem{2019ApJ...872..101B}
E.~{Boera}, G.D.~{Becker}, J.S.~{Bolton} and F.~{Nasir}, \emph{{Revealing
  Reionization with the Thermal History of the Intergalactic Medium: New
  Constraints from the Ly{\ensuremath{\alpha}} Flux Power Spectrum}},
  \href{https://doi.org/10.3847/1538-4357/aafee4}{\emph{ApJ} {\bfseries 872}
  (2019) 101} [\href{https://arxiv.org/abs/1809.06980}{{\ttfamily
  1809.06980}}].

\bibitem{2019MNRAS.490.3177W}
X.~{Wu}, M.~{McQuinn}, R.~{Kannan}, A.~{D'Aloisio}, S.~{Bird}, F.~{Marinacci}
  et~al., \emph{{Imprints of temperature fluctuations on the z
  {\ensuremath{\sim}} 5 Lyman-{\ensuremath{\alpha}} forest: a view from
  radiation-hydrodynamic simulations of reionization}},
  \href{https://doi.org/10.1093/mnras/stz2807}{\emph{MNRAS} {\bfseries 490}
  (2019) 3177} [\href{https://arxiv.org/abs/1907.04860}{{\ttfamily
  1907.04860}}].

\bibitem{2021MNRAS.506.4389G}
P.~{Gaikwad}, R.~{Srianand}, M.G.~{Haehnelt} and T.R.~{Choudhury}, \emph{{A
  consistent and robust measurement of the thermal state of the IGM at 2
  {\ensuremath{\leq}} z {\ensuremath{\leq}} 4 from a large sample of Ly
  {\ensuremath{\alpha}} forest spectra: evidence for late and rapid He II
  reionization}}, \href{https://doi.org/10.1093/mnras/stab2017}{\emph{MNRAS}
  {\bfseries 506} (2021) 4389}
  [\href{https://arxiv.org/abs/2009.00016}{{\ttfamily 2009.00016}}].

\bibitem{2022ApJ...933...59V}
B.~{Villasenor}, B.~{Robertson}, P.~{Madau} and E.~{Schneider},
  \emph{{Inferring the Thermal History of the Intergalactic Medium from the
  Properties of the Hydrogen and Helium Ly{\ensuremath{\alpha}} Forest}},
  \href{https://doi.org/10.3847/1538-4357/ac704e}{\emph{ApJ} {\bfseries 933}
  (2022) 59} [\href{https://arxiv.org/abs/2111.00019}{{\ttfamily 2111.00019}}].

\bibitem{2005PhRvD..71f3534V}
M.~{Viel}, J.~{Lesgourgues}, M.G.~{Haehnelt}, S.~{Matarrese} and A.~{Riotto},
  \emph{{Constraining warm dark matter candidates including sterile neutrinos
  and light gravitinos with WMAP and the Lyman-{\ensuremath{\alpha}} forest}},
  \href{https://doi.org/10.1103/PhysRevD.71.063534}{\emph{Phys. Rev. D}
  {\bfseries 71} (2005) 063534}
  [\href{https://arxiv.org/abs/astro-ph/0501562}{{\ttfamily
  astro-ph/0501562}}].

\bibitem{2013PhRvD..88d3502V}
M.~{Viel}, G.D.~{Becker}, J.S.~{Bolton} and M.G.~{Haehnelt}, \emph{{Warm dark
  matter as a solution to the small scale crisis: New constraints from high
  redshift Lyman-{\ensuremath{\alpha}} forest data}},
  \href{https://doi.org/10.1103/PhysRevD.88.043502}{\emph{Phys. Rev. D}
  {\bfseries 88} (2013) 043502}
  [\href{https://arxiv.org/abs/1306.2314}{{\ttfamily 1306.2314}}].

\bibitem{2017PhRvD..96b3522I}
V.~{Ir{\v{s}}i{\v{c}}}, M.~{Viel}, M.G.~{Haehnelt}, J.S.~{Bolton},
  S.~{Cristiani}, G.D.~{Becker} et~al., \emph{{New constraints on the
  free-streaming of warm dark matter from intermediate and small scale
  Lyman-{\ensuremath{\alpha}} forest data}},
  \href{https://doi.org/10.1103/PhysRevD.96.023522}{\emph{Phys. Rev. D}
  {\bfseries 96} (2017) 023522}
  [\href{https://arxiv.org/abs/1702.01764}{{\ttfamily 1702.01764}}].

\bibitem{2021MNRAS.502.2356G}
A.~{Garzilli}, A.~{Magalich}, O.~{Ruchayskiy} and A.~{Boyarsky}, \emph{{How to
  constrain warm dark matter with the Lyman-{\ensuremath{\alpha}} forest}},
  \href{https://doi.org/10.1093/mnras/stab192}{\emph{MNRAS} {\bfseries 502}
  (2021) 2356} [\href{https://arxiv.org/abs/1912.09397}{{\ttfamily
  1912.09397}}].

\bibitem{2021PhRvL.126g1302R}
K.K.~{Rogers} and H.V.~{Peiris}, \emph{{Strong Bound on Canonical Ultralight
  Axion Dark Matter from the Lyman-Alpha Forest}},
  \href{https://doi.org/10.1103/PhysRevLett.126.071302}{\emph{Phys. Rev. L}
  {\bfseries 126} (2021) 071302}
  [\href{https://arxiv.org/abs/2007.12705}{{\ttfamily 2007.12705}}].

\bibitem{2022arXiv220914220V}
B.~{Villasenor}, B.~{Robertson}, P.~{Madau} and E.~{Schneider}, \emph{{New
  constraints on warm dark matter from the Lyman-{\ensuremath{\alpha}} forest
  power spectrum}},
  \href{https://doi.org/10.1103/PhysRevD.108.023502}{\emph{\prd} {\bfseries
  108} (2023) 023502} [\href{https://arxiv.org/abs/2209.14220}{{\ttfamily
  2209.14220}}].

\bibitem{2019JCAP...07..017C}
S.~{Chabanier}, N.~{Palanque-Delabrouille}, C.~{Y{\`e}che}, J.-M.~{Le Goff},
  E.~{Armengaud}, J.~{Bautista} et~al., \emph{{The one-dimensional power
  spectrum from the SDSS DR14 Ly{\ensuremath{\alpha}} forests}},
  \href{https://doi.org/10.1088/1475-7516/2019/07/017}{\emph{JCAP} {\bfseries
  2019} (2019) 017} [\href{https://arxiv.org/abs/1812.03554}{{\ttfamily
  1812.03554}}].

\bibitem{2022AJ....164..207A}
B.~{Abareshi}, J.~{Aguilar}, S.~{Ahlen}, S.~{Alam}, D.M.~{Alexander},
  R.~{Alfarsy} et~al., \emph{{Overview of the Instrumentation for the Dark
  Energy Spectroscopic Instrument}},
  \href{https://doi.org/10.3847/1538-3881/ac882b}{\emph{AJ} {\bfseries 164}
  (2022) 207} [\href{https://arxiv.org/abs/2205.10939}{{\ttfamily
  2205.10939}}].

\bibitem{2023arXiv230606316G}
N.~{G{\"o}ksel Kara{\c{c}}ayl{\i}}, P.~{Martini}, J.~{Guy}, C.~{Ravoux},
  M.L.A.~{Karim}, E.~{Armengaud} et~al., \emph{{Optimal 1D Ly$\alpha$ Forest
  Power Spectrum Estimation -- III. DESI early data}}, {\emph{arXiv e-prints}
  (2023) arXiv:2306.06316} [\href{https://arxiv.org/abs/2306.06316}{{\ttfamily
  2306.06316}}].

\bibitem{2023arXiv230606311R}
C.~{Ravoux}, M.L.A.~{Karim}, E.~{Armengaud}, M.~{Walther}, N.~{G{\"o}ksel
  Kara{\c{c}}ayl{\i}}, P.~{Martini} et~al., \emph{{The Dark Energy
  Spectroscopic Instrument: One-dimensional power spectrum from first
  Lyman-$\alpha$ forest samples with Fast Fourier Transform}}, {\emph{arXiv
  e-prints} (2023) arXiv:2306.06311}
  [\href{https://arxiv.org/abs/2306.06311}{{\ttfamily 2306.06311}}].

\bibitem{2016arXiv161100036D}
{DESI Collaboration}, A.~{Aghamousa}, J.~{Aguilar}, S.~{Ahlen}, S.~{Alam},
  L.E.~{Allen} et~al., \emph{{The DESI Experiment Part I: Science,Targeting,
  and Survey Design}}, {\emph{arXiv e-prints} (2016) arXiv:1611.00036}
  [\href{https://arxiv.org/abs/1611.00036}{{\ttfamily 1611.00036}}].

\bibitem{2017MNRAS.466.4332I}
V.~{Ir{\v{s}}i{\v{c}}}, M.~{Viel}, T.A.M.~{Berg}, V.~{D'Odorico},
  M.G.~{Haehnelt}, S.~{Cristiani} et~al., \emph{{The Lyman
  {\ensuremath{\alpha}} forest power spectrum from the XQ-100 Legacy Survey}},
  \href{https://doi.org/10.1093/mnras/stw3372}{\emph{MNRAS} {\bfseries 466}
  (2017) 4332} [\href{https://arxiv.org/abs/1702.01761}{{\ttfamily
  1702.01761}}].

\bibitem{2019MNRAS.489.2536D}
A.~{Day}, D.~{Tytler} and B.~{Kambalur}, \emph{{Power spectrum of the flux in
  the Lyman-alpha forest from high-resolution spectra of 87 QSOs}},
  \href{https://doi.org/10.1093/mnras/stz2214}{\emph{MNRAS} {\bfseries 489}
  (2019) 2536}.

\bibitem{2022MNRAS.509.2842K}
N.G.~{Kara{\c{c}}ayl{\i}}, N.~{Padmanabhan}, A.~{Font-Ribera},
  V.~{Ir{\v{s}}i{\v{c}}}, M.~{Walther}, D.~{Brooks} et~al., \emph{{Optimal 1D
  Ly {\ensuremath{\alpha}} forest power spectrum estimation - II. KODIAQ,
  SQUAD, and XQ-100}},
  \href{https://doi.org/10.1093/mnras/stab3201}{\emph{MNRAS} {\bfseries 509}
  (2022) 2842} [\href{https://arxiv.org/abs/2108.10870}{{\ttfamily
  2108.10870}}].

\bibitem{2022MNRAS.515..857E}
M.~{Esposito}, V.~{Ir{\v{s}}i{\v{c}}}, M.~{Costanzi}, S.~{Borgani}, A.~{Saro}
  and M.~{Viel}, \emph{{Weighing cosmic structures with clusters of galaxies
  and the intergalactic medium}},
  \href{https://doi.org/10.1093/mnras/stac1825}{\emph{\mnras} {\bfseries 515}
  (2022) 857} [\href{https://arxiv.org/abs/2202.00974}{{\ttfamily
  2202.00974}}].

\bibitem{2023simsuite}
S.~{Bird}, M.~{Fernandez}, M.-F.~{Ho}, M.~{Qezlou}, R.~{Monadi}, Y.~{Ni}
  et~al., \emph{{PRIYA: a new suite of Lyman-{\ensuremath{\alpha}} forest
  simulations for cosmology}},
  \href{https://doi.org/10.1088/1475-7516/2023/10/037}{\emph{\jcap} {\bfseries
  2023} (2023) 037} [\href{https://arxiv.org/abs/2306.05471}{{\ttfamily
  2306.05471}}].

\bibitem{2018MNRAS.475..676S}
V.~{Springel}, R.~{Pakmor}, A.~{Pillepich}, R.~{Weinberger}, D.~{Nelson},
  L.~{Hernquist} et~al., \emph{{First results from the IllustrisTNG
  simulations: matter and galaxy clustering}},
  \href{https://doi.org/10.1093/mnras/stx3304}{\emph{\mnras} {\bfseries 475}
  (2018) 676} [\href{https://arxiv.org/abs/1707.03397}{{\ttfamily
  1707.03397}}].

\bibitem{2022MNRAS.512.3703B}
S.~{Bird}, Y.~{Ni}, T.~{Di Matteo}, R.~{Croft}, Y.~{Feng} and N.~{Chen},
  \emph{{The ASTRID simulation: galaxy formation and reionization}},
  \href{https://doi.org/10.1093/mnras/stac648}{\emph{MNRAS} {\bfseries 512}
  (2022) 3703} [\href{https://arxiv.org/abs/2111.01160}{{\ttfamily
  2111.01160}}].

\bibitem{2022MNRAS.513..670N}
Y.~{Ni}, T.~{Di Matteo}, S.~{Bird}, R.~{Croft}, Y.~{Feng}, N.~{Chen} et~al.,
  \emph{{The ASTRID simulation: the evolution of supermassive black holes}},
  \href{https://doi.org/10.1093/mnras/stac351}{\emph{MNRAS} {\bfseries 513}
  (2022) 670} [\href{https://arxiv.org/abs/2110.14154}{{\ttfamily
  2110.14154}}].

\bibitem{2014JCAP...07..005B}
A.~{Borde}, N.~{Palanque-Delabrouille}, G.~{Rossi}, M.~{Viel}, J.S.~{Bolton},
  C.~{Y{\`e}che} et~al., \emph{{New approach for precise computation of
  Lyman-{\ensuremath{\alpha}} forest power spectrum with hydrodynamical
  simulations}},
  \href{https://doi.org/10.1088/1475-7516/2014/07/005}{\emph{JCAP} {\bfseries
  2014} (2014) 005} [\href{https://arxiv.org/abs/1401.6472}{{\ttfamily
  1401.6472}}].

\bibitem{2019JCAP...02..050B}
S.~{Bird}, K.K.~{Rogers}, H.V.~{Peiris}, L.~{Verde}, A.~{Font-Ribera} and
  A.~{Pontzen}, \emph{{An emulator for the Lyman-{\ensuremath{\alpha}}
  forest}}, \href{https://doi.org/10.1088/1475-7516/2019/02/050}{\emph{JCAP}
  {\bfseries 2019} (2019) 050}
  [\href{https://arxiv.org/abs/1812.04654}{{\ttfamily 1812.04654}}].

\bibitem{2022MNRAS.509.2551H}
M.-F.~{Ho}, S.~{Bird} and C.R.~{Shelton}, \emph{{Multifidelity emulation for
  the matter power spectrum using Gaussian processes}},
  \href{https://doi.org/10.1093/mnras/stab3114}{\emph{MNRAS} {\bfseries 509}
  (2022) 2551} [\href{https://arxiv.org/abs/2105.01081}{{\ttfamily
  2105.01081}}].

\bibitem{2022MNRAS.517.3200F}
M.A.~{Fernandez}, M.-F.~{Ho} and S.~{Bird}, \emph{{A multifidelity emulator for
  the Lyman-{\ensuremath{\alpha}} forest flux power spectrum}},
  \href{https://doi.org/10.1093/mnras/stac2435}{\emph{MNRAS} {\bfseries 517}
  (2022) 3200} [\href{https://arxiv.org/abs/2207.06445}{{\ttfamily
  2207.06445}}].

\bibitem{10.1093/biomet/87.1.1}
M.~Kennedy and A.~O'Hagan, \emph{{Predicting the output from a complex computer
  code when fast approximations are available}},
  \href{https://doi.org/10.1093/biomet/87.1.1}{\emph{Biometrika} {\bfseries 87}
  (2000) 1}
  [\href{https://arxiv.org/abs/https://academic.oup.com/biomet/article-pdf/87/1/1/590577/870001.pdf}{{\ttfamily
  https://academic.oup.com/biomet/article-pdf/87/1/1/590577/870001.pdf}}].

\bibitem{Heitmann:2009}
K.~{Heitmann}, D.~{Higdon}, M.~{White}, S.~{Habib}, B.J.~{Williams},
  E.~{Lawrence} et~al., \emph{{The Coyote Universe. II. Cosmological Models and
  Precision Emulation of the Nonlinear Matter Power Spectrum}},
  \href{https://doi.org/10.1088/0004-637X/705/1/156}{\emph{ApJ} {\bfseries 705}
  (2009) 156} [\href{https://arxiv.org/abs/0902.0429}{{\ttfamily 0902.0429}}].

\bibitem{Heitmann:2014}
K.~{Heitmann}, E.~{Lawrence}, J.~{Kwan}, S.~{Habib} and D.~{Higdon}, \emph{{The
  Coyote Universe Extended: Precision Emulation of the Matter Power Spectrum}},
  \href{https://doi.org/10.1088/0004-637X/780/1/111}{\emph{ApJ} {\bfseries 780}
  (2014) 111} [\href{https://arxiv.org/abs/1304.7849}{{\ttfamily 1304.7849}}].

\bibitem{Lawrence:2017}
E.~{Lawrence}, K.~{Heitmann}, J.~{Kwan}, A.~{Upadhye}, D.~{Bingham}, S.~{Habib}
  et~al., \emph{{The Mira-Titan Universe. II. Matter Power Spectrum
  Emulation}}, \href{https://doi.org/10.3847/1538-4357/aa86a9}{\emph{Astrophys.
  J. Let.} {\bfseries 847} (2017) 50}
  [\href{https://arxiv.org/abs/1705.03388}{{\ttfamily 1705.03388}}].

\bibitem{Giblin:2019}
B.~{Giblin}, M.~{Cataneo}, B.~{Moews} and C.~{Heymans}, \emph{{On the road to
  per cent accuracy - II. Calibration of the non-linear matter power spectrum
  for arbitrary cosmologies}},
  \href{https://doi.org/10.1093/mnras/stz2659}{\emph{MNRAS} {\bfseries 490}
  (2019) 4826} [\href{https://arxiv.org/abs/1906.02742}{{\ttfamily
  1906.02742}}].

\bibitem{Euclid:2021}
{Euclid Collaboration}, M.~{Knabenhans}, J.~{Stadel}, D.~{Potter}, J.~{Dakin},
  S.~{Hannestad} et~al., \emph{{Euclid preparation: IX. EuclidEmulator2 - power
  spectrum emulation with massive neutrinos and self-consistent dark energy
  perturbations}}, \href{https://doi.org/10.1093/mnras/stab1366}{\emph{MNRAS}
  {\bfseries 505} (2021) 2840}
  [\href{https://arxiv.org/abs/2010.11288}{{\ttfamily 2010.11288}}].

\bibitem{Arico:2021}
G.~{Aric{\`o}}, R.E.~{Angulo}, S.~{Contreras}, L.~{Ondaro-Mallea},
  M.~{Pellejero-Iba{\~n}ez} and M.~{Zennaro}, \emph{{The BACCO simulation
  project: a baryonification emulator with neural networks}},
  \href{https://doi.org/10.1093/mnras/stab1911}{\emph{MNRAS} {\bfseries 506}
  (2021) 4070} [\href{https://arxiv.org/abs/2011.15018}{{\ttfamily
  2011.15018}}].

\bibitem{Giri:2021}
S.K.~{Giri} and A.~{Schneider}, \emph{{Emulation of baryonic effects on the
  matter power spectrum and constraints from galaxy cluster data}},
  \href{https://doi.org/10.1088/1475-7516/2021/12/046}{\emph{JCAP} {\bfseries
  2021} (2021) 046} [\href{https://arxiv.org/abs/2108.08863}{{\ttfamily
  2108.08863}}].

\bibitem{Harnois:2019}
J.~{Harnois-D{\'e}raps}, B.~{Giblin} and B.~{Joachimi}, \emph{{Cosmic shear
  covariance matrix in wCDM: Cosmology matters}},
  \href{https://doi.org/10.1051/0004-6361/201935912}{\emph{Astron. Astrophys.}
  {\bfseries 631} (2019) A160}
  [\href{https://arxiv.org/abs/1905.06454}{{\ttfamily 1905.06454}}].

\bibitem{Davies:2021}
C.T.~{Davies}, M.~{Cautun}, B.~{Giblin}, B.~{Li}, J.~{Harnois-D{\'e}raps} and
  Y.-C.~{Cai}, \emph{{Constraining cosmology with weak lensing voids}},
  \href{https://doi.org/10.1093/mnras/stab2251}{\emph{MNRAS} {\bfseries 507}
  (2021) 2267} [\href{https://arxiv.org/abs/2010.11954}{{\ttfamily
  2010.11954}}].

\bibitem{McClintock:2019}
T.~{McClintock}, E.~{Rozo}, M.R.~{Becker}, J.~{DeRose}, Y.-Y.~{Mao},
  S.~{McLaughlin} et~al., \emph{{The Aemulus Project. II. Emulating the Halo
  Mass Function}}, \href{https://doi.org/10.3847/1538-4357/aaf568}{\emph{ApJ}
  {\bfseries 872} (2019) 53}
  [\href{https://arxiv.org/abs/1804.05866}{{\ttfamily 1804.05866}}].

\bibitem{Nishimichi:2019}
T.~{Nishimichi}, M.~{Takada}, R.~{Takahashi}, K.~{Osato}, M.~{Shirasaki},
  T.~{Oogi} et~al., \emph{{Dark Quest. I. Fast and Accurate Emulation of Halo
  Clustering Statistics and Its Application to Galaxy Clustering}},
  \href{https://doi.org/10.3847/1538-4357/ab3719}{\emph{ApJ} {\bfseries 884}
  (2019) 29} [\href{https://arxiv.org/abs/1811.09504}{{\ttfamily 1811.09504}}].

\bibitem{Bocquet:2022}
S.~{Bocquet}, K.~{Heitmann}, S.~{Habib}, E.~{Lawrence}, T.~{Uram},
  N.~{Frontiere} et~al., \emph{{The Mira-Titan Universe. III. Emulation of the
  Halo Mass Function}},
  \href{https://doi.org/10.3847/1538-4357/abac5c}{\emph{ApJ} {\bfseries 901}
  (2020) 5} [\href{https://arxiv.org/abs/2003.12116}{{\ttfamily 2003.12116}}].

\bibitem{Kern:2017}
N.S.~{Kern}, A.~{Liu}, A.R.~{Parsons}, A.~{Mesinger} and B.~{Greig},
  \emph{{Emulating Simulations of Cosmic Dawn for 21 cm Power Spectrum
  Constraints on Cosmology, Reionization, and X-Ray Heating}},
  \href{https://doi.org/10.3847/1538-4357/aa8bb4}{\emph{ApJ} {\bfseries 848}
  (2017) 23} [\href{https://arxiv.org/abs/1705.04688}{{\ttfamily 1705.04688}}].

\bibitem{Cohen:2020}
A.~{Cohen}, A.~{Fialkov}, R.~{Barkana} and R.A.~{Monsalve}, \emph{{Emulating
  the global 21-cm signal from Cosmic Dawn and Reionization}},
  \href{https://doi.org/10.1093/mnras/staa1530}{\emph{MNRAS} {\bfseries 495}
  (2020) 4845} [\href{https://arxiv.org/abs/1910.06274}{{\ttfamily
  1910.06274}}].

\bibitem{Bevins:2021}
H.T.J.~{Bevins}, W.J.~{Handley}, A.~{Fialkov}, E.~{de Lera Acedo} and
  K.~{Javid}, \emph{{GLOBALEMU: a novel and robust approach for emulating the
  sky-averaged 21-cm signal from the cosmic dawn and epoch of reionization}},
  \href{https://doi.org/10.1093/mnras/stab2737}{\emph{MNRAS} {\bfseries 508}
  (2021) 2923} [\href{https://arxiv.org/abs/2104.04336}{{\ttfamily
  2104.04336}}].

\bibitem{Bye:2022}
C.H.~{Bye}, S.K.N.~{Portillo} and A.~{Fialkov}, \emph{{21cmVAE: A Very Accurate
  Emulator of the 21 cm Global Signal}},
  \href{https://doi.org/10.3847/1538-4357/ac6424}{\emph{ApJ} {\bfseries 930}
  (2022) 79} [\href{https://arxiv.org/abs/2107.05581}{{\ttfamily 2107.05581}}].

\bibitem{Rogers:2019}
K.K.~{Rogers}, H.V.~{Peiris}, A.~{Pontzen}, S.~{Bird}, L.~{Verde} and
  A.~{Font-Ribera}, \emph{{Bayesian emulator optimisation for cosmology:
  application to the Lyman-alpha forest}},
  \href{https://doi.org/10.1088/1475-7516/2019/02/031}{\emph{JCAP} {\bfseries
  2019} (2019) 031} [\href{https://arxiv.org/abs/1812.04631}{{\ttfamily
  1812.04631}}].

\bibitem{2021JCAP...05..033P}
C.~{Pedersen}, A.~{Font-Ribera}, K.K.~{Rogers}, P.~{McDonald}, H.V.~{Peiris},
  A.~{Pontzen} et~al., \emph{{An emulator for the Lyman-{\ensuremath{\alpha}}
  forest in beyond-{\ensuremath{\Lambda}}CDM cosmologies}},
  \href{https://doi.org/10.1088/1475-7516/2021/05/033}{\emph{JCAP} {\bfseries
  2021} (2021) 033} [\href{https://arxiv.org/abs/2011.15127}{{\ttfamily
  2011.15127}}].

\bibitem{2021JCAP...04..059W}
M.~{Walther}, E.~{Armengaud}, C.~{Ravoux}, N.~{Palanque-Delabrouille},
  C.~{Y{\`e}che} and Z.~{Luki{\'c}}, \emph{{Simulating intergalactic gas for
  DESI-like small scale Lyman{\ensuremath{\alpha}} forest observations}},
  \href{https://doi.org/10.1088/1475-7516/2021/04/059}{\emph{JCAP} {\bfseries
  2021} (2021) 059} [\href{https://arxiv.org/abs/2012.04008}{{\ttfamily
  2012.04008}}].

\bibitem{Rogers:2021a}
K.K.~{Rogers} and H.V.~{Peiris}, \emph{{General framework for cosmological dark
  matter bounds using N -body simulations}},
  \href{https://doi.org/10.1103/PhysRevD.103.043526}{\emph{Phys. Rev. D}
  {\bfseries 103} (2021) 043526}
  [\href{https://arxiv.org/abs/2007.13751}{{\ttfamily 2007.13751}}].

\bibitem{2023MNRAS.tmp.2406C}
L.~{Cabayol-Garcia}, J.~{Chaves-Montero}, A.~{Font-Ribera} and C.~{Pedersen},
  \emph{{A neural network emulator for the Lyman-{\ensuremath{\alpha}} forest
  1D flux power spectrum}},
  \href{https://doi.org/10.1093/mnras/stad2512}{\emph{\mnras} (2023) }
  [\href{https://arxiv.org/abs/2305.19064}{{\ttfamily 2305.19064}}].

\bibitem{2019ApJ...874..154D}
A.~{D'Aloisio}, M.~{McQuinn}, O.~{Maupin}, F.B.~{Davies}, H.~{Trac},
  S.~{Fuller} et~al., \emph{{Heating of the Intergalactic Medium by Hydrogen
  Reionization}}, \href{https://doi.org/10.3847/1538-4357/ab0d83}{\emph{ApJ}
  {\bfseries 874} (2019) 154}
  [\href{https://arxiv.org/abs/1807.09282}{{\ttfamily 1807.09282}}].

\bibitem{Springel:2021}
V.~{Springel}, R.~{Pakmor}, O.~{Zier} and M.~{Reinecke}, \emph{{Simulating
  cosmic structure formation with the GADGET-4 code}},
  \href{https://doi.org/10.1093/mnras/stab1855}{\emph{\mnras} {\bfseries 506}
  (2021) 2871} [\href{https://arxiv.org/abs/2010.03567}{{\ttfamily
  2010.03567}}].

\bibitem{2020JCAP...06..002B}
S.~{Bird}, Y.~{Feng}, C.~{Pedersen} and A.~{Font-Ribera}, \emph{{More accurate
  simulations with separate initial conditions for baryons and dark matter}},
  \href{https://doi.org/10.1088/1475-7516/2020/06/002}{\emph{JCAP} {\bfseries
  2020} (2020) 002} [\href{https://arxiv.org/abs/2002.00015}{{\ttfamily
  2002.00015}}].

\bibitem{2020MNRAS.496.4372U}
P.~{Upton Sanderbeck} and S.~{Bird}, \emph{{Inhomogeneous He II reionization in
  hydrodynamic simulations}},
  \href{https://doi.org/10.1093/mnras/staa1850}{\emph{MNRAS} {\bfseries 496}
  (2020) 4372} [\href{https://arxiv.org/abs/2002.05733}{{\ttfamily
  2002.05733}}].

\bibitem{2007MNRAS.382.1657K}
T.S.~{Kim}, J.S.~{Bolton}, M.~{Viel}, M.G.~{Haehnelt} and R.F.~{Carswell},
  \emph{{An improved measurement of the flux distribution of the
  Ly{\ensuremath{\alpha}} forest in QSO absorption spectra: the effect of
  continuum fitting, metal contamination and noise properties}},
  \href{https://doi.org/10.1111/j.1365-2966.2007.12406.x}{\emph{MNRAS}
  {\bfseries 382} (2007) 1657}
  [\href{https://arxiv.org/abs/0711.1862}{{\ttfamily 0711.1862}}].

\bibitem{2017ascl.soft10012B}
S.~{Bird}, \emph{{FSFE: Fake Spectra Flux Extractor}},  Oct., 2017.

\bibitem{2015MNRAS.447.1834B}
S.~{Bird}, M.~{Haehnelt}, M.~{Neeleman}, S.~{Genel}, M.~{Vogelsberger} and
  L.~{Hernquist}, \emph{{Reproducing the kinematics of damped Lyman
  {\ensuremath{\alpha}} systems}},
  \href{https://doi.org/10.1093/mnras/stu2542}{\emph{MNRAS} {\bfseries 447}
  (2015) 1834} [\href{https://arxiv.org/abs/1407.7858}{{\ttfamily 1407.7858}}].

\bibitem{2021arXiv211013293P}
A.~{Paleyes}, M.~{Pullin}, M.~{Mahsereci}, C.~{McCollum}, N.D.~{Lawrence} and
  J.~{Gonzalez}, \emph{{Emulation of physical processes with Emukit}},
  {\emph{arXiv e-prints} (2021) arXiv:2110.13293}
  [\href{https://arxiv.org/abs/2110.13293}{{\ttfamily 2110.13293}}].

\bibitem{2021JCAP...05..057T}
J.~{Torrado} and A.~{Lewis}, \emph{{Cobaya: code for Bayesian analysis of
  hierarchical physical models}},
  \href{https://doi.org/10.1088/1475-7516/2021/05/057}{\emph{JCAP} {\bfseries
  2021} (2021) 057} [\href{https://arxiv.org/abs/2005.05290}{{\ttfamily
  2005.05290}}].

\bibitem{2019ascl.soft10019T}
J.~{Torrado} and A.~{Lewis}, ``{Cobaya: Bayesian analysis in cosmology}.''
  Astrophysics Source Code Library, record ascl:1910.019, Oct., 2019.

\bibitem{2013AJ....145...10D}
K.S.~{Dawson}, D.J.~{Schlegel}, C.P.~{Ahn}, S.F.~{Anderson}, {\'E}.~{Aubourg},
  S.~{Bailey} et~al., \emph{{The Baryon Oscillation Spectroscopic Survey of
  SDSS-III}}, \href{https://doi.org/10.1088/0004-6256/145/1/10}{\emph{AJ}
  {\bfseries 145} (2013) 10} [\href{https://arxiv.org/abs/1208.0022}{{\ttfamily
  1208.0022}}].

\bibitem{2016AJ....151...44D}
K.S.~{Dawson}, J.-P.~{Kneib}, W.J.~{Percival}, S.~{Alam}, F.D.~{Albareti},
  S.F.~{Anderson} et~al., \emph{{The SDSS-IV Extended Baryon Oscillation
  Spectroscopic Survey: Overview and Early Data}},
  \href{https://doi.org/10.3847/0004-6256/151/2/44}{\emph{AJ} {\bfseries 151}
  (2016) 44} [\href{https://arxiv.org/abs/1508.04473}{{\ttfamily 1508.04473}}].

\bibitem{2018MNRAS.474.3032R}
K.K.~{Rogers}, S.~{Bird}, H.V.~{Peiris}, A.~{Pontzen}, A.~{Font-Ribera} and
  B.~{Leistedt}, \emph{{Simulating the effect of high column density absorbers
  on the one-dimensional Lyman {\ensuremath{\alpha}} forest flux power
  spectrum}}, \href{https://doi.org/10.1093/mnras/stx2942}{\emph{MNRAS}
  {\bfseries 474} (2018) 3032}
  [\href{https://arxiv.org/abs/1706.08532}{{\ttfamily 1706.08532}}].

\bibitem{2017AJ....154..114O}
J.M.~{O'Meara}, N.~{Lehner}, J.C.~{Howk}, J.X.~{Prochaska}, A.J.~{Fox},
  M.S.~{Peeples} et~al., \emph{{The Second Data Release of the KODIAQ Survey}},
  \href{https://doi.org/10.3847/1538-3881/aa82b8}{\emph{AJ} {\bfseries 154}
  (2017) 114} [\href{https://arxiv.org/abs/1707.07905}{{\ttfamily
  1707.07905}}].

\bibitem{2023ApJ...944..223P}
C.~{Pedersen}, A.~{Font-Ribera} and N.Y.~{Gnedin}, \emph{{Compressing the
  Cosmological Information in One-dimensional Correlations of the
  Lyman-{\ensuremath{\alpha}} Forest}},
  \href{https://doi.org/10.3847/1538-4357/acb433}{\emph{\apj} {\bfseries 944}
  (2023) 223} [\href{https://arxiv.org/abs/2209.09895}{{\ttfamily
  2209.09895}}].

\bibitem{2013PhRvD..87j3529L}
A.~{Lewis}, \emph{{Efficient sampling of fast and slow cosmological
  parameters}}, \href{https://doi.org/10.1103/PhysRevD.87.103529}{\emph{Phys.
  Rev. D} {\bfseries 87} (2013) 103529}
  [\href{https://arxiv.org/abs/1304.4473}{{\ttfamily 1304.4473}}].

\bibitem{2002PhRvD..66j3511L}
A.~{Lewis} and S.~{Bridle}, \emph{{Cosmological parameters from CMB and other
  data: A Monte Carlo approach}},
  \href{https://doi.org/10.1103/PhysRevD.66.103511}{\emph{Phys. Rev. D}
  {\bfseries 66} (2002) 103511}
  [\href{https://arxiv.org/abs/astro-ph/0205436}{{\ttfamily
  astro-ph/0205436}}].

\bibitem{Montero:2019}
P.~{Montero-Camacho}, C.M.~{Hirata}, P.~{Martini} and K.~{Honscheid},
  \emph{{Impact of inhomogeneous reionization on the
  Lyman-{\ensuremath{\alpha}} forest}},
  \href{https://doi.org/10.1093/mnras/stz1388}{\emph{\mnras} {\bfseries 487}
  (2019) 1047} [\href{https://arxiv.org/abs/1902.02892}{{\ttfamily
  1902.02892}}].

\bibitem{2013MNRAS.430.2067B}
G.D.~{Becker}, P.C.~{Hewett}, G.~{Worseck} and J.X.~{Prochaska}, \emph{{A
  refined measurement of the mean transmitted flux in the
  Ly{\ensuremath{\alpha}} forest over \ensuremath{2 < z < 5} using composite
  quasar spectra}}, \href{https://doi.org/10.1093/mnras/stt031}{\emph{MNRAS}
  {\bfseries 430} (2013) 2067}
  [\href{https://arxiv.org/abs/1208.2584}{{\ttfamily 1208.2584}}].

\bibitem{2020A&A...641A...6P}
{Planck Collaboration}, N.~{Aghanim}, Y.~{Akrami}, M.~{Ashdown}, J.~{Aumont},
  C.~{Baccigalupi} et~al., \emph{{Planck 2018 results. VI. Cosmological
  parameters}},
  \href{https://doi.org/10.1051/0004-6361/201833910}{\emph{Astron. Astrophys.}
  {\bfseries 641} (2020) A6}
  [\href{https://arxiv.org/abs/1807.06209}{{\ttfamily 1807.06209}}].

\bibitem{2011arXiv1104.2932L}
J.~{Lesgourgues}, \emph{{The Cosmic Linear Anisotropy Solving System (CLASS) I:
  Overview}}, {\emph{arXiv e-prints} (2011) arXiv:1104.2932}
  [\href{https://arxiv.org/abs/1104.2932}{{\ttfamily 1104.2932}}].

\bibitem{2023arXiv230405202Q}
F.J.~{Qu}, B.D.~{Sherwin}, M.S.~{Madhavacheril}, D.~{Han}, K.T.~{Crowley},
  I.~{Abril-Cabezas} et~al., \emph{{The Atacama Cosmology Telescope: A
  Measurement of the DR6 CMB Lensing Power Spectrum and its Implications for
  Structure Growth}},
  \href{https://doi.org/10.48550/arXiv.2304.05202}{\emph{arXiv e-prints} (2023)
  arXiv:2304.05202} [\href{https://arxiv.org/abs/2304.05202}{{\ttfamily
  2304.05202}}].

\bibitem{2022PhRvD.105b3520A}
T.M.C.~{Abbott}, M.~{Aguena}, A.~{Alarcon}, S.~{Allam}, O.~{Alves}, A.~{Amon}
  et~al., \emph{{Dark Energy Survey Year 3 results: Cosmological constraints
  from galaxy clustering and weak lensing}},
  \href{https://doi.org/10.1103/PhysRevD.105.023520}{\emph{\prd} {\bfseries
  105} (2022) 023520} [\href{https://arxiv.org/abs/2105.13549}{{\ttfamily
  2105.13549}}].

\bibitem{2020JCAP...05..042I}
M.M.~{Ivanov}, M.~{Simonovi{\'c}} and M.~{Zaldarriaga}, \emph{{Cosmological
  parameters from the BOSS galaxy power spectrum}},
  \href{https://doi.org/10.1088/1475-7516/2020/05/042}{\emph{\jcap} {\bfseries
  2020} (2020) 042} [\href{https://arxiv.org/abs/1909.05277}{{\ttfamily
  1909.05277}}].

\bibitem{2022JHEAp..34...49A}
E.~{Abdalla}, G.F.~{Abell{\'a}n}, A.~{Aboubrahim}, A.~{Agnello},
  {\"O}.~{Akarsu}, Y.~{Akrami} et~al., \emph{{Cosmology intertwined: A review
  of the particle physics, astrophysics, and cosmology associated with the
  cosmological tensions and anomalies}},
  \href{https://doi.org/10.1016/j.jheap.2022.04.002}{\emph{Journal of High
  Energy Astrophysics} {\bfseries 34} (2022) 49}
  [\href{https://arxiv.org/abs/2203.06142}{{\ttfamily 2203.06142}}].

\bibitem{2023JCAP...04..057Y}
B.~{Yu}, U.~{Seljak}, Y.~{Li} and S.~{Singh}, \emph{{RSD measurements from BOSS
  galaxy power spectrum using the halo perturbation theory model}},
  \href{https://doi.org/10.1088/1475-7516/2023/04/057}{\emph{\jcap} {\bfseries
  2023} (2023) 057} [\href{https://arxiv.org/abs/2211.16794}{{\ttfamily
  2211.16794}}].

\bibitem{2016ApJ...825..144W}
G.~{Worseck}, J.X.~{Prochaska}, J.F.~{Hennawi} and M.~{McQuinn}, \emph{{Early
  and Extended Helium Reionization over More Than 600 Million Years of Cosmic
  Time}}, \href{https://doi.org/10.3847/0004-637X/825/2/144}{\emph{\apj}
  {\bfseries 825} (2016) 144}
  [\href{https://arxiv.org/abs/1405.7405}{{\ttfamily 1405.7405}}].

\bibitem{2021ApJ...912...38M}
K.~{Makan}, G.~{Worseck}, F.B.~{Davies}, J.F.~{Hennawi}, J.X.~{Prochaska} and
  P.~{Richter}, \emph{{New Evidence for Extended He II Reionization at z
  {\ensuremath{\gtrsim}} 3.5 from He II Lyman Alpha and Beta Transmission
  Spikes}}, \href{https://doi.org/10.3847/1538-4357/abee17}{\emph{\apj}
  {\bfseries 912} (2021) 38}
  [\href{https://arxiv.org/abs/2012.07876}{{\ttfamily 2012.07876}}].

\bibitem{2009ApJ...704L..89M}
M.~{McQuinn}, \emph{{The Implications of Gunn-Peterson Troughs in the He II
  Ly{\ensuremath{\alpha}} Forest}},
  \href{https://doi.org/10.1088/0004-637X/704/2/L8910.48550/arXiv.0905.0481}{\emph{ApJL}
  {\bfseries 704} (2009) L89}
  [\href{https://arxiv.org/abs/0905.0481}{{\ttfamily 0905.0481}}].

\bibitem{2011ApJ...733L..24W}
G.~{Worseck}, J.X.~{Prochaska}, M.~{McQuinn}, A.~{Dall'Aglio}, C.~{Fechner},
  J.F.~{Hennawi} et~al., \emph{{The End of Helium Reionization at \ensuremath{z
  \sim 2.7} Inferred from Cosmic Variance in HST/COS He II
  Ly{\ensuremath{\alpha}} Absorption Spectra}},
  \href{https://doi.org/10.1088/2041-8205/733/2/L2410.48550/arXiv.1103.5752}{\emph{ApJL}
  {\bfseries 733} (2011) L24}
  [\href{https://arxiv.org/abs/1103.5752}{{\ttfamily 1103.5752}}].

\bibitem{2019ApJ...875..111W}
G.~{Worseck}, F.B.~{Davies}, J.F.~{Hennawi} and J.X.~{Prochaska}, \emph{{The
  Evolution of the He II-ionizing Background at Redshifts \ensuremath{2.3 < z <
  3.8} Inferred from a Statistical Sample of 24 HST/COS He II
  Ly{\ensuremath{\alpha}} Absorption Spectra}},
  \href{https://doi.org/10.3847/1538-4357/ab0fa1}{\emph{ApJ]} {\bfseries 875}
  (2019) 111} [\href{https://arxiv.org/abs/1808.05247}{{\ttfamily
  1808.05247}}].

\bibitem{2021ApJ...919..120M}
A.M.~{Morales}, C.A.~{Mason}, S.~{Bruton}, M.~{Gronke}, F.~{Haardt} and
  C.~{Scarlata}, \emph{{The Evolution of the Lyman-alpha Luminosity Function
  during Reionization}},
  \href{https://doi.org/10.3847/1538-4357/ac110410.48550/arXiv.2101.01205}{\emph{ApJ}
  {\bfseries 919} (2021) 120}
  [\href{https://arxiv.org/abs/2101.01205}{{\ttfamily 2101.01205}}].

\bibitem{Ho:2020}
M.-F.~{Ho}, S.~{Bird} and R.~{Garnett}, \emph{{Detecting Multiple DLAs per
  Spectrum in SDSS DR12 with Gaussian Processes}},
  \href{https://doi.org/10.1093/mnras/staa1806}{\emph{\mnras} (2020) }
  [\href{https://arxiv.org/abs/2003.11036}{{\ttfamily 2003.11036}}].

\bibitem{Parks:2018}
D.~{Parks}, J.X.~{Prochaska}, S.~{Dong} and Z.~{Cai}, \emph{{Deep learning of
  quasar spectra to discover and characterize damped Ly{\ensuremath{\alpha}}
  systems}}, \href{https://doi.org/10.1093/mnras/sty196}{\emph{\mnras}
  {\bfseries 476} (2018) 1151}
  [\href{https://arxiv.org/abs/1709.04962}{{\ttfamily 1709.04962}}].

\bibitem{2022arXiv220100827W}
B.~{Wang}, J.~{Zou}, Z.~{Cai}, J.X.~{Prochaska}, Z.~{Sun}, J.~{Ding} et~al.,
  \emph{{Deep Learning of Dark Energy Spectroscopic Instrument Mock Spectra to
  Find Damped Ly{\ensuremath{\alpha}} Systems}},
  \href{https://doi.org/10.3847/1538-4365/ac4504}{\emph{\apjs} {\bfseries 259}
  (2022) 28}.

\bibitem{2013A&A...559A..85P}
N.~{Palanque-Delabrouille}, C.~{Y{\`e}che}, A.~{Borde}, J.-M.~{Le Goff},
  G.~{Rossi}, M.~{Viel} et~al., \emph{{The one-dimensional
  Ly{\ensuremath{\alpha}} forest power spectrum from BOSS}},
  \href{https://doi.org/10.1051/0004-6361/201322130}{\emph{Astron. Astrophys.}
  {\bfseries 559} (2013) A85}
  [\href{https://arxiv.org/abs/1306.5896}{{\ttfamily 1306.5896}}].

\bibitem{2015JCAP...11..011P}
N.~{Palanque-Delabrouille}, C.~{Y{\`e}che}, J.~{Baur}, C.~{Magneville},
  G.~{Rossi}, J.~{Lesgourgues} et~al., \emph{{Neutrino masses and cosmology
  with Lyman-alpha forest power spectrum}},
  \href{https://doi.org/10.1088/1475-7516/2015/11/011}{\emph{JCAP} {\bfseries
  2015} (2015) 011} [\href{https://arxiv.org/abs/1506.05976}{{\ttfamily
  1506.05976}}].

\bibitem{2022MNRAS.516.5355A}
A.~{Amon} and G.~{Efstathiou}, \emph{{A non-linear solution to the S$_{8}$
  tension?}}, \href{https://doi.org/10.1093/mnras/stac2429}{\emph{\mnras}
  {\bfseries 516} (2022) 5355}
  [\href{https://arxiv.org/abs/2206.11794}{{\ttfamily 2206.11794}}].

\bibitem{2010MNRAS.402.1536S}
J.~{Schaye}, C.~{Dalla Vecchia}, C.M.~{Booth}, R.P.C.~{Wiersma}, T.~{Theuns},
  M.R.~{Haas} et~al., \emph{{The physics driving the cosmic star formation
  history}},
  \href{https://doi.org/10.1111/j.1365-2966.2009.16029.x}{\emph{\mnras}
  {\bfseries 402} (2010) 1536}
  [\href{https://arxiv.org/abs/0909.5196}{{\ttfamily 0909.5196}}].

\bibitem{Prochaska:2009a}
J.X.~{Prochaska}, G.~{Worseck} and J.M.~{O'Meara}, \emph{{A Direct Measurement
  of the Intergalactic Medium Opacity to H I Ionizing Photons}},
  \href{https://doi.org/10.1088/0004-637X/705/2/L113}{\emph{\apjl} {\bfseries
  705} (2009) L113} [\href{https://arxiv.org/abs/0910.0009}{{\ttfamily
  0910.0009}}].

\bibitem{Worseck:2011}
G.~{Worseck} and J.X.~{Prochaska}, \emph{{GALEX Far-ultraviolet Color Selection
  of UV-bright High-redshift Quasars}},
  \href{https://doi.org/10.1088/0004-637X/728/1/23}{\emph{\apj} {\bfseries 728}
  (2011) 23} [\href{https://arxiv.org/abs/1004.3347}{{\ttfamily 1004.3347}}].

\bibitem{Fumagalli:2013}
M.~{Fumagalli}, J.M.~{O'Meara}, J.X.~{Prochaska} and G.~{Worseck},
  \emph{{Dissecting the Properties of Optically Thick Hydrogen at the Peak of
  Cosmic Star Formation History}},
  \href{https://doi.org/10.1088/0004-637X/775/1/78}{\emph{\apj} {\bfseries 775}
  (2013) 78} [\href{https://arxiv.org/abs/1308.1101}{{\ttfamily 1308.1101}}].

\bibitem{2015JCAP...02..045P}
N.~{Palanque-Delabrouille}, C.~{Y{\`e}che}, J.~{Lesgourgues}, G.~{Rossi},
  A.~{Borde}, M.~{Viel} et~al., \emph{{Constraint on neutrino masses from
  SDSS-III/BOSS Ly{\ensuremath{\alpha}} forest and other cosmological probes}},
  \href{https://doi.org/10.1088/1475-7516/2015/02/045}{\emph{\jcap} {\bfseries
  2015} (2015) 045} [\href{https://arxiv.org/abs/1410.7244}{{\ttfamily
  1410.7244}}].

\bibitem{2011MNRAS.413.1717B}
S.~{Bird}, H.V.~{Peiris}, M.~{Viel} and L.~{Verde}, \emph{{Minimally parametric
  power spectrum reconstruction from the Lyman {\ensuremath{\alpha}} forest}},
  \href{https://doi.org/10.1111/j.1365-2966.2011.18245.x}{\emph{\mnras}
  {\bfseries 413} (2011) 1717}
  [\href{https://arxiv.org/abs/1010.1519}{{\ttfamily 1010.1519}}].

\bibitem{2023arXiv231116377R}
K.K.~{Rogers} and V.~{Poulin}, \emph{{$5 \sigma$ tension between Planck cosmic
  microwave background and eBOSS Lyman-alpha forest and constraints on physics
  beyond $\Lambda$CDM}},
  \href{https://doi.org/10.48550/arXiv.2311.16377}{\emph{arXiv e-prints} (2023)
  arXiv:2311.16377} [\href{https://arxiv.org/abs/2311.16377}{{\ttfamily
  2311.16377}}].

\bibitem{2020JCAP...04..025P}
C.~{Pedersen}, A.~{Font-Ribera}, T.D.~{Kitching}, P.~{McDonald}, S.~{Bird},
  A.~{Slosar} et~al., \emph{{Massive neutrinos and degeneracies in Lyman-alpha
  forest simulations}},
  \href{https://doi.org/10.1088/1475-7516/2020/04/025}{\emph{JCAP} {\bfseries
  2020} (2020) 025} [\href{https://arxiv.org/abs/1911.09596}{{\ttfamily
  1911.09596}}].

\bibitem{2023arXiv230402038G}
P.~{Gaikwad}, M.G.~{Haehnelt}, F.B.~{Davies}, S.E.I.~{Bosman}, M.~{Molaro},
  G.~{Kulkarni} et~al., \emph{{Measuring the photoionization rate, neutral
  fraction, and mean free path of H I ionizing photons at 4.9
  {\ensuremath{\leq}} z {\ensuremath{\leq}} 6.0 from a large sample of XShooter
  and ESI spectra}},
  \href{https://doi.org/10.1093/mnras/stad2566}{\emph{\mnras} {\bfseries 525}
  (2023) 4093} [\href{https://arxiv.org/abs/2304.02038}{{\ttfamily
  2304.02038}}].

\end{thebibliography}\endgroup

\end{document}